\documentclass[12pt,twoside,a4paper]{book}

\usepackage[T1]{fontenc}
\usepackage[english]{babel}
\usepackage[latin1]{inputenc}
\usepackage[pdftex]{graphicx}           
\usepackage{setspace}                   
\usepackage{indentfirst}                
\usepackage{makeidx}                    
\usepackage[nottoc]{tocbibind}          
\usepackage{courier}                    
\usepackage{float}
\usepackage{type1cm}                    
\usepackage{listings}                   
\usepackage{titletoc}
\usepackage{bm}
\usepackage[fixlanguage]{babelbib}
\usepackage[font=small,format=plain,labelfont=bf,up,textfont=it,up]{caption}
\usepackage[font=small,format=plain]{subcaption}
\usepackage[usenames,svgnames,dvipsnames]{xcolor}
\usepackage[a4paper,top=4cm,bottom=4.0cm,left=4.0cm,right=4cm]{geometry} 
\usepackage[pdftex,plainpages=false,pdfpagelabels,pagebackref,colorlinks=true,citecolor=DarkGreen,linkcolor=NavyBlue,urlcolor=DarkRed,filecolor=green,bookmarksopen=true]{hyperref} 
\usepackage[all]{hypcap}                    
\usepackage[round,sort,nonamebreak]{natbib} 

\fontsize{60}{62}\usefont{OT1}{cmr}{m}{n}{\selectfont}

\def\Eset{{\textnormal{E}}}

\def\Pr{{\textnormal{Pr}}}

\def\d{{\textnormal{d}}}
\newcommand{\Cint}[2]{\displaystyle \subset\hspace{-0.4cm}\int\limits_{#1}^{#2}}
\newcommand{\Dprodaux}[1]{\displaystyle \textnormal{D}\hspace{-0.55cm}\prod\limits_{#1}}
\newcommand{\Dprodauxa}[1]{\displaystyle \textnormal{D}\hspace{-0.50cm}\prod\limits_{#1}}
\newcommand{\Dprod}[2]{\displaystyle \hspace{0.35cm}{\mathop{\Dprodaux{#1}}}^{#2}}
\newcommand{\Dproda}[2]{\displaystyle \hspace{0.35cm}{\mathop{\Dprodauxa{#1}}}^{#2}}

\newtheorem{Theorem}{Theorem}
\newtheorem{Definition}{Definition}

\newtheorem{Property}{Property}

\usepackage{fancyhdr}
\graphicspath{{./figuras/}}             
\makeindex                              
\fontsize{60}{62}\usefont{OT1}{cmr}{m}{n}{\selectfont}
\cleardoublepage
\normalsize

\begin{document}
	
\frontmatter 
\fancyhead[RO]{{\footnotesize\rightmark}\hspace{2em}\thepage}
\setcounter{tocdepth}{2}
\fancyhead[LE]{\thepage\hspace{2em}\footnotesize{\leftmark}}
\fancyhead[RE,LO]{}
\fancyhead[RO]{{\footnotesize\rightmark}\hspace{2em}\thepage}

\onehalfspacing  

\thispagestyle{empty}
\begin{center}
    \vspace*{5cm}
    \textbf{\Large{Reliability Estimation in Coherent Systems}}\\
    
    \vspace*{8cm}
    \Large{Agatha Sacramento Rodrigues \\
    	Carlos Alberto de Bragan\c{c}a Pereira \\
    	Adriano Polpo.}
    
    \vskip 10cm
    
    \vskip 1.5cm

   	\vskip 1cm

    \vskip 0.5cm
\end{center}

%

\tableofcontents    

%


\mainmatter

\fancyhead[RE,LO]{\thesection}

\doublespacing             
\chapter*{Preface} 
\label{preface}

Usually, methods evaluating system reliability require engineers to quantify the reliability of each of the system components. For series and parallel systems, there are some options to handle the estimation of each component's reliability. We will treat the reliability estimation of complex problems of two classes of coherent systems: series-parallel, and parallel-series. In both of the cases, the component reliabilities may be unknown. We will present estimators for reliability functions at all levels of the system (component and system reliabilities). Nonparametric Bayesian estimators of all sub-distribution and distribution functions are derived, and a Dirichlet multivariate process as a prior distribution is presented. Parametric estimator of the component's reliability based on Weibull model is presented for any kind of system. Also, some ideas in systems with masked data are discussed.

\newpage 

\vspace{20cm}

This is a first version of the manuscript. We are sure that is necessary many improvements to this text. We are improving, and probably we will have a new version soon. 
												 \begin{flushright}
                                                May 23, 2018.\\
                                                
                                                     Agatha Sacramento Rodrigues \\
                                                     Carlos Alberto de Bragan\c{c}a Pereira \\
                                                     Adriano Polpo.
                                                 \end{flushright}                 
                                                                
\chapter{Introduction} 
\label{introduction}

In engineering, the quality of the produced system is of great interest. In this sense, the reliability study has been object of research in last years. 
Among important works, \citet{BProschan} is highlight about theory of reliability. Barlow and Proschan's book shows concepts and theories about reliability, presenting important results when components and system reliabilities are known. In situations where component's reliability is unknown, statistical inference can be suitable considered to estimate system and components reliabilities and it will be discussed in the sequel of this book. 

As a motivation, the reliability estimation of an automatic coffee machine is presented. The machine has two causes of failure: 1. the failure of the component that grinds the grain; or 2. the failure of the heating water component. Clearly, the failure of any component, 1 or 2, leads the coffee machine failure. The failure of a component implies that the possible future failure time of the other becomes invisible, i.e. a censored data. Statistical inference for the reliability of the machine depends on both marginal components models. The reliability study of components allows for one-off actions on the components that need to be improved in order to maximize system performance, rather than changing the entire system, generating less costs, time and unnecessary effort. Hence, inferences for both components are needed.

Statistical inference of component reliability is not an easy task: censorship, dependence and unequal distributions are some of the troubles.  Considering a sample of the coffee machine example for which all $n$ sample units are observed up to death. Every sample unit will produce a component failure time and a censored failure time to the other component.  Both components failing at the same time is considered unlikely in such situations. In this example, the sample will produce $n$ failure times observations and $n$ censored times for the two components in test.  Relative to component failure time, it is reasonable to say that the two components are not identically distributed: probably one of the components may suffer more censors than the other.  It is common that only one component is responsible for the system failure at time $t$, implying that all the remaining components are censored also at time $t$, although the types of censor could be different. In general, the number of censored observations should be higher than the uncensored ones. 

\begin{figure}[!h]
	\centering
	\includegraphics[width=0.5\linewidth]{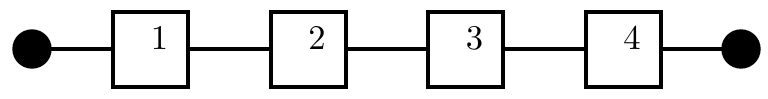}
	\caption{Series system with four components.}
	\label{series}
\end{figure}

The reliabilities of a system and its components also depend on the structure of the system, that is, the way that components are interconnected. The coffee machine is a series system of two components, a simple case known as competing risks problem. The illustration of a system structure can be considered by what is known as the block diagram, where each component of the system is illustrated by a block.  
Figure \ref{series} is the block diagram of a series system with four components -  at the time the system fails only one component is uncensored and the other three components are right-censored at the system failure time, that is, they still could continue to work after system fail.

Suppose a system of $m$ components and $X_j$ denoting the failure time of the $j$th component, $j=1,\ldots,m$. Let $T$ the random variable that represents the system failure time. The function that defines $T$ in relation to $X_1,\ldots,X_m$ depends on the system structure. For a system with the structure represented in Figure \ref{series}, for instance, $m=4$ and $T=\min\{X_1,X_2,X_3,X_4\}$.  

Consider initially that a random sample of $n$ systems with the structure in Figure \ref{series} is observed and ${\bf t}=(t_1,t_2,\ldots,t_{n})$ being a sample of the random variable $T$. 
The goal is to estimate the reliabilities of components involved in this series structure. At the system failure, however, not all components would have their failure time observed. In addition, a particular component may be responsible for system failures in some sample units and not in the remaining ones, cases of right-censored observations.

When a system fails, the failure time of a given component $j$ may not be observed, but its censored time of failure is. For all sample units, the system failure times $t_1,\ldots$ and $t_n$ are recorded. Associated to each sample unit, let $\delta_{i}$ be the indicator of the component whose failure produced the system to fail, with $i=1,\ldots,n$. At the time a series system fails, a given component can only be uncensored (responsible for system failure), that is $\delta_{i}=j$, or right-censored at the system failure time, that is $\delta_{i}\neq j$, for $j=1,\ldots,m$.

The data of $n=10$ observed systems are presented in Table \ref{apeA.serie}. For instance, system ID=1 failed at time $1.92$ and component 1 is the first to fail, that is, $\delta_{1}=1$ and the others components are right-censored at $1.92$.

\begin{table}[h!]
	\centering
	\caption{Observed data of $n=10$ series system with $m=4$ components.}
	\begin{tabular}{ ccc }
		\hline
		System ID & t     & $\delta$  \\
		\hline
		1 &	1.92 & 1    \\
		2 &	1.85 & 2   \\
		3 & 2.00 & 4   \\
		4 &	1.74 & 3  \\
		5 &	1.41 & 1     \\
		6 &	1.97 & 2   \\
		7 &	1.65 & 3   \\
		8 &	2.08 & 1   \\
		9 &	1.74 & 4   \\
		10 &	2.40 & 2    \\
		\hline
	\end{tabular}%
	\label{apeA.serie}%
\end{table}%

Considering a parametric model, let $R(x|\mbox{\boldmath{$\theta_j$}})={\rm P}(X_j>x |\mbox{\boldmath{$\theta_j$}})$ and $f(x|\mbox{\boldmath{$\theta_j$}})$ the reliability and density functions, respectively, and $\mbox{\boldmath{$\theta_j$}}$ is the parameter that can be either a scalar or a vector. For $j$th component, the likelihood function can be writen as 
\begin{eqnarray}
{\rm L}(\mbox{\boldmath{$\theta_j$}} \mid {\bf t},\mbox{\boldmath{$\delta$}}) = \prod_{i=1}^n  {\Big[f(t_{i}|\mbox{\boldmath{$\theta_j$}})\Big]}^{{\rm I}_{\{\delta_{i}=j\}}} {\Big[R(t_{i}|\mbox{\boldmath{$\theta_j$}})\Big]}^{1-{\rm I}_{\{\delta_{i}=j\}}},  \label{veros_serie}
\end{eqnarray}
where  ${\rm I}_{\{TRUE\}}=1$ or ${\rm I}_{\{FALSE\}}=0$,  $\mbox{\boldmath{$\delta$}}=(\delta_{1},\ldots,\delta_{n})$ and $j=1,\ldots,m$. 

\begin{figure}[!h]
	\centering
	\includegraphics[width=0.3\linewidth]{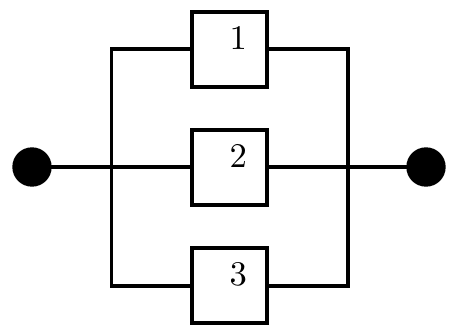}
	\caption{Parallel system with three components.}
	\label{parallel}
\end{figure}

A parallel system as in Figure \ref{parallel} works whenever at least one component is working. Again, only one component has its failure time uncensored, the other components are left-censored observations, that is, they had failed before system failure.  For a system with the structure represented in Figure \ref{parallel},  $m=3$ and $T=\max\{X_1,X_2,X_3\}$.  

Consider that a random sample of $n=7$ systems with the structure in \ref{parallel} is observed and ${\bf t}=(t_1,t_2,\ldots,t_{7})$ being a sample of the random variable $T$.  Associated to each sample unit, let $\delta_{i}$ be the indicator of the component whose failure produced the $i$th system to fail, for $j=1,2,3$ and $i=1,\ldots,7$. At the time a parallel system fails, a given component can only be uncensored (the last component to fail), that is $\delta_{i}=j$, or left-censored at the system failure time, that is, $\delta_{i}\neq j$, for $j=1,\ldots,m$. 

The parallel systems data are presented in Table \ref{apeA.paralelo}. For instance, system ID=3 failed at time $4.93$ and component 2 is the last to fail, that is, $\delta_{3}=2$ and the others components are left-censored at $4.93$.

\begin{table}[h!]
	\centering
	\caption{Observed data of $n=7$ parallel system with $m=3$ components.}
	\begin{tabular}{ ccc }
		\hline
		System ID & t     & $\delta$ \\
		\hline
		1 & 0.18 & 1        \\
		2 & 	1.15 & 3         \\
		3 & 4.93 & 2    \\ 
		4 & 0.01 & 2           \\
		5 & 1.01 & 1     \\
		6 & 1.51 & 3         \\
		7 & 1.74 & 2         \\
		\hline
	\end{tabular}%
	\label{apeA.paralelo}%
\end{table}%

For $j$th component and considering a parametric model, the likelihood function can be writen as
\begin{eqnarray}
{\rm L}(\mbox{\boldmath{$\theta_j$}} \mid {\bf t},\mbox{\boldmath{$\delta$}}) = \prod_{i=1}^n  {\Big[f(t_{i}|\mbox{\boldmath{$\theta_j$}})\Big]}^{{\rm I}_{\{\delta_{i}=j\}}} {\Big[1- R(t_{i}|\mbox{\boldmath{$\theta_j$}})\Big]}^{1-{\rm I}_{\{\delta_{i}=j\}}},  \label{veros_paralelo}
\end{eqnarray}
where  ${\rm I}_{\{TRUE\}}=1$ or ${\rm I}_{\{FALSE\}}=0$,  $\mbox{\boldmath{$\delta$}}=(\delta_{1},\ldots,\delta_{n})$ and $j=1,\ldots,m$.  

The literature on reliability of either parallel or series systems is abundant; different solutions have been presented. \citet{VHS1997}, \citet{SalinasTorres}, \citet{PolpoCarlinhos} and \\ \citet{PolpoSinha} discussed the Bayesian nonparametric statistics for series and parallel systems. Under Weibull probability distributions, Bayesian inferences for system and component reliabilities were introduced by \citet{CoqueJr} and \citet{Bhering} presented a hierarchical Bayesian Weibull model for components' reliability estimation in series and parallel systems, proposing an useful computational approach. Using simulation for series systems, \citet{AgathaIC}, considering Weibull families, compared three estimation types: Kaplan-Meier, Maximum Likelihood and Bayesian Plug-in Estimators. \citet{PolpoSinhaSimoniCAB} performed a comparative study about Bayesian estimation of a survival curve. 

Series and parallel systems are particular cases of a class of system called coherent. A system is said to be coherent if all components are relevant, that is, 
all components play any role in the functional capacity of the system, and the structure function is nondecreasing in each component, that is, the system will not become worse than before if a failed component is replaced by another that works. 

%

\begin{figure}[!h]\centering
	\begin{minipage}[H]{0.26\linewidth}
		\includegraphics[width=\linewidth]{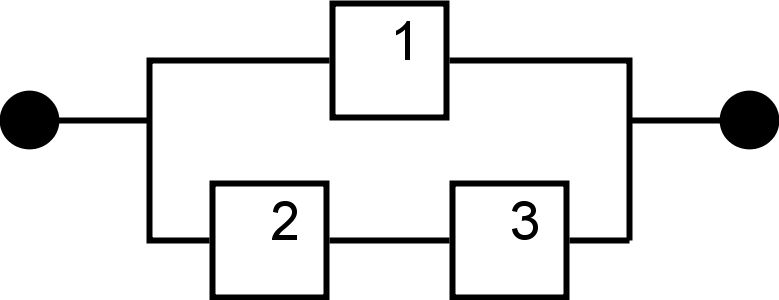}
		\subcaption{ $\mbox{   }$ }  \label{Fig_1_para_SPS}
	\end{minipage} 
	\begin{minipage}[H]{0.32\linewidth}
		\includegraphics[width=\linewidth]{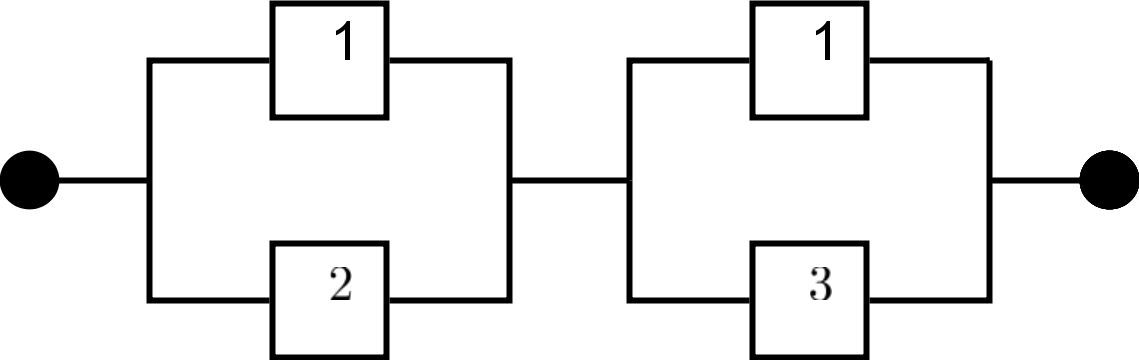}
		\subcaption{ $\mbox{   }$ } \label{SPS_para_Fig1}
	\end{minipage}
	\caption{(a) PSS  (b) SPS representation of system in (a).}
	\label{Fig1_SPS}
\end{figure}

\begin{figure}[!h]\centering
	\begin{minipage}[H]{0.28\linewidth}
		\includegraphics[width=\linewidth]{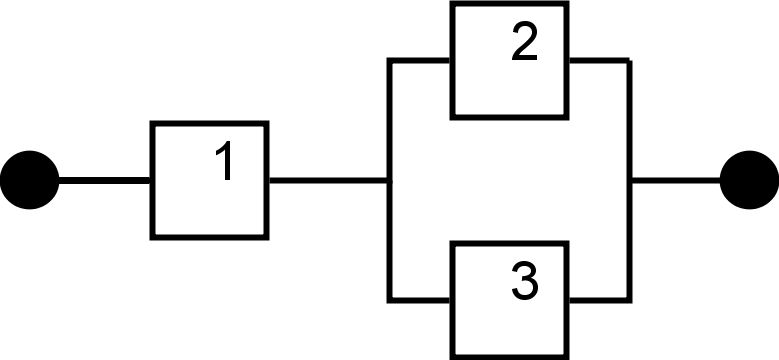}
		\subcaption{ $\mbox{   }$ }  \label{Fig_2_para_PSS}
	\end{minipage} 
	\begin{minipage}[H]{0.3\linewidth}
		\includegraphics[width=\linewidth]{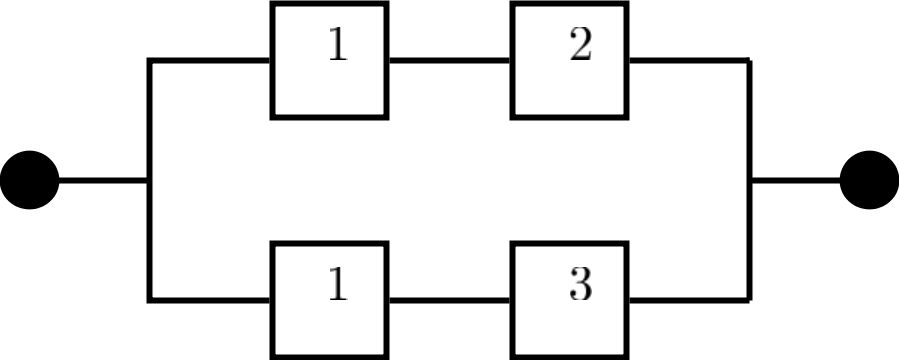}
		\subcaption{ $\mbox{   }$ } \label{PSS_para_Fig2}
	\end{minipage}
	\caption{(a) SPS (b) PSS representation of system in (a).}
	\label{Fig2_PSS}
\end{figure}

An important property is presented by \citet{BProschan} that every coherent system can be written as series-parallel system (SPS) representation and as parallel-series system (PSS) representation. Consider the PSS in Figure \ref{Fig_1_para_SPS} and Figure \ref{SPS_para_Fig1} presents its SPS representation. The SPS in Figure \ref{Fig_2_para_PSS} has its PSS representation in Figure \ref{PSS_para_Fig2}.

Considering this celebrated property, \citet{PolpoSiCar} introduced Bayesian nonparametric statistics for a class of coherent system in order to estimate components reliabilities. They restricted themselves to cases for which no component appears more than twice in parallel-series and series-parallel representations, under assumption that two components or more can not fail at the same instant of time. However, it is common that, in the representation of the system, some components appear in two different places within it. For instance, consider again the representation in Figure \ref{SPS_para_Fig1} (or Figure \ref{PSS_para_Fig2}). We have the reliabilities of four components to estimate. However, two of them are in fact the same component (component $1$), and they will fail at the same time, which violates the assumption. For this reason, it is important to have the estimators for both SPS and PSS that give a wide variety of representations. If one of these representations does not violate the assumptions, then the proposed Bayesian nonparametric estimator can be used. This nonparametric approach is presented with details in Chapter \ref{nonparametric}. 

\begin{figure}[!h]\centering
	\centering
	\includegraphics[width=0.4\linewidth]{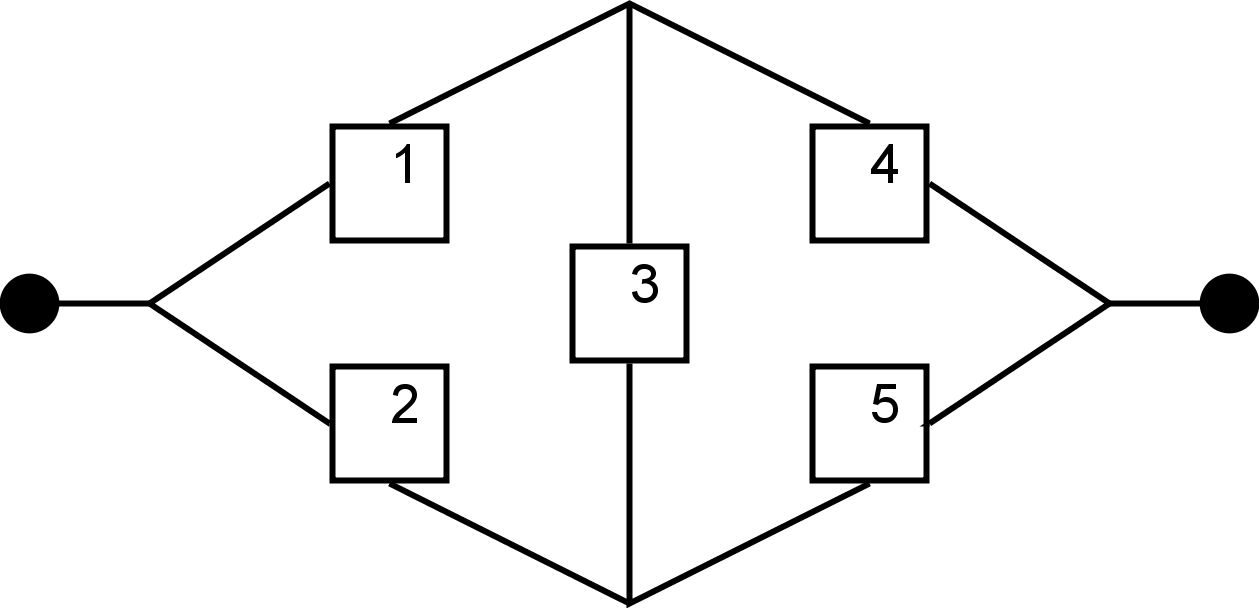}
	\caption{Bridge structure.}
	\label{fig:bridgenew}
\end{figure}

\begin{figure}[!h]\centering
	\begin{minipage}[H]{0.45\linewidth}
		\includegraphics[width=\linewidth]{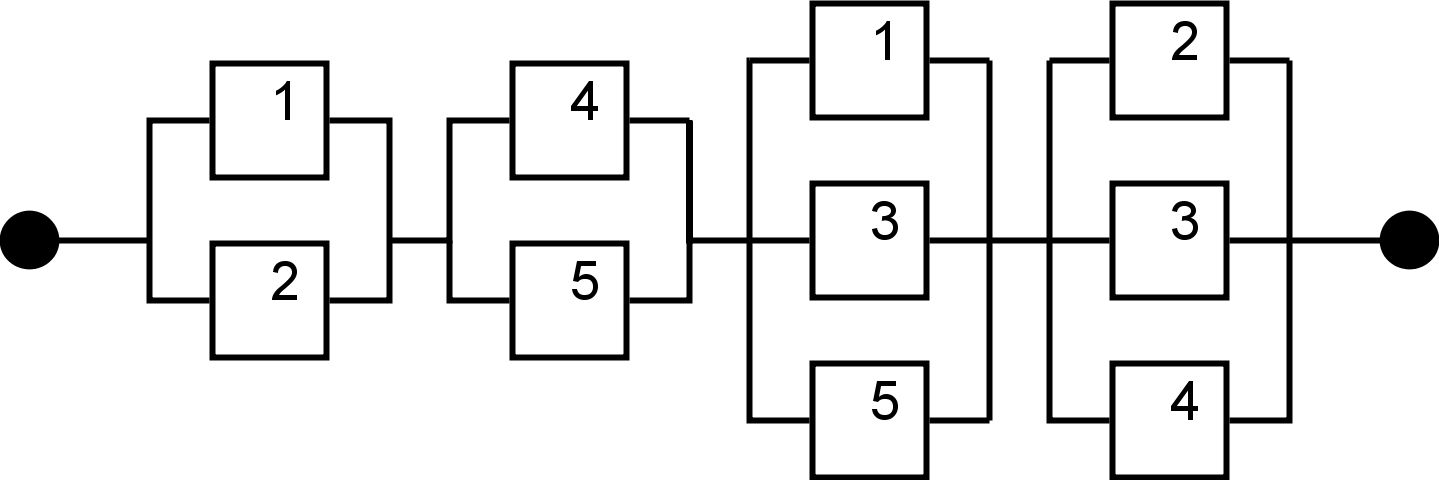}
		\subcaption{ $\mbox{   }$ } \label{bridge_SPS}
	\end{minipage}  
	\begin{minipage}[H]{0.3\linewidth}
		\includegraphics[width=\linewidth]{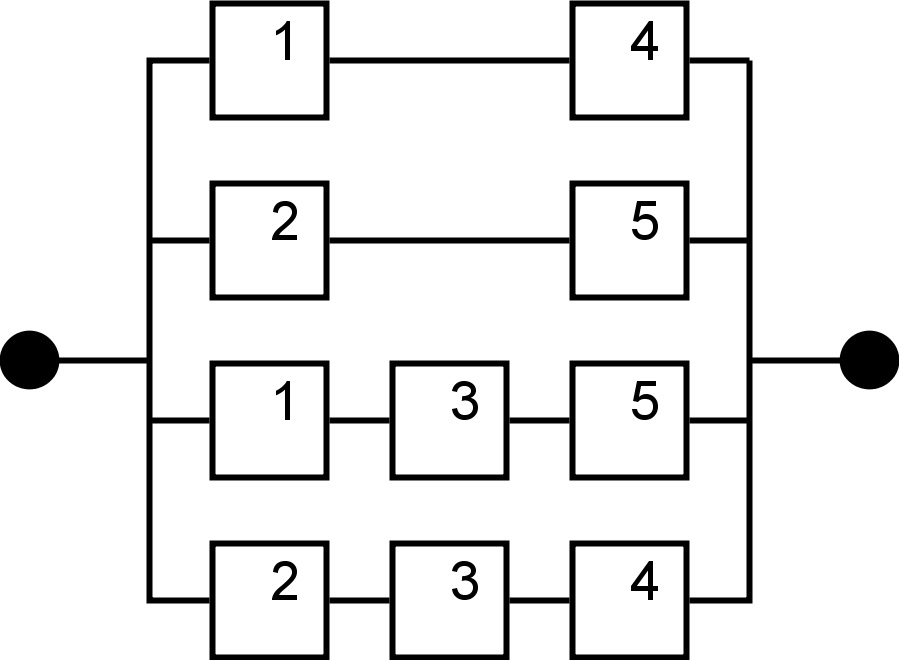}
		\subcaption{ $\mbox{   }$ } \label{bridge_PSS}
	\end{minipage}
	\caption{(a) SPS bridge representation (b) PSS bridge representation.}
	\label{system_bridge_SPS_PSS}
\end{figure}

\begin{figure}[!h]\centering
	\begin{minipage}[H]{0.35\linewidth}
		\includegraphics[width=\linewidth]{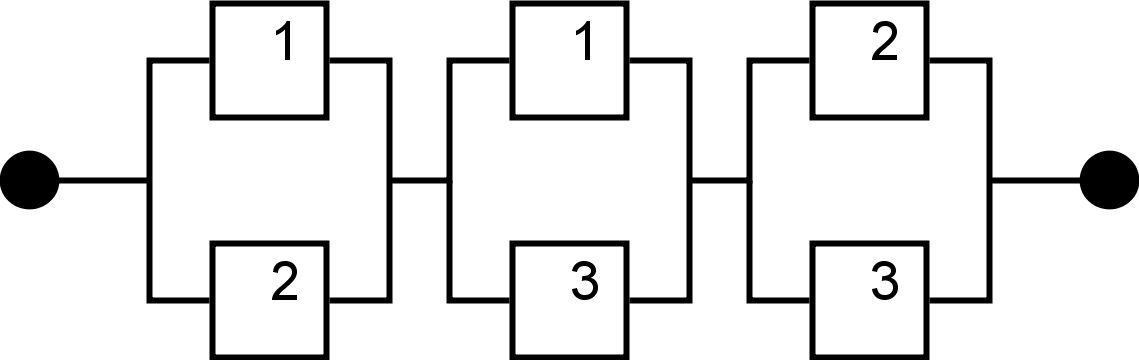}
		\subcaption{ $\mbox{   }$ }  \label{2out3_SPS}
	\end{minipage} 
	\begin{minipage}[H]{0.30\linewidth}
		\includegraphics[width=\linewidth]{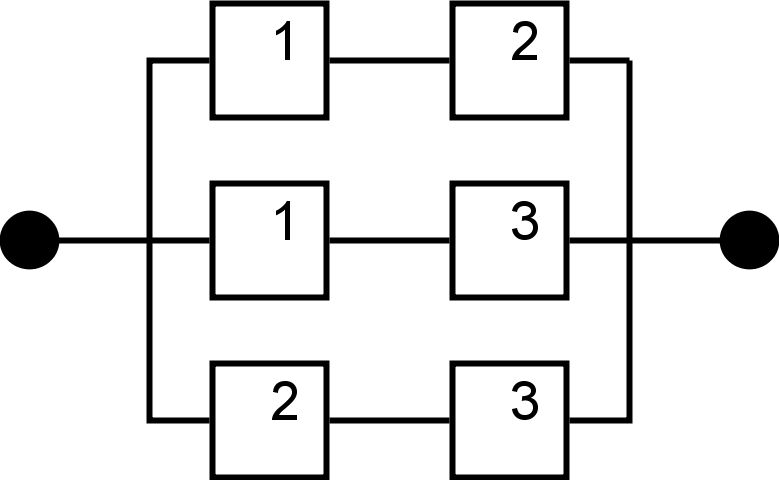}
		\subcaption{ $\mbox{   }$ } \label{2out3_PSS}
	\end{minipage}
	\caption{(a) SPS $2$-out-of-$3$ representation (b) PSS $2$-out-of-$3$ representation.}
	\label{system_2de3_SPS_PSS}
\end{figure}

Figure \ref{fig:bridgenew} is the bridge system described in the literature \\ \citep{BProschan} and Figure \ref{system_bridge_SPS_PSS} illustrates its SPS and PSS representations. Note that each of the five components appears twice for both representations. Another interesting structure is the $k$-out-of-$m$ system - work only if at least $k$ out of the $m$ components work. For instance, Figure \ref{system_2de3_SPS_PSS} considers the simple $2$-out-of-$3$ case into SPS and PSS representations. Note that each of the three components also appears twice in both combinations. Situations like these violate \citet{PolpoSiCar} assumption and their approach is not suitable anymore both for the PSS representation and for SPS representation. Thus, a solution to estimate the reliability of components in systems such as Figures \ref{fig:bridgenew} and \ref{system_2de3_SPS_PSS} is to consider the parametric approach and the general likelihood function is developed in the sequel. 

In likelihood functions (\ref{veros_serie}) and (\ref{veros_paralelo}) the $j$th component is susceptible only to right-censored or only to left-censored data, respectively. For a more general case, a component can be susceptible to both side of censoring. For instance, system $2$-out-of-$3$ (Figure \ref{system_2de3_SPS_PSS}) - system work if at least 2 out of 3 components work. Consider that a system $2$-out-of-$3$ is observed and component 1 failed first and component 3 failed in the sequence. At the moment of component 3 failure, the system failed. Thus, component 3 is uncensored, component 1 is left-censored data and component 2 is right-censored observation. Another $2$-out-of-$3$ system is observed but for this system, component 2 was the first to fail (left-censored), component 1 was the last to fail (uncensored) and component 3 yet worked in system failure (right-censored). Note that each component is susceptible to be uncensored, left or right-censored failure time.

Another kind of censoring could also occur:  suppose a machine failure time (a sample unit) is in an interval $(l,u) ~ -$  $l$ for the observed lower limit and $u$ for the upper limit.  If two or more components failed, they are all interval censored in $(l,u)$. 

To generalize the notation for all cases of component failure and censoring, consider the following notation: for a specific component $j$ of the system unit $i$, let $(l_{ji}, u_{ji})$ be a general interval of time, in which
\begin{itemize}
	\item $l_{ji}=u_{ji}=t_i$, if the $j$-th component failure time causes the $i$-th system failure time;
	\item $l_{ji}=t_i$ and $u_{ji}=\infty$, if the $j$-th component is right-censored at $t_i$;
	\item $l_{ji}=0$ and $u_{ji}=t_i$ , if the $j$-th component is left-censored at $t_i$;
	\item $0 < l_{ji} < u_{ji} < \infty$, if the $j$-th component is interval-censored.
\end{itemize} 

Consider that a random sample of $n=10$ systems with the structure in Figure \ref{system_2de3_SPS_PSS} is observed. The data are presented in Table \ref{apeA.2de3}.
For instance, system ID=2 failed at time $2.09$ and the failure of component 2 causes the system failure ($l_{22}=2.09$ and $u_{22}=2.09$), component 1 had failed before time $2.09$ ($l_{12}=0$ and $u_{12}=2.09$) and component 3 is right-censored at time $2.09$ ($l_{32}=2.09$ and $u_{32}=\infty$).

\begin{table}[htbp]
	\centering
	\caption{Observed data of $n=10$ $2$-out-of-$3$ systems.}
	\begin{tabular}{c|cc|cc|cc}
		\hline
		System ID  & \multicolumn{2}{c|}{Component 1} & \multicolumn{2}{c|}{Component 2} & \multicolumn{2}{c}{Component 3} \\
		& l     & u     & l     & u     & l     & u \\
		\hline
		1     & 1.95  & 1.95  & 1.95  & $\infty$ & 0     & 1.95 \\
		2     & 0     & 2.09  & 2.09  & 2.09  & 2.09  &   $\infty$ \\
		3     & 3.56  &   $\infty$ & 3.56  & 3.56  & 0     & 3.56 \\
		4     & 2.55  &   $\infty$ & 0     & 2.55  & 2.55  & 2.55 \\
		5     & 1.89  & 1.89  & 1.89  &   $\infty$ & 0     & 1.89 \\
		6     & 3.01  &   $\infty$ & 0     & 3.01  & 3.01  & 3.01 \\
		7     & 2.43  & 2.43  & 0     & 2.43  & 2.43  &   $\infty$\\
		8     & 0     & 1.51  & 1.51  & 1.51  & 1.51  &   $\infty$ \\
		9     & 3.55  & 3.55  & 3.55  &   $\infty$ & 0     & 3.55 \\
		10    & 2.35  &   $\infty$ & 0     & 2.35  & 2.35  & 2.35 \\
		\hline
	\end{tabular}%
	\label{apeA.2de3}%
\end{table}%

 To complete the theoretical environment, let $f(\cdot|\mbox{\boldmath{$\theta_j$}})$ and $R(\cdot|\mbox{\boldmath{$\theta_j$}})$ the density and reliability functions, respectively, and $\mbox{\boldmath{$\theta_j$}}$ is the parameter that can be either a scalar or a vector. Using the above notation, the likelihood function is as follow:
\begin{eqnarray}
{\rm L}(\mbox{\boldmath{$\theta_j$}} \mid {\bf l_{j}},{\bf u_{j}}) = \prod_{i=1}^n  {\Big[f(l_{ji}|\mbox{\boldmath{$\theta_j$}})\Big]}^{{\rm I}_{\{l_{ji}=u_{ji}\}}} {\Big[R(l_{ji}|\mbox{\boldmath{$\theta_j$}})
	-R(u_{ji}|\mbox{\boldmath{$\theta_j$}})\Big]}^{1-{\rm I}_{\{l_{ji}=u_{ji}\}}} \label{vero}
\end{eqnarray}
where ${\rm I}_{\{TRUE\}}=1$ or ${\rm I}_{\{FALSE\}}=0$; ${\bf l_{j}}=(l_{j1},\ldots,l_{jn})$ and \\ ${\bf u_{j}}=(u_{j1},\ldots,u_{jn})$.

A parametric approach which considers the likelihood function (\ref{vero}) for estimation of components' reliabilities involved in any kind of coherent system, from the simplest to the most complex structures, is presented in Chapter \ref{weibull_model}. The available information are the failure time of system and the status of each component at system failure instant. This approach does not need the supposition of identically distributed components lifetimes. The main assumption is that components' lifetimes are mutually independent and the components lifetime distributions are the three-parameter Weibull, a very general distribution that can approximate most of the lifetimes distributions. The paradigm is the Bayesian one. Another advantage of the Weibull is that, in our paradigm, even with improper priors, the posterior distributions turn out to be proper. The presented mechanism of calculus can well be used for any other family of distributions whenever proper priors are used.


Chapters \ref{nonparametric} and \ref{weibull_model} will address the problem of component estimation in coherent systems under the Bayesian paradigm in nonparametric and parametric approaches, respectively. In both chapter the status of each component at the time of system failure is considered to be known. However, identifying which component fail caused the failure of a given system can be a difficult task. In special situations, we can only establish that failed components belong to small sets of components. Cases like this are known as masked data failure cause and it is usually due to limited resources for the diagnosis of the cause of the failure. 
As an example of masked data problem, \cite{BasuBasuMuk} cited situations of failures of large computer systems, where the analysis is often performed in such a way that a small subset of components is identified as the cause of failure. In an attempt to repair the system as quickly as possible, the entire subset of components is replaced and the component responsible for the failure can not be investigated.

In Chapter \ref{masked_failure} the masked data problem formulation is developed and a Bayesian three-parameter Weibull model for components' reliabilities in masked data scenario is presented. This model is general and it can be considered for components involved in any coherent system. 

Finally, the datasets considered in Chapters \ref{nonparametric}, \ref{weibull_model} and \ref{masked_failure} are presented in Appendixs \ref{apeD}, \ref{apenB} and \ref{apeC}, respectively.

\chapter{Nonparametric} 
\label{nonparametric}

A nonparametric estimator for all the reliability functions involved in the series-parallel system (SPS) and parallel-series system (PSS) under the assumptions that the components reliabilities are unknown is presented; the only available information are the failure times of the system and the component that produced the failure. The required assumptions are mutually independent components failure times and that two or more components cannot fail at same instant of time. 


In Section \ref{sec-prob} are presented the probability results necessary for the development of the estimator. Section \ref{sec_bayes} is devoted to the construction of the nonparametric Bayesian estimator for SPS and PSS with three components ($m = 3$). In Section \ref{sec_general} the results are extended to a more general case of $m \geq 4$; and in Section \ref{sec_num-ex} the estimator is used in simulated datasets and illustrated its qualities. 


\section{Probability relations}
\label{sec-prob}
In this section, important results and properties of the PSS and SPS are presented. Before it, we present these results for series and parallel systems that will facilitate the understanding of the results for PSS and SPS, once series and parallel systems are simpler.

\subsection{Series and parallel systems}
First consider a parallel system with $m$ components. Let $X_j$ be the failure time of $j$th component with marginal distribution function (DF) $F_j$ and $T=\max\{X_1,\ldots,X_m\}$ be the system failure time. The indicator of the component whose failure produced the system to fail is $\delta = j$ when $T = X_j$, $j=1,\ldots,m$. The $j$th sub-distribution function evaluated at a time $t$ is the probability that the system survives at most to time $t$ and the last component to fail is the $j$th component, that is, $F_j^*(t)=\rm{P}(T\leq t,\delta=j)$. 

Let $F(t_1,\ldots,t_m)=P(X_1\leq t_1,\ldots,X_m \leq t_m)$ be the joint distribution function, in which continuous partial derivatives are assumed over all arguments.
The following theorem establishes the relation between the joint distribution function with the $j$th sub-distribution $F_j^*(t)$. 

\begin{Theorem}
	\label{teo_relation}
	The derivative of $F_j^{*}(t)$, $d F_j^{*}(t)/d t$,
	 is equal to the partial derivative of $F(t_1,\ldots,t_m)$ at the $j$th component, evaluated at $t_1 = t_2 =\ldots=t_m= t$.
\end{Theorem}

Because the life of the components are assumed to be mutually $s$-independent, 
\begin{eqnarray}
F(t_1,\ldots,t_m)=\prod_{j=1}^{m}F_j(t_j).  \label{F_conjunta}
\end{eqnarray}
Using the fact in (\ref{F_conjunta}) and the Theorem \ref{teo_relation}, 
\begin{eqnarray}
\frac{d}{d t}F_j^{*}(t)=u_j(t)\prod_{j=1}^{m}F_j(t), \label{deriv_sub_dist}
\end{eqnarray}
where $u_j$ is the reversed hazard rate (RHR) of the $j$th component:
\begin{eqnarray}
u_j(t)=\frac{f_j(t)}{F_j(t)}=\frac{d}{d t}\ln F_j(t). \label{RHR}
\end{eqnarray}
From (\ref{RHR}) one can write
\begin{eqnarray}
F_j(t)=\exp\Bigg\{-\int_{t}^{\infty}u_j(y)d y \Bigg\}. \label{relacao1}
\end{eqnarray}
Letting $u(y)=\sum_{j=1}^m u_j(y)$, (\ref{deriv_sub_dist}) becomes
\begin{eqnarray}
\frac{d}{d t}F_j^{*}(t)=u_j(t)\exp\Bigg\{-\int_{t}^{\infty}u(y)d y \Bigg\}, \label{deriv_sub_dist2}
\end{eqnarray}
Taking now the sum for $j=1,\ldots,m$ in both sides of (\ref{deriv_sub_dist2}), we obtain
\begin{eqnarray}
\sum_{j=1}^m\frac{d}{d t}F_j^{*}(t) &=& u(t)\exp\Bigg\{-\int_{t}^{\infty}u(y)d y \Bigg\} \nonumber \\
                                   &=& \frac{d}{dt}\exp\Bigg\{-\int_{t}^{\infty}u(y)d y \Bigg\}. \label{deriv_sub_dist_soma}
\end{eqnarray}
Consequently,
\begin{eqnarray}
\sum_{j=1}^mF_j^{*}(t) = \exp\Bigg\{-\int_{t}^{\infty}u(y)d y \Bigg\}, \nonumber
\end{eqnarray}
which combined with (\ref{deriv_sub_dist2}) leads to
\begin{eqnarray}
u_j(t)=\frac{dF_j^{*}(t)/dt}{\sum_{j=1}^m F_{j}^{*}(t)}.
\end{eqnarray}
Finally, (\ref{relacao1}) implies
\begin{eqnarray}
F_j(t)=\exp\Bigg\{-\int_{t}^{\infty}\frac{dF_j^{*}(y)}{\sum_{j=1}^m F_{j}^{*}(y)} \Bigg\}, \label{relacao2}
\end{eqnarray}
that is, the relationship of interest between marginal distribution functions and sub-distribution functions.

Unfortunately, the expression in (\ref{relacao2}) does not work for the case with jump points. To obtain a version of (\ref{relacao2}) in the presence of jumps, we introduce the following definition and theorem. 

\begin{Definition}
	\label{def_paralelo}
For simplicity, consider the case of $m=2$. The function $\Phi_{p}\left(F_1^*,F_2^*,t \right)$ based on the sub-distributions $F_1^*$ and $F_2^*$ is
\begin{eqnarray*}
\Phi_{p2}\left(F_1^*,F_2^*,t \right)  \equiv   \exp\hspace{-0.1cm}\left\{ \hspace{-0.05cm}\Cint{t}{\infty} \hspace{-0.05cm}\frac{-\d F_1^*(v)}{\sum\limits_{j=1}^2 F_j^*(v)} \hspace{-0.05cm}\right\} \hspace{-0.2cm} \Dprod{v > t}{F_1^*} \hspace{-0.2cm} \left[\frac{\sum\limits_{j=1}^2 F_j^*(v^{-})}{\sum\limits_{j=1}^2 F_j^*(v^{+})} \right]\hspace{-0.1cm},
\end{eqnarray*}
where $\displaystyle \subset\hspace{-0,5cm}\int g(s) \d s$ is integration over disjoint open intervals that do not include the jump points of $g(\cdot)$ and $\Dproda{~}{G}$ is product over jump points of $G(\cdot)$.
\end{Definition}
The next result, although restricted to $m=2$, extends expression (\ref{relacao2}) in the sense that it can include disjoint jump points. 
\begin{Theorem}
	\label{teo_paralelo}
	The sub-distribution functions $F_1^*$ and $F_2^*$ determine (uniquely) the distribution function $F_1$ for $t \leq t^*$ by $F_1(t) = \Phi_{p2}\hspace{-0.1cm}\left(F_1^*,F_2^*,t \right)$.
	
\end{Theorem}

An analogous development can be performed for a series system with $m$ components, in which $T=\min\{X_1,\ldots,X_m\}$ and $\delta=j$, if $T=X_j$. The version of (\ref{relacao2}) for a series system is given by \citep{SalinasTorres}:
\begin{eqnarray}
F_j(t)=1-\exp\Bigg\{\int_{0}^{t}\frac{dR_j^{*}(y)}{\sum_{j=1}^m R_{j}^{*}(y)} \Bigg\}, \label{relacao_serie}
\end{eqnarray}
in which $R_j^*(t)=\rm{P}(T > t,\delta=j)$ is the sub-reliability function for $j$th component. 

Unfortunately, the expression in (\ref{relacao_serie}) does not work for the case with jump points. To obtain a version of (\ref{relacao_serie}) in the presence of jumps, we introduce the following definition and theorem. 
\begin{Definition}
	\label{def_serie}
	For simplicity, consider the case of $m=2$. The function $\Phi_{s2}\left(R_1^*,R_2^*,t \right)$ based on the sub-reliability functions $R_1^*$ and $R_2^*$ is
	\begin{eqnarray*}
		\Phi_{s2}\left(R_1^*,R_2^*,t \right)  \equiv  1- \exp\hspace{-0.1cm}\left\{ \hspace{-0.05cm}\Cint{0}{t} \hspace{-0.05cm}\frac{\d R_1^*(v)}{\sum\limits_{j=1}^2 R_j^*(v)} \hspace{-0.05cm}\right\} \hspace{-0.2cm} \Dprod{v > t}{F_1^*} \hspace{-0.2cm} \left[\frac{\sum\limits_{j=1}^2 R_j^*(v^{+})}{\sum\limits_{j=1}^2 R_j^*(v^{-})} \right]\hspace{-0.1cm},
	\end{eqnarray*}
	where $\displaystyle \subset\hspace{-0,5cm}\int g(s) \d s$ is integration over disjoint open intervals that do not include the jump points of $g(\cdot)$ and $\Dproda{~}{G}$ is product over jump points of $G(\cdot)$.
\end{Definition}
The next result, although restricted to $m=2$, extends expression (\ref{relacao2}) in the sense that it can include disjoint jump points. 
\begin{Theorem}
	\label{teo_serie}
	The sub-reliability functions $R_1^*$ and $R_2^*$ determine (uniquely) the distribution function $F_1$ for $t \leq t^*$ by $F_1(t) = \Phi_{s2}\hspace{-0.1cm}\left(R_1^*,R_2^*,t \right)$.
	
\end{Theorem}

For more details about relations among the distributions and sub-distribution functions (sub-reliability functions) can be found in \\ \citet{SalinasTorres} and \citet{PolpoSinha} for series system and \citet{PolpoCarlinhos} for parallel system.

In next Sub-section the relations among the distributions and sub-distribution functions are presented for a more general class of systems - SPS and PSS.

\subsection{PSS and SPS}
 We restrict ourselves to a system with three components ($m=3$), given in Figure \ref{Fig_1_para_SPS} and in Figure \ref{Fig_2_para_PSS}.
 

Let $X_1, ~ X_2$ and $X_3$ be the lifetimes of three components of an PSS and SPS with marginal distribution functions (DF) $F_1$, $F_2$, and $F_3$, respectively. The restriction here is that the three sets of jump points of $F_1$, $F_2$, and $F_3$ must be disjoint. The indicator of the component whose failure produced the system to fail is $\delta = 1$ when $T = X_1$, $\delta = 2$ when $T = X_2$, and $\delta = 3$ when $T = X_3$. Let  $F_j^*(t)=\rm{P}(T\leq t,\delta=j)$ be the sub-distribution function of the $j$th component and $F(\cdot)$ the distribution function of the system.  The following properties can be proved.

\begin{Property}
	\label{prop_system-dist}
	The sub-distribution functions (SDF) $F_1^*$, $F_2^*$, and $F_3^*$ determine the DF of the system,
	\begin{equation}
	\label{eq_system_dist-sub}
	F(t) = F_1^*(t) + F_2^*(t) + F_3^*(t).
	\end{equation}
\end{Property}

\begin{Property}
	\label{prop_sub-infty}
	\vspace{0.1ex}
	\begin{enumerate}
		\item $F_1^*(+\infty) = \Pr(\delta=1)$;
		\item $F_2^*(+\infty) = \Pr(\delta=2)$;
		\item $F_3^*(+\infty) = \Pr(\delta=3)$;
		\item $F_1^*(+\infty) + F_2^*(+\infty) + F_3^*(+\infty) = 1$.
	\end{enumerate}
\end{Property}

\begin{Property}
	\label{prop_jp}
	The set of jump points $F_j^*$ and $F_j$ are the same, where $j=1,2,3$. Because $F_1$, $F_2$, and $F_3$ have disjoint set of jump points, so have $F_1^*$, $F_2^*$, and $F_3^*$.
\end{Property}

\begin{Property}
	\label{prop_support}
	If $\min(F_1(t), F_2(t), F_3(t)) < 1$ for $t < t^*$, and 1 for $t \geq t^*$, then $t^*$ is the largest support point of the system.
\end{Property}

The lifetime of the SPS is \mbox{$T = $} \mbox{$\min(X_1, \max(X_2, X_3))$} and the system reliability of {\it s}-independent components is
\begin{equation}
\label{eq_sp_system-dist}
R(t) = [1-F_1(t)][1-F_2(t)F_3(t)].
\end{equation}
The lifetime of the PSS is \mbox{$T = $} \mbox{$\max(X_1, \min(X_2, X_3))$} and the system reliability of {\it s}-independent components is
\begin{equation}
\label{eq_ps_system-dist}
R(t) = 1-F_1(t)\left\{1-[1-F_2(t)][1-F_3(t)]\right\}.
\end{equation}

\begin{Property}
	\label{prop_sdf}
	The SDF of the SPS can be expressed using the marginal DF of the components by
	\begin{eqnarray}
	\label{eq_sp_sub-dist}
	F_1^*(t) & = & \int\limits_0^t [1 - F_2(t) F_3(t)] \d F_1(t), \nonumber \\
	F_2^*(t) & = & \int\limits_0^t [1 - F_1(t)] F_3(t) \d F_2(t), \nonumber \\
	F_3^*(t) & = & \int\limits_0^t [1 - F_1(t)] F_2(t) \d F_3(t), 
	\end{eqnarray}
	and the SDF of the PSS can be expressed using the marginal DF of the components by
	\begin{eqnarray}
	\label{eq_ps_sub-dist}
	F_1^*(t) & = & \int\limits_0^t \{1-[1-F_2(t)][1-F_3(t)]\} \d F_1(t), \nonumber \\
	F_2^*(t) & = & \int\limits_0^t F_1(t) [1-F_3(t)] \d F_2(t), \nonumber \\
	F_3^*(t) & = & \int\limits_0^t F_1(t) [1-F_2(t)] \d F_3(t).
	\end{eqnarray}
\end{Property}

Our interest is to obtain the inverse of (\ref{eq_sp_sub-dist}) and (\ref{eq_ps_sub-dist}); that is, to express the DFs $F_1$, $F_2$ and $F_3$ as a function of the SDF ($F_1^*$, $F_2^*$, $F_3^*$). These inverses are presented with the following definitions and theorems.

\begin{Definition}
	\label{def_f-s}
	The functions $\Phi_{s}\hspace{-0.1cm}\left(F_1^*,F_2^*,F_3^*,t \right)$, and $\Phi_{p}\hspace{-0.1cm}\left(F_1^*,F_2^*,F_3^*,t \right)$ based on sub-distributions $F_1^*$, $F_2^*$, and $F_3^*$ are 
	\begin{eqnarray*}
		\Phi_{s}\hspace{-0.1cm}\left(F_1^*,F_2^*,F_3^*,t \right) \hspace{-0.3cm} & \equiv & \hspace{-0.3cm} 1-\left\{\exp\left[ \Cint{0}{t} \frac{-\d F_1^*(v)}{1-\sum\limits_{j=1}^3 F_j^*(v)} \right]\right. \\
		&& \hspace{0.7cm}\left. \Dprod{v \leq t}{F_1^*} \left[ \frac{ 1 -\sum\limits_{j=1}^3 F_j^*(v^{+}) }{ 1 - \sum\limits_{j=1}^3 F_j^*(v^{-})}. \right]\right\}, \\
		\Phi_{p}\hspace{-0.1cm}\left(F_1^*,F_2^*,F_3^*,t \right) \hspace{-0.3cm} & \equiv & \hspace{-0.3cm} \exp\hspace{-0.1cm}\left\{ \hspace{-0.05cm}\Cint{t}{\infty} \hspace{-0.05cm}\frac{-\d F_1^*(v)}{\sum\limits_{j=1}^3 F_j^*(v)} \hspace{-0.05cm}\right\} \hspace{-0.2cm} \Dprod{v > t}{F_1^*} \hspace{-0.2cm} \left[\frac{\sum\limits_{j=1}^3 F_j^*(v^{-})}{\sum\limits_{j=1}^3 F_j^*(v^{+})} \right]\hspace{-0.1cm}.
	\end{eqnarray*}
\end{Definition}
The functions $\Phi_{s}$ (for a series system), and $\Phi_{p}$ (for a parallel system) are the versions with three components for those presented in  \citet{PolpoSinha} and \citet{PolpoCarlinhos}, respectively. First, Theorem \ref{teo_peterson} states the relation between $F_1$ and $F_1^*$, $F_2^*$, and $F_3^*$. 
\begin{Theorem}
	\label{teo_peterson}
	The SDF $F_1^*$, $F_2^*$, and $F_3^*$ determine (uniquely) the DF $F_1$ of an SPS for $t \leq t^*$ by $F_1(t) = \Phi_{s}\hspace{-0.1cm}\left(F_1^*,F_2^*,F_3^*,t \right)$, and the DF $F_1$ of a PSS for $t \leq t^*$ by $F_1(t) = \Phi_{p}\hspace{- 0.1cm}\left(F_1^*,F_2^*,F_3^*,t \right)$.
\end{Theorem}

The next definition gives the functions $\Phi_{sp}$ (for the SPS), and $\Phi_{ps}$ (for the PSS).

\begin{Definition}
	\label{def_f-sp}
	The functions $\Phi_{sp}\hspace{-0.1cm}\left(F_1^*,F_2^*,F_3^*,t \right)$, and $\Phi_{ps}\hspace{-0.1cm}\left(F_1^*,F_2^*,F_3^*,t \right)$, based on sub-distributions $F_1^*$, $F_2^*$, and $F_3^*$, are
	\begin{eqnarray*}
		\Phi_{sp}\hspace{-0.1cm}\left(F_1^*,F_2^*,F_3^*,t \right) & \equiv & \\
		&& \hspace{-2.2cm} \exp\hspace{-0.05cm}\left\{ \Cint{t}{\infty} \frac{-\d F_2^*(v)}{\sum\limits_{j=1}^3 F_j^*(v)-\Phi_{s}\hspace{-0.1cm}\left(F_1^*,F_2^*,F_3^*,v \right)} \right\} \hspace{-0.2cm} \\
		&& \hspace{-1.4cm} \Dprod{v > t}{F_2^*} \hspace{-0.15cm} \left[\frac{\sum\limits_{j=1}^3 F_j^*(v^{-}) - \Phi_{s}\hspace{-0.1cm}\left(F_1^*,F_2^*,F_3^*,v^{-} \right)}{\sum\limits_{j=1}^3 F_j^*(v^{+}) - \Phi_{s}\hspace{-0.1cm}\left(F_1^*,F_2^*,F_3^*,v^{+} \right)} \right]\hspace{-0.1cm}, \\
		\Phi_{ps}\hspace{-0.1cm}\left(F_1^*,F_2^*,F_3^*,t \right) & \equiv & \\ 
		&& \hspace{-2.0cm} 1-\left\{ \exp\hspace{-0.05cm}\left[ \Cint{0}{t} \frac{-\d F_2^*(v)}{\Phi_{p}\hspace{-0.1cm}\left(F_1^*,F_2^*,F_3^*,v \right)-\hspace{-0.2cm}\sum\limits_{j=1}^3 F_j^*(v)} \right] \right. \\
		&& \hspace{-1.2cm} \Dprod{v \leq t}{F_2^*} \hspace{-0.15cm} \left[\frac{\Phi_{p}\hspace{-0.1cm} \left(F_1^*, F_2^*, F_3^*, v^{+} \right) - \hspace{-0.2cm}\sum\limits_{j=1}^3 F_j^*(v^{+})}{\Phi_{p}\hspace{-0.1cm} \left(F_1^*, F_2^*, F_3^*, v^{-} \right) - \hspace{-0.2cm}\sum\limits_{j=1}^3 F_j^*(v^{-})} \right]\hspace{-0.1cm}.
	\end{eqnarray*}
\end{Definition}

\begin{Theorem}
	\label{teo_rel_sp}
	The SDF $F_1^*$, $F_2^*$, and $F_3^*$ determine (uniquely) the DF $F_2$ of an SPS for $t \leq t^*$ by $F_2(t) = \Phi_{sp}\hspace{-0.1cm}\left(F_1^*,F_2^*,F_3^*,t \right)$, and the DF $F_2$ of a PSS for $t \leq t^*$ by $F_2(t) = \Phi_{ps}\hspace{-0.1cm} \left(F_1^*,F_2^*,F_3^*,t \right)$.
\end{Theorem}
Note that Theorem \ref{teo_rel_sp} can be easily rewritten to obtain the relation of DF $F_3$ and the SDF. 

The proofs of Theorems \ref{teo_peterson} and \ref{teo_rel_sp} can be seen in \citet {PolpoSiCar}.

Theorem \ref{teo_rel_sp} provides an important relation between the SDF and DF, for both SPS and PSS. Using this result, in the next section, we have developed the nonparametric Bayesian estimator for the DF of the system's components.

\section{Bayesian Analysis}
\label{sec_bayes}

This section describes a Bayesian reliability approach to SPS and PSS. We have derived a nonparametric Bayesian estimator of the distribution function using the multivariate Dirichlet process \\ \citep{SalinasTorres}. From Property \ref{prop_system-dist}, we have that the sub-distribution functions are related to the system distribution function by a sum. Considering that $F_1^*(t) + F_2^*(t) + F_3^*(t) + (1-F(t)) = 1$, we have the restriction that these four quantities have a sum equal to $1$, and that the set of possible points for $\{F_1^*(t), F_2^*(t), F_3^*(t), 1-F(t)\}$ is the four-dimensional simplex, or $\{F_1^*(t), F_2^*(t), F_3^*(t)\}$ for the non-singular form. In this case, for a fixed $t$, we have that a natural prior choice is the Dirichlet distribution, and for any $t$, we have the Dirichlet multivariate process. 
In this Section a nonparametric estimator for the distribution function of the components in an SPS or in a PSS, and using the Dirichlet process, we have a complete distribution for the set $\{F_1^*(t), F_2^*(t), F_3^*(t)\}$. In this case, our parameters are the functions that we want to estimate, giving us a nonparametric framework. 

Consider a sample of size $n$ and the observed data are $(T_1,\delta_1),\ldots,$ $(T_n,\delta_n)$, in which $\mbox{$T_i = $} \mbox{$\min(X_{1i}, \max(X_{2i}, X_{3i}))$}$ for SPS and \\ $\mbox{$T_i = $} \mbox{$\max(X_{1i}, \min(X_{2i}, X_{3i}))$}$ for PSS. Besides, $\delta_i=j$ if $T_i=X_{ji}$, for $i=1,\ldots,n$ and $j=1,2,3$. Equivalently, for each $t$,the random variables are observed:
\begin{eqnarray}
nF^{*}_{jn}(t)=\sum_{i=1}^n {\rm I}(T_i\leq t,\delta_i=j), ~~ \mbox{for} ~ j=1,2,3, \nonumber 
\end{eqnarray}
in which ${\rm I}(A)$ is a indicator function of set $A$.

The function $F^{*}_{jn}$ is empirical sub-distribution function of $j$th component. If $F_n(\cdot)$ is the empirical distribution function corresponding to the observations $T_1,\ldots,T_n$, thus for each $t$, 
\begin{eqnarray}
nF_n(t)=nF^{*}_{1n}(t)+nF^{*}_{2n}(t)+nF^{*}_{3n}(t). \nonumber 
\end{eqnarray}

For each $t$, let $k_j(t)$ the realization of $nF^{*}_{jn}(t)$, in which
\begin{eqnarray}
k_j(t)=\sum_{i=1}^n {\rm I}(t_i\leq t,\delta_i=j), ~~ \mbox{for} ~ j=1,2,3. \nonumber 
\end{eqnarray}

In this context, for each $t$, the likelihood function corresponds to the likelihood of a multinomial model $M(n; p_1(t), p_2(t), p_3(t), p_4(t))$ being
$p_j(t)=F_j^{*}(t)$, for $j=1,2,3$, and $p_4(t)=1-F(t)=1-\sum_{j=1}^3F_j^{*}(t)$, that is, 
\begin{eqnarray}
L&=&{\rm P}(nF_{1n}^{*}=k_1(t),nF_{2n}^{*}=k_2(t),nF_{3n}^{*}=k_3(t))  \nonumber \\
  && \propto [F_1^{*}(t)]^{k_1(t)}[F_2^{*}(t)]^{k_2(t)}[F_3^{*}(t)]^{k_3(t)}[1-F(t)]^{n-\sum_{j=1}^3 k_j(t)}  \label{vero_npar}
\end{eqnarray}

The prior distribution for $(F_1^{*}(\cdot),F_2^{*}(\cdot),F_3^{*}(\cdot))$ is constructed from the characterization of the multivariate Dirichlet process, defined in \\ \citet{VHS1997} and it may have the following simplified version. 

\begin{Definition}
	\label{def_processo_dirichlet}
Let $\Omega$ be a sample space, $\alpha_1,\ldots,\alpha_m$ be finite positive measures defined over $\Omega$, and $\rho=(\rho_1,\ldots,\rho_m)$ be a random vector having a Dirichlet distribution with parameters $(\alpha_1(\Omega),\ldots,\alpha_m(\Omega))$. Consider $m$ Dirichlet processes, $P_1,\ldots,P_m$, with $P_j \sim \mathcal{D}(\alpha_j)$, $j=1,\ldots,m$. All these processes and $\rho$ are mutually $s$-independent random quantities. Define $P^{*}=(P_1^{*},\ldots,P_m^{*})=(\rho_1P_1,\ldots,\rho_mP_m)$. The $P^{*}$ is a Dirichlet multivariate process with parameter measures $\alpha_1,\ldots,\alpha_m$.
\end{Definition}

In the context of SPS and PSS, consider $\Omega=(0,\infty)$, $\rho_j={\rm P}(\delta=j)$ and $P_j(t)={\rm P}(T\leq t | \delta=j)$, for $j=1,2,3$. Then, the vector of components sub-distribution functions is $F^{*}=(F_1^{*},F_2^{*},F_3^{*})=(\rho_1P_1,\rho_2P_2,\rho_3P_3)$ and the prior distribution is given by
\begin{eqnarray}
 F^{*}(t)\sim \mathcal{D}(\alpha_1(0,t],\alpha_2(0,t],\alpha_3(0,t]; \sum_{j=1}^3\alpha_j(t,\infty)). \label{priori_NP}
\end{eqnarray}
Combining the prior distribution (\ref{priori_NP}) and the likelihood function in (\ref{vero_npar}), the posterior distribution of $F^{*}(t)=(F_1^{*}(t),F_2^{*}(t),F_3^{*}(t))$ is, for each $t$, 
\begin{eqnarray}
(F_1^{*}(t),F_2^{*}(t),F_3^{*}(t))|Data \sim \mathcal{D}(\alpha_1(0,t]+nF_{1n}^{*}(t),\alpha_2(0,t]+nF_{2n}^{*}(t),\nonumber \\
\alpha_3(0,t]+nF_{3n}^{*}(t); \sum_{j=1}^3\alpha_j(t,\infty)+n-\sum_{j=1}^3 nF_{jn}^{*}(t)). \nonumber
\end{eqnarray}

Thus, the posterior means of $F_j^{*}(t)$ and $F(t)$ are given by
\begin{equation}
\label{eq_est-sub}
\widehat{F}_j^*(t) = p_{\alpha} \frac{\alpha_j(0,t]}{\sum\limits_{\ell=1}^3 \alpha_\ell(0,\infty)}+(1-p_{\alpha}) F_{jn}^*(t),
\end{equation}
where $p_{\alpha} = \Big(\sum\limits_{j=1}^3 \alpha_j(0,\infty)\Big)\Big/\Big(n + \sum\limits_{j=1}^3 \alpha_j(0,\infty)\Big)$, and
\begin{equation}
\label{eq_estima_f}
\widehat{F}(t) = \sum\limits_{j=1}^3 \widehat{F}_j^*(t).
\end{equation}

These Bayesian estimators are strongly {\it s-}consistent. For instance, using the Glivenko Cantelli Theorem \citep{Billingsley1985}, it can be shown that $\widehat{F}_j^*$ converges to $F_j^*$ uniformly with probability 1.

If $\alpha_j(0,\infty) < \infty$, the Bayesian estimator of $\rho_j = {\rm P}(\delta = j)$ is given by
\begin{equation}
\label{est_equ3_22}
\widehat{\rho}_j = \lim_{t \uparrow \infty} \widehat{F}_j^*(t) = \frac{\alpha_j(0,\infty)}{n + \sum\limits_{\ell=1}^3 \alpha_\ell(0,\infty)} + \frac{\sum\limits_{i=1}^n {\rm I}(\delta_i = j)}{n + \sum\limits_{\ell=1}^3 \alpha_\ell(0,\infty)}.
\end{equation}

Let the ~$v$ $(\leq n)$ distinct order statistics ~of $T$ be $\mbox{$T_{(1)}^{\bullet} <$} ~ \mbox{$\ldots <$} ~ \mbox{$T_{(v)}^\bullet$}$. Set $\mbox{$N_i =$} \mbox{$\sum\limits_{\ell=1}^n {\rm I} (T_\ell < T_{(i)}^\bullet)$}$, and $\mbox{$d_{j i} =$} \mbox{$\sum\limits_{\ell=1}^n {\rm I} (T_\ell = T_{(i)}^\bullet, \delta_\ell = j)$}$, $i=1,\ldots,v$. Define

\begin{equation}
\label{eq_integral-s}
I_{s}(t) = \exp \left\{ \frac{1}{\sum\limits_{j=1}^3 \alpha_{j}(0,\infty) + n} \int\limits_0^t \frac{-\d \alpha_1(0,s]}{1-\widehat{F}(s)} \right\},
\end{equation}

\begin{equation}
\label{eq_produto-s}
\Pi_{s}(t) = \prod\limits_{i: T_{(i)}^\bullet \leq t} \frac{\sum\limits_{j=1}^3 \alpha_{j}(T_{(i)}^\bullet,\infty) + n - N_i - d_{1i}}{\sum\limits_{j=1}^3 \alpha_{j}(T_{(i)}^\bullet,\infty) + n - N_i}.
\end{equation}

\begin{equation}
\label{eq_integral-sp}
I_{sp}(t) = \exp \left\{ \frac{1}{\sum\limits_{j=1}^3 \alpha_{j}(0,\infty) + n} \int\limits_t^\infty \frac{-\d \alpha_2(0,s]}{\widehat{F}(s)-\widehat{F}_1(s)} \right\},
\end{equation}

\begin{equation}
\label{eq_produto-sp}
\Pi_{sp}(t) = \prod\limits_{i: T_{(i)}^\bullet > t} \frac{\frac{\sum\limits_{j=1}^3 \alpha_{j}(0,T_{(i)}^\bullet] + N_i}{n + \sum\limits_{j=1}^3 \alpha_{j}(0,\infty)} - \widehat{F}_1(T_{(i)}^\bullet)}{\frac{ \sum\limits_{j=1}^3 \alpha_{j}(0,T_{(i)}^\bullet] + N_i + d_{2i} }{n + \sum\limits_{j=1}^3 \alpha_{j}(0,\infty)} - \widehat{F}_1(T_{(i)}^\bullet)}.
\end{equation}

\begin{equation}
\label{eq_integral-p}
I_{p}(t) = \exp \left\{ \frac{1}{\sum\limits_{j=1}^3 \alpha_{j}(0,\infty) + n} \int\limits_t^\infty \frac{-\d \alpha_1(0,s]}{\widehat{F}(s)} \right\},
\end{equation}

\begin{equation}
\label{eq_produto-p}
\Pi_{p}(t) = \prod\limits_{i: T_{(i)}^\bullet > t} \frac{\sum\limits_{j=1}^3 \alpha_{j}(0,T_{(i)}^\bullet] + N_i}{\sum\limits_{j=1}^3 \alpha_{j}(0,T_{(i)}^\bullet] + N_i + d_{1i}}.
\end{equation}

\begin{equation}
\label{eq_integral-ps}
I_{ps}(t) = \exp \left\{ \frac{1}{\sum\limits_{j=1}^3 \alpha_{j}(0,\infty) + n} \int\limits_0^t \frac{-\d \alpha_2(0,s]}{\widehat{F}_1(s)-\widehat{F}(s)} \right\},
\end{equation}
and
\begin{equation}
\label{eq_produto-ps}
\Pi_{ps}(t) = \prod\limits_{i: T_{(i)}^\bullet \leq t} \frac{\widehat{F}_1(T_{(i)}^\bullet) - \frac{\sum\limits_{j=1}^3 \alpha_{j}(0,T_{(i)}^\bullet] + N_i + d_{2i}}{n + \sum\limits_{j=1}^3 \alpha_{j}(0,\infty)}}{\widehat{F}_1(T_{(i)}^\bullet) - \frac{ \sum\limits_{j=1}^3 \alpha_{j}(0,T_{(i)}^\bullet] + N_i }{n + \sum\limits_{j=1}^3 \alpha_{j}(0,\infty)}}.
\end{equation}

The main result of this study is given in the following paragraph.
\begin{Theorem}
	\label{teo_main-bayes}
	Suppose that $\alpha_1(0,\cdot),\alpha_2(0,\cdot),\alpha_3(0,\cdot)$ are continuous on $(t,\infty)$, for each $t > 0$, and $F_1$, $F_2$, and $F_3$ have no common discontinuities. Then, for $t \leq T_{(n)}$, and SPS, we have that
	\begin{eqnarray}
	\label{eq_main-s} \widehat{F}_1(t) \hspace{-0.2cm}& = &\hspace{-0.2cm} \Eset[F_1(t) | data] \\ 
	& = & \hspace{-0.2cm}\Phi_{s}(\widehat{F}_1^*,\widehat{F}_2^*,\widehat{F}_3^*,t) = 1-I_{s}(t)\Pi_{s}(t), \nonumber \\
	\label{eq_main-sp} \widehat{F}_2(t) \hspace{-0.2cm}& = &\hspace{-0.2cm} \Eset[F_2(t) | data] \\
	& = & \hspace{-0.2cm}\Phi_{sp}(\widehat{F}_1^*,\widehat{F}_2^*,\widehat{F}_3^*,t) = I_{sp}(t)\Pi_{sp}(t); \nonumber
	\end{eqnarray}
	and, for PSS,
	\begin{eqnarray}
	\label{eq_main-p} \widehat{F}_1(t) \hspace{-0.2cm}& = &\hspace{-0.2cm} \Eset[F_1(t) | data] \\
	& = & \hspace{-0.2cm}\Phi_{p}(\widehat{F}_1^*,\widehat{F}_2^*,\widehat{F}_3^*,t) = I_{p}(t)\Pi_{p}(t), \nonumber \\
	\label{eq_main-ps} \widehat{F}_2(t) \hspace{-0.2cm}& = &\hspace{-0.2cm} \Eset[F_2(t) | data] \\
	& = & \hspace{-0.2cm}\Phi_{ps}(\widehat{F}_1^*,\widehat{F}_2^*,\widehat{F}_3^*,t) = 1-I_{ps}(t)\Pi_{ps}(t). \nonumber
	\end{eqnarray}
	$\widehat{F}_1(t)$, and $\widehat{F}_2(t)$ are the nonparametric estimators of $F_1(t)$, and $F_2(t)$, respectively, based on posterior means.
\end{Theorem}

As in Theorem \ref{teo_rel_sp}, it is straightforward to express the nonparametric estimator of $F_3(t)$. In the next section, we extend the estimators to a general case of $m \geq 4$.

\section{Bayesian estimator for $m \geq 4$}
\label{sec_general}

\begin{figure}[!h]\centering
	\begin{minipage}[H]{0.4\linewidth}
		\includegraphics[width=\linewidth]{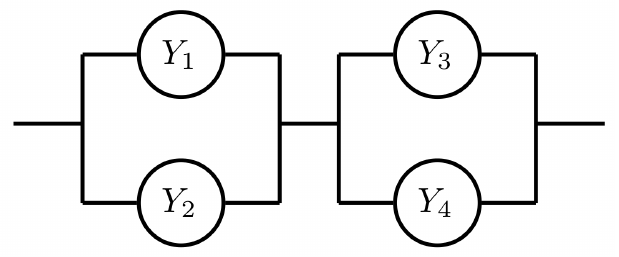}
		\subcaption{ $\mbox{   }$ }  \label{SPS_general}
	\end{minipage} 
	\begin{minipage}[H]{0.4\linewidth}
		\includegraphics[width=\linewidth]{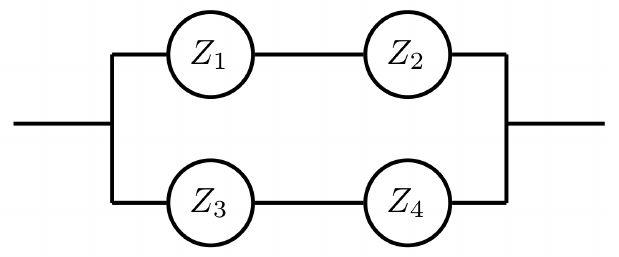}
		\subcaption{ $\mbox{   }$ } \label{PSS_general}
	\end{minipage}
	\caption{(a) SPS (b) PSS}
	\label{fig_general}
\end{figure}

The extension of the nonparametric Bayesian estimator for SPS and PSS, given in Section \ref{sec_bayes}, is based on rewriting the system representation in a proper simplified version of the general case ($m \geq 4$) to the one given with $m = 3$, which has a solution given in Theorem \ref{teo_main-bayes}. Considering the SPS and the PSS presented in Fig.~\ref{fig_general}, we specify how to rewrite the system representation and estimation of their components reliabilities in the following.

\begin{figure}[!h]\centering
	\begin{minipage}[H]{0.4\linewidth}
		\includegraphics[width=\linewidth]{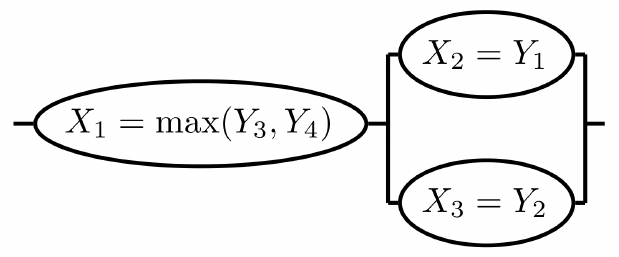}
		\subcaption{ $\mbox{   }$ }  \label{fig_ext-sp}
	\end{minipage} 
	\begin{minipage}[H]{0.4\linewidth}
		\includegraphics[width=\linewidth]{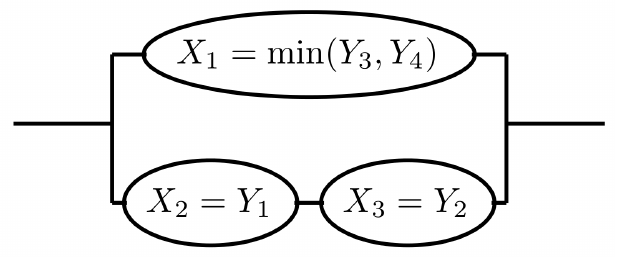}
		\subcaption{ $\mbox{   }$ } \label{fig_ext-ps}
	\end{minipage}
	\caption{(a) SPS (b) PSS}
	\label{fig_ext}
\end{figure}

We provided how to estimate $Y_1$ (for the SPS), and $Z_1$ (for the PSS), because the reliability estimation of the other components are straightforward once these two are given. The idea of the extension is to represent the systems in a simple version with three components (Figures \ref{Fig_1_para_SPS} and \ref{Fig_2_para_PSS}). In this case, to estimate the reliability of $Y_1$, we use the SPS solution considering $X_1 = \max(Y_3, Y_4)$, $X_2 = Y_1$, and $X_3 = Y_2$ (Figure \ref{fig_ext-sp}); and for the estimation of $Z_1$, we use the PSS solution considering $X_1 = \min(Y_3, Y_4)$, $X_2 = Y_1$, and $X_3 = Y_2$ (Figure \ref{fig_ext-ps}). It must be noted that other more complex systems can also be considered, but the task is only to simplify the representation of the system as one of either the PSS or SPS given in Figures \ref{Fig_1_para_SPS} and \ref{Fig_2_para_PSS}.

\begin{figure}[!h]\centering
	\centering
	\includegraphics[width=0.7\linewidth]{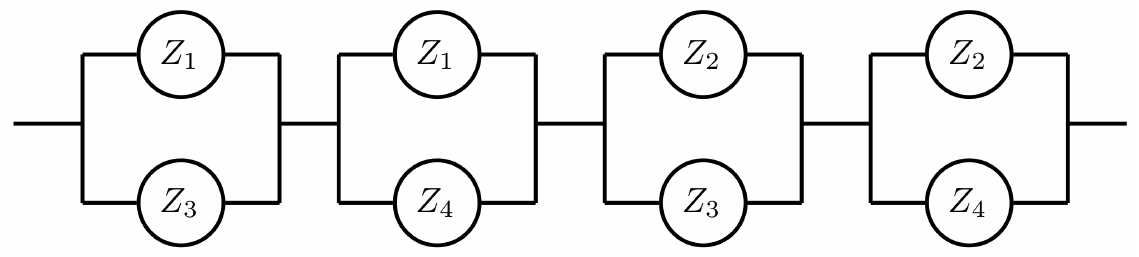}
	\caption{ The SPS representation of the PSS in Figure \ref{PSS_general}.}
	\label{fig_g-ps}
\end{figure}

Furthermore, both the classes (SPS and PSS) are important so as to have a more general solution, because we have the restriction that two different components cannot have the same failure time, which in turn would result in different representations giving more options to the reliability estimation problem. Considering the PSS given in Figure \ref{PSS_general}, we can write their SPS representation as that presented in Figure \ref{fig_g-ps}. The component's reliability of the original PSS (Figure \ref{PSS_general}) can be estimated using the PSS result of Theorem \ref{teo_main-bayes}, which has a simple solution. However, as the SPS representation (Figure \ref{fig_g-ps}) has some components repeated, the SPS result of Theorem \ref{teo_main-bayes} is not applicable. Thus, the solutions for both SPS and PSS are important and can be used in different situations.

\section{Simulated datasets}
\label{sec_num-ex}

This section presents two examples to demonstrate the estimation steps and to show the quality of the Bayesian nonparametric estimator. The estimation steps for the PSS are very similar to those for the SPS, and for the sake of brevity, we have omitted them. The estimation steps for SPS are as follows. 

\begin{enumerate}
	\item {\it Defining priors:} The prior measures ($\alpha_1, \alpha_2, \alpha_3$) are prior guesses of the SDF ($F_1^*, F_2^*, F_3^*$), but it is not simple to elicit these measures. It is easier to elicit the priors for the DF ($F_1, F_2, F_3$), and use (\ref{eq_sp_sub-dist}) for the SPS to evaluate the prior measures (for PSS we can use (\ref{eq_ps_sub-dist})). We chose the exponential distribution with mean 1 as the prior guess for each of the three components DF. By evaluating the prior measures using (\ref{eq_sp_sub-dist}), we have \mbox{$\alpha_1(0,v] =$} \mbox{$(e^{-3v}-3e^{-2v}+2)/3$}, and \mbox{$\alpha_2(0,v] =$} \mbox{$\alpha_3(0,v] =$} \mbox{$(2e^{-3v}-3e^{-2v}+1)/6$}. Note that this prior is not very informative because the measure of the whole parameter space is only one ($\alpha_1(\Omega) + \alpha_2(\Omega) + \alpha_3(\Omega) =1$). Also, we have that \mbox{$\d \alpha_1(0,v] =$} \mbox{$2e^{-2v}-e^{-3v} \d v$}, and \mbox{$\d \alpha_2(0,v] =$} \mbox{$\d \alpha_3(0,v] =$} \mbox{$e^{-2v}-e^{-3v} \d v$}.
	
	\item {\it Obtaining Posteriors:} The ~posterior ~processes ~for ~the ~SDF ~functions ~are ~\mbox{$\mathcal{D}((e^{-3t}-3e^{-2t}+2)/3 +$} \mbox{$nF_{1n}^*(t),$} ~\mbox{$(2e^{-3t}-3e^{-2t}+1)/6 +$} \mbox{$nF_{2n}^*(t),$} \mbox{$(2e^{-3t}-3e^{-2t}+1)/6 +$} \mbox{$nF_{3n}^*(t))$}; and from (\ref{eq_est-sub}), we have
	\begin{eqnarray*}
		\widehat{F}_1^*(t) & = & \frac{nF_{1n}^*(t)+(e^{-3t}-3e^{-2t}+2)/3}{n+1}, \\
		\widehat{F}_2^*(t) & = & \frac{nF_{2n}^*(t)+(2e^{-3t}-3e^{-2t}+1)/6}{n+1}, \\
		\widehat{F}_3^*(t) & = & \frac{nF_{3n}^*(t)+(2e^{-3t}-3e^{-2t}+1)/6}{n+1},
	\end{eqnarray*}
	which are the estimators of SDF.
	
	\item {\it Computing system's reliability:} (\ref{eq_estima_f}) provides the estimator of the system distribution function. For the prior defined earlier, we have
	\[ \widehat{F}(t) = \frac{n F_{n}^*(t) + e^{-3t}-2e^{-2t}+1 }{n+1}. \]
	
	\item {\it Computing components reliabilities:} Theorem \ref{teo_main-bayes} gives the estimators of the components DF. Using (\ref{eq_main-s}), we obtain the estimate for component 1 DF; and from (\ref{eq_main-sp}), we obtain the estimate for component 2 DF. For Component 3 DF, we substitute $\d \alpha_2(0,v]$ by $\d \alpha_3(0,v]$ (in the integral part $I_{sp}$), and $d_{2i}$ by $d_{3i}$ (in the product part $\Pi_{sp}$) in (\ref{eq_main-sp}). Also, the integral part of the estimator can be solved by using a numerical procedure, such as the Simpson's rule. For more details and other numerical integration methods, see \citet{DavisRabinowitz1984}.
\end{enumerate}

\subsection{Simulated dataset 1}
	\label{ex-sps4}
	We obtained $100$ observations of SPS presented in Figure \ref{SPS_general}, where all the components had gamma distributions; the first component ($Y_1$) had a mean of $4$ and a standard deviation (SD) of $2.83$, the second component ($Y_2$) had a mean of $6$ and a SD of $4.9$, the third component ($Y_3$) had a mean of $8$ and a SD of $5.67$, and the fourth component ($Y_4$) had a mean of $3$ and a SD of $2.45$. The Bayesian estimators are based on $(T_1, \delta_1),\ldots,(T_{100}, \delta_{100})$. The simulated values are listed in Table \ref{tabela_data1} in Appendix \ref{apeD}.
	
	To estimate components 1--4, we rewrote the representation of the system as follows. For component 1, we considered that $X_1 = \max(Y_3,Y_4)$, $X_2 = Y_1$, and $X_3 = Y_2$; then $\widehat{F}_{Y_1}(t) = \widehat{F}_2(t)$, where $\widehat{F}_{Y_1}$ is the DF estimate of component 1, and $\widehat{F}_2$ is the proposed estimator for SPS (\ref{eq_main-sp}). In a similar way, for component 2, we considered $X_1 = \max(Y_3,Y_4)$, $X_2 = Y_2$, and $X_3 = Y_1$; for component 3, we considered $X_1 = \max(Y_1,Y_2)$, $X_2 = Y_3$, and $X_3 = Y_4$; and for component 4, we considered $X_1 = \max(Y_1,Y_2)$, $X_2 = Y_4$, and $X_3 = Y_3$ (see Section \ref{sec_general}). We found that the proportions of censored data for components 1--4 are $77\%$, $64\%$, $73\%$, and $86\%$, respectively.
	
\begin{figure}[h]\centering
	\begin{minipage}[b]{0.32\linewidth}
		\includegraphics[width=\linewidth]{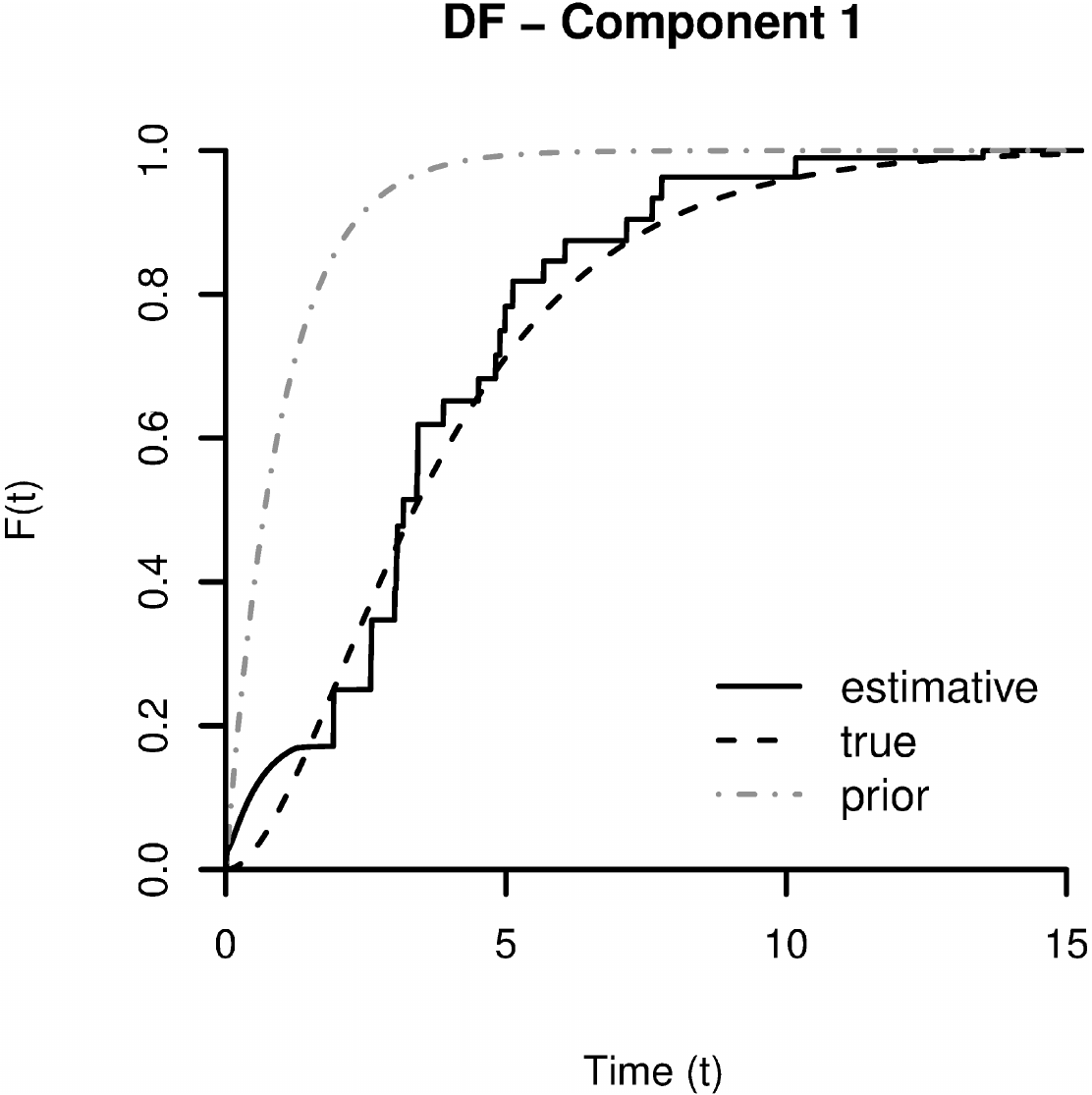}
		\subcaption{  }
	\end{minipage} 
	\begin{minipage}[b]{0.32\linewidth}
		\includegraphics[width=\linewidth]{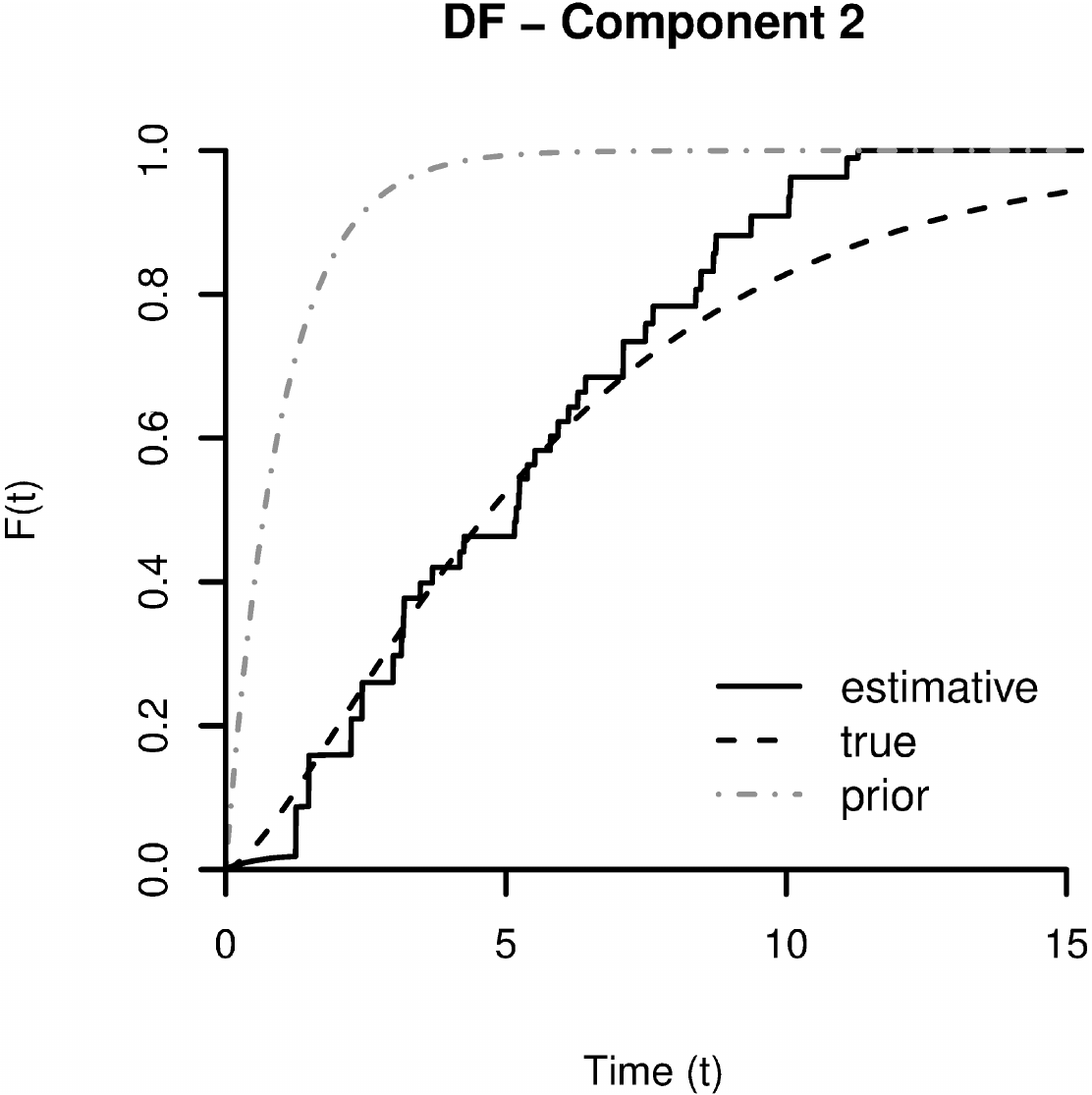}
		\subcaption{  }
	\end{minipage}
	\begin{minipage}[b]{0.32\linewidth}
		\includegraphics[width=\linewidth]{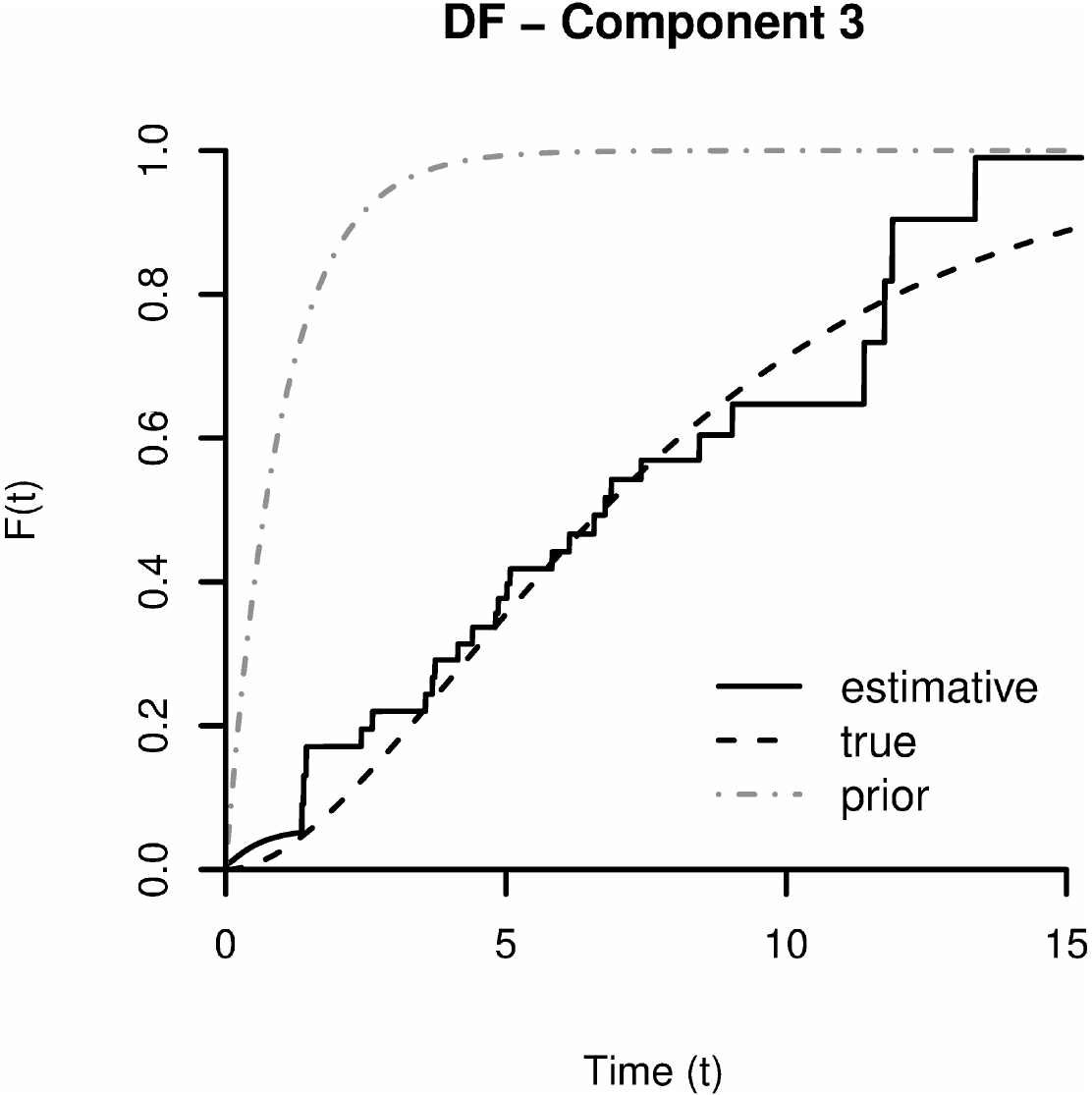}
		\subcaption{ }
	\end{minipage}	
	\begin{minipage}[b]{0.32\linewidth}
		\includegraphics[width=\linewidth]{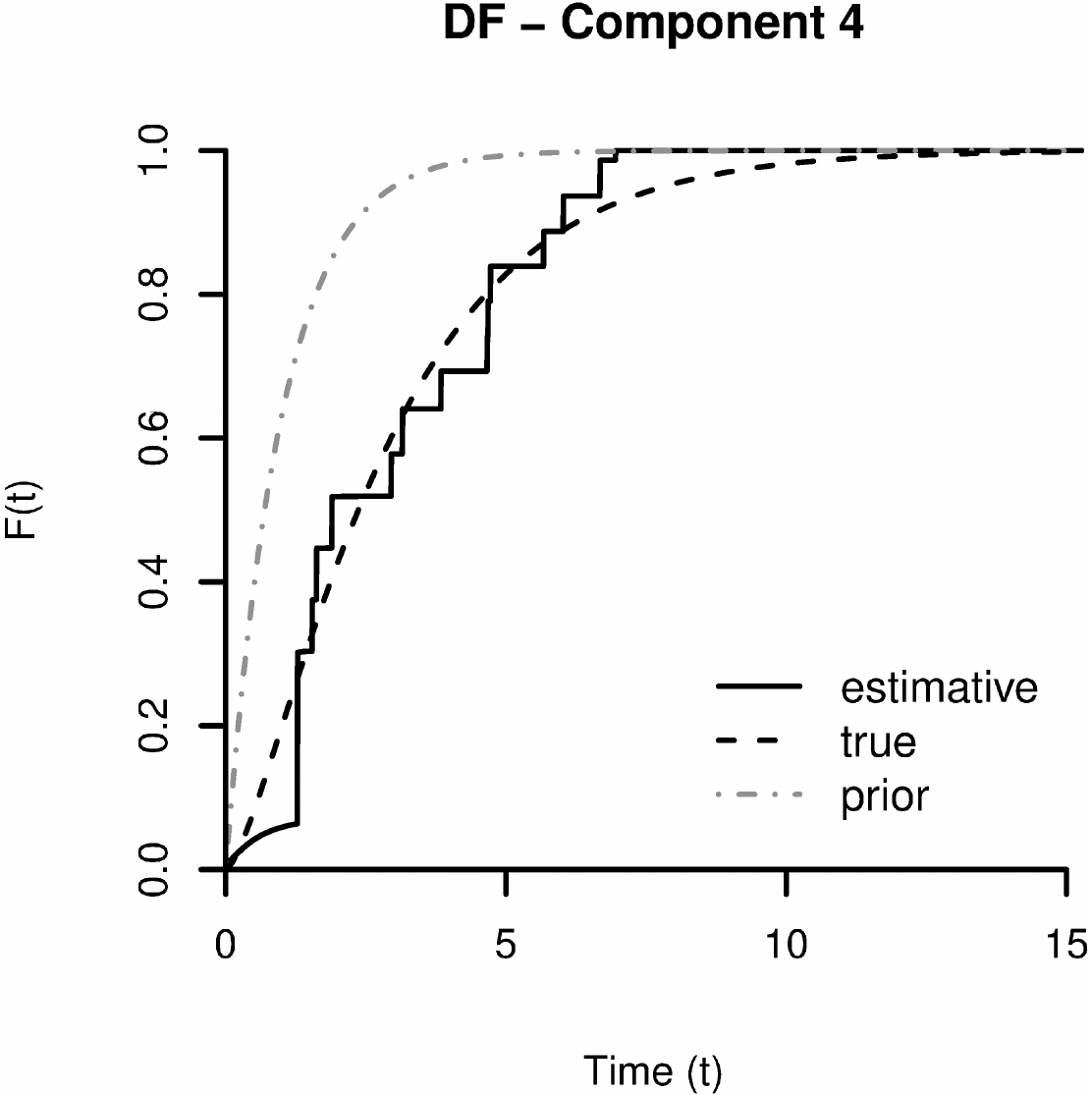}
		\subcaption{ }
	\end{minipage}
	\begin{minipage}[b]{0.32\linewidth}
		\includegraphics[width=\linewidth]{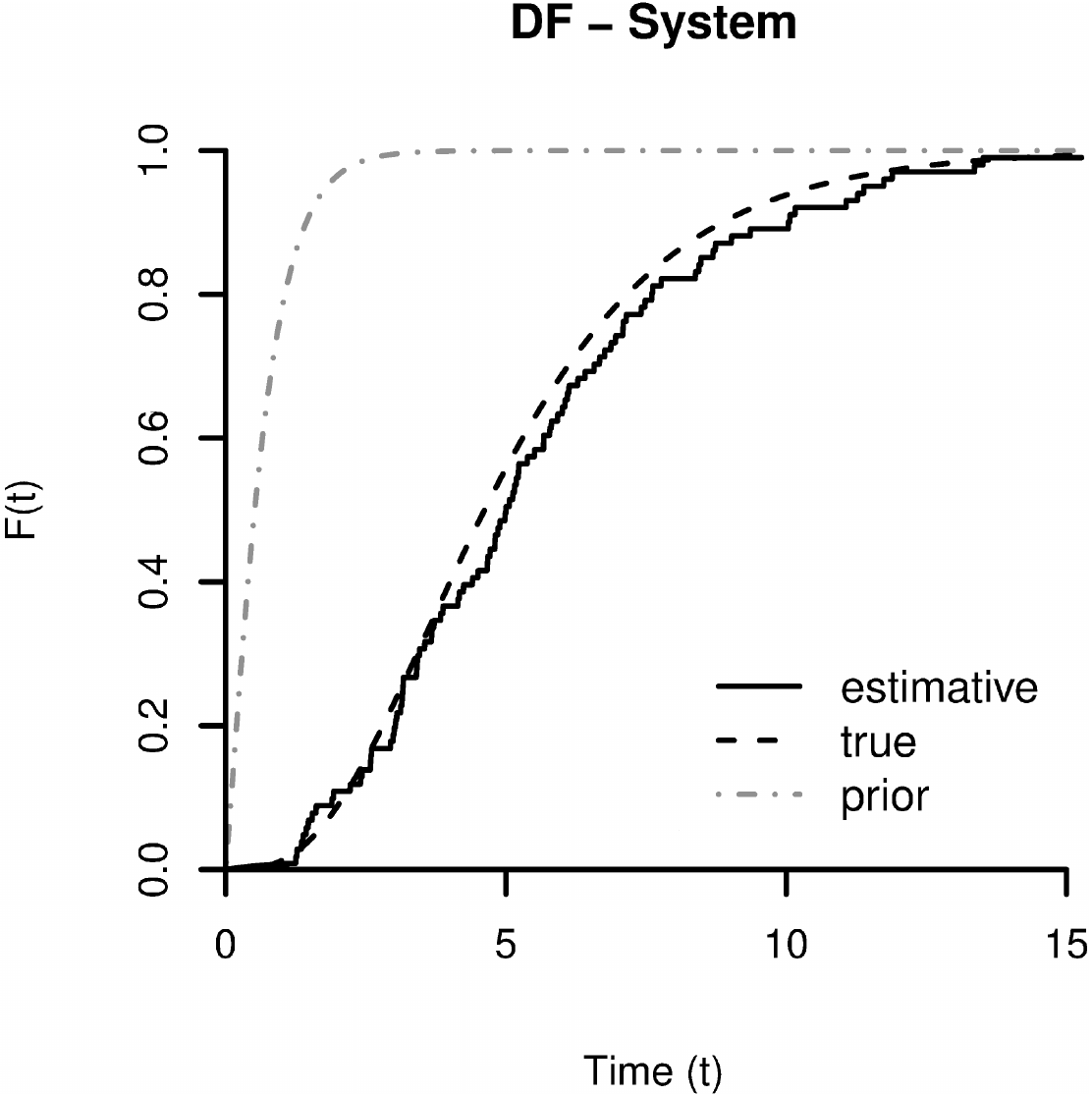}
		\subcaption{ }
	\end{minipage}
	\caption{Estimates for the Example \ref{ex-sps4}.}
	\label{est_graf:sub}
  \end{figure}

	Figure \ref{est_graf:sub} presents the estimates of the five distribution functions associated with components 1--4 and the system. In all plots, the {\it true} distribution functions (dashed lines) and the prior mean (dashed-dot line) are also illustrated. The conditional reliabilities of the components relative to the system are $\widehat{\rho}_1 \cong 0.2294$, $\widehat{\rho}_2 \cong 0.3581$, $\widehat{\rho}_3 \cong 0.2690$, and $\widehat{\rho}_4 \cong 0.1403$.
                 
\subsection{Simulated dataset 2}
	\label{ex-pss4}
	
	In this example, we considered a PSS represented in Figure \ref{PSS_general}. One of the components had the distribution function of a mixture of an exponential distribution and a discrete distribution, with positive probability to fail at times $1$ and $3$, and is given by
	\begin{equation}
	\label{eq_ex2-dist3}
	F(t) = \left\{ \begin{array}{ll}    0, & \textnormal{if}~ t \leq 0, \\
	0.6(1-e^{-t/4}), & \textnormal{if}~ t < 1, \\
	0.6(1-e^{-t/4})+0.25, & \textnormal{if}~ t < 3, \\
	0.6(1-e^{-t/4})+0.4, & \textnormal{if}~ t \geq 3. \\ \end{array} \right.
	\end{equation}
	
	We obtained $100$ observed systems, where the first component ($Z_1$) had a gamma distribution with a mean of $4$ and a SD of $2.83$, the second component ($Z_2$) had a Weibull distribution with a mean of $4.51$ and a SD of $3.06$, the third component ($Z_3$) had a mixture of an exponential and a discrete distribution (\ref{eq_ex2-dist3}) with a mean of $3.1$ and a SD of $3.34$, and the fourth component ($Z_4$) had a log-normal distribution with a mean of $4.59$ and a SD of $2.45$. The Bayesian estimators are based on $(T_1, \delta_1),\ldots,(T_{100}, \delta_{100})$. The simulated values are listed in Table \ref{tabela_data2} in Appendix \ref{apeD}.
	
	To estimate the parameters for components 1 through 4, we rewrote the representation of the system as follows. For component 1, we considered that $X_1 = \min(Z_3,Z_4)$, $X_2 = Z_1$, and $X_3 = Z_2$; then, $\widehat{F}_{Z_1}(t) = \widehat{F}_2(t)$, where $\widehat{F}_{Z_1}$ is the DF estimate of component 1, and $\widehat{F}_2$ is the proposed estimator for PSS (\ref{eq_main-ps}). In a similar way, for component 2, we considered $X_1 = \min(Z_3,Z_4)$, $X_2 = Z_2$, and $X_3 = Z_1$; for component 3, we considered $X_1 = \min(Z_1,Z_2)$, $X_2 = Z_3$, and $X_3 = Z_4$; and for component 4, we considered $X_1 = \min(Z_1,Z_2)$, $X_2 = Z_4$, and $X_3 = Z_3$ (see Section \ref{sec_general}). We found that the proportion of the censored data for components 1 through 3 is $71\%$, and that for component 4 is $87\%$.
	
\begin{figure}[h]\centering
	\begin{minipage}[b]{0.32\linewidth}
		\includegraphics[width=\linewidth]{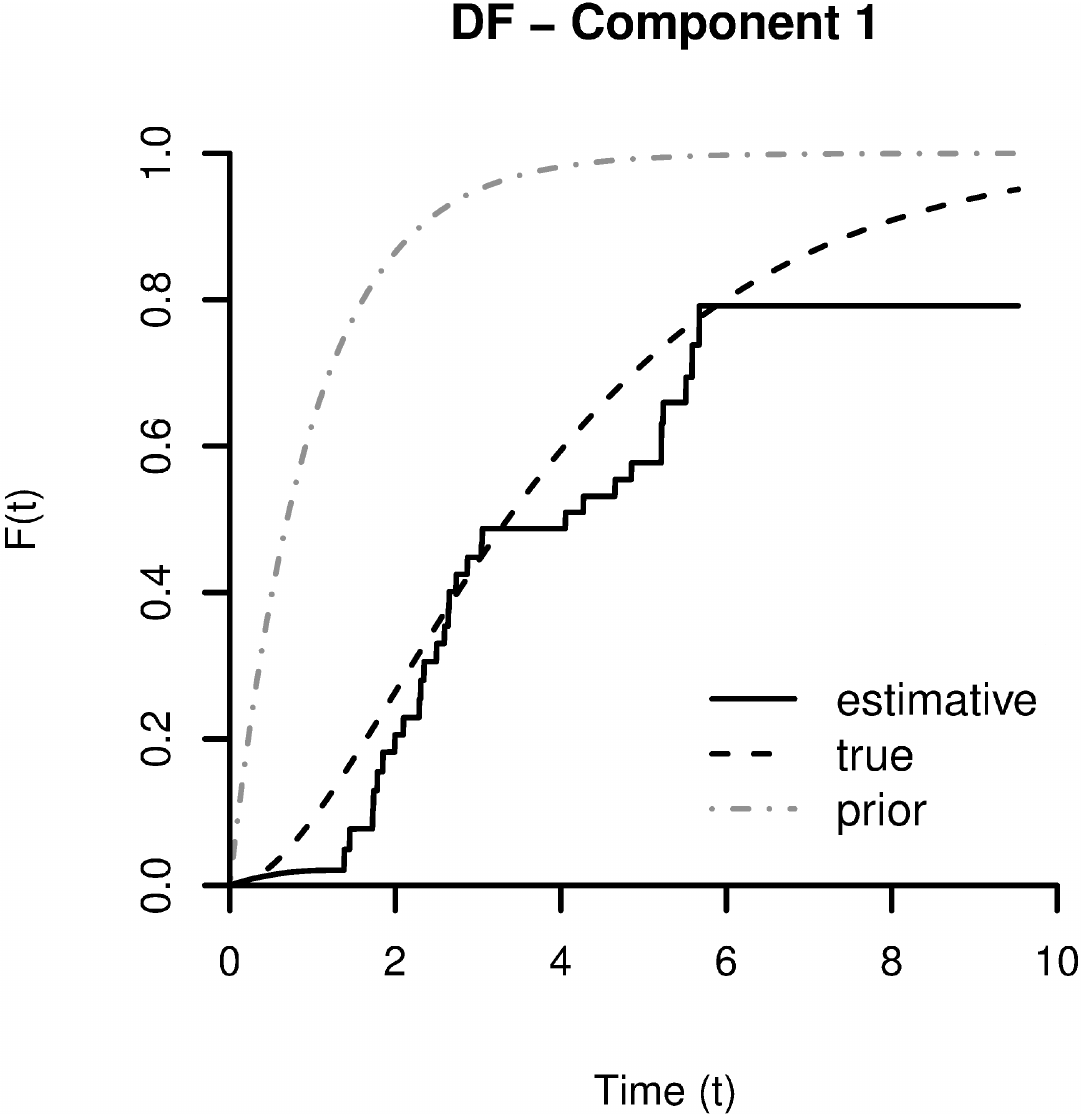}
		\subcaption{  }
	\end{minipage} 
	\begin{minipage}[b]{0.32\linewidth}
		\includegraphics[width=\linewidth]{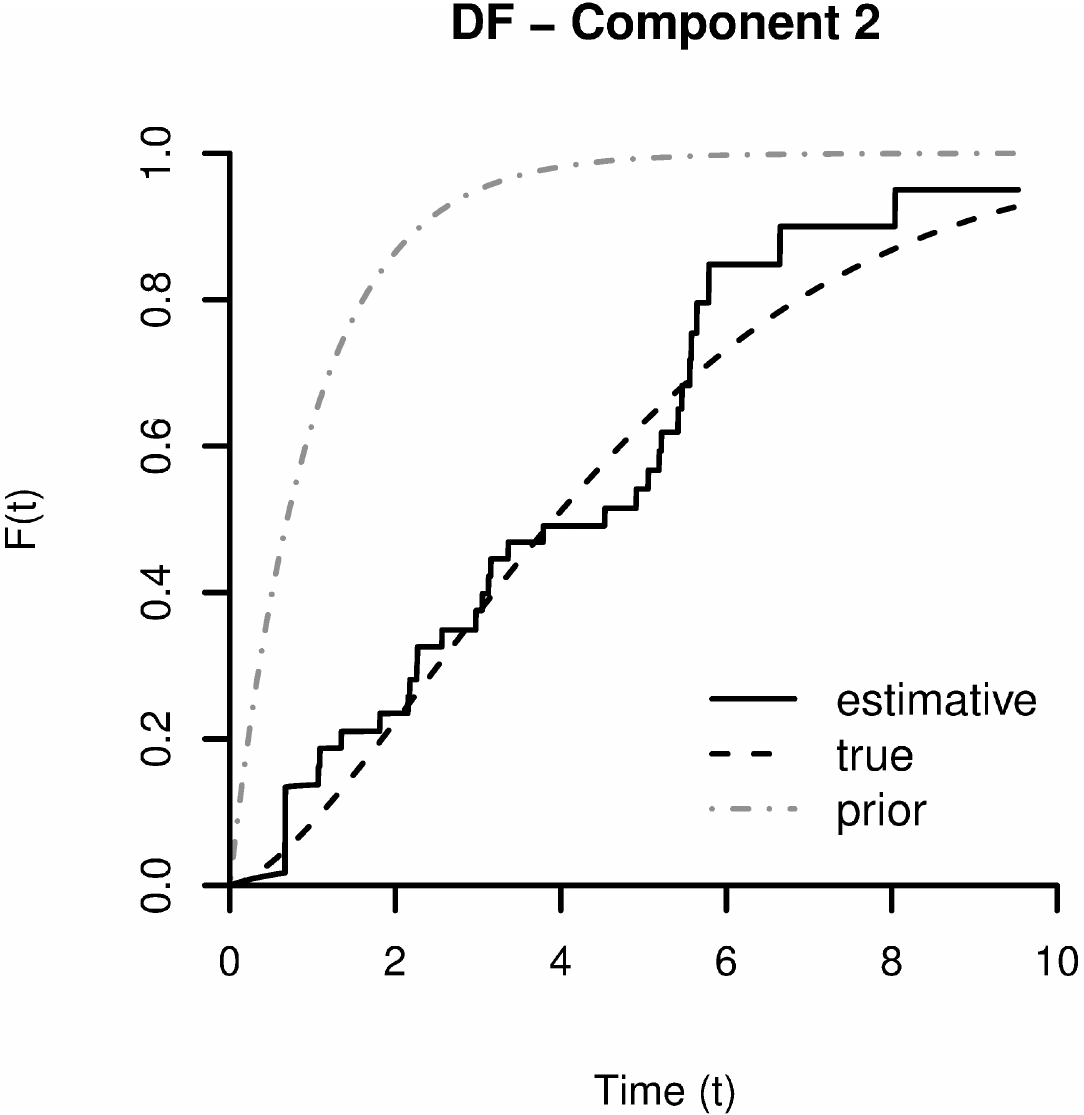}
		\subcaption{  }
	\end{minipage}
	\begin{minipage}[b]{0.32\linewidth}
		\includegraphics[width=\linewidth]{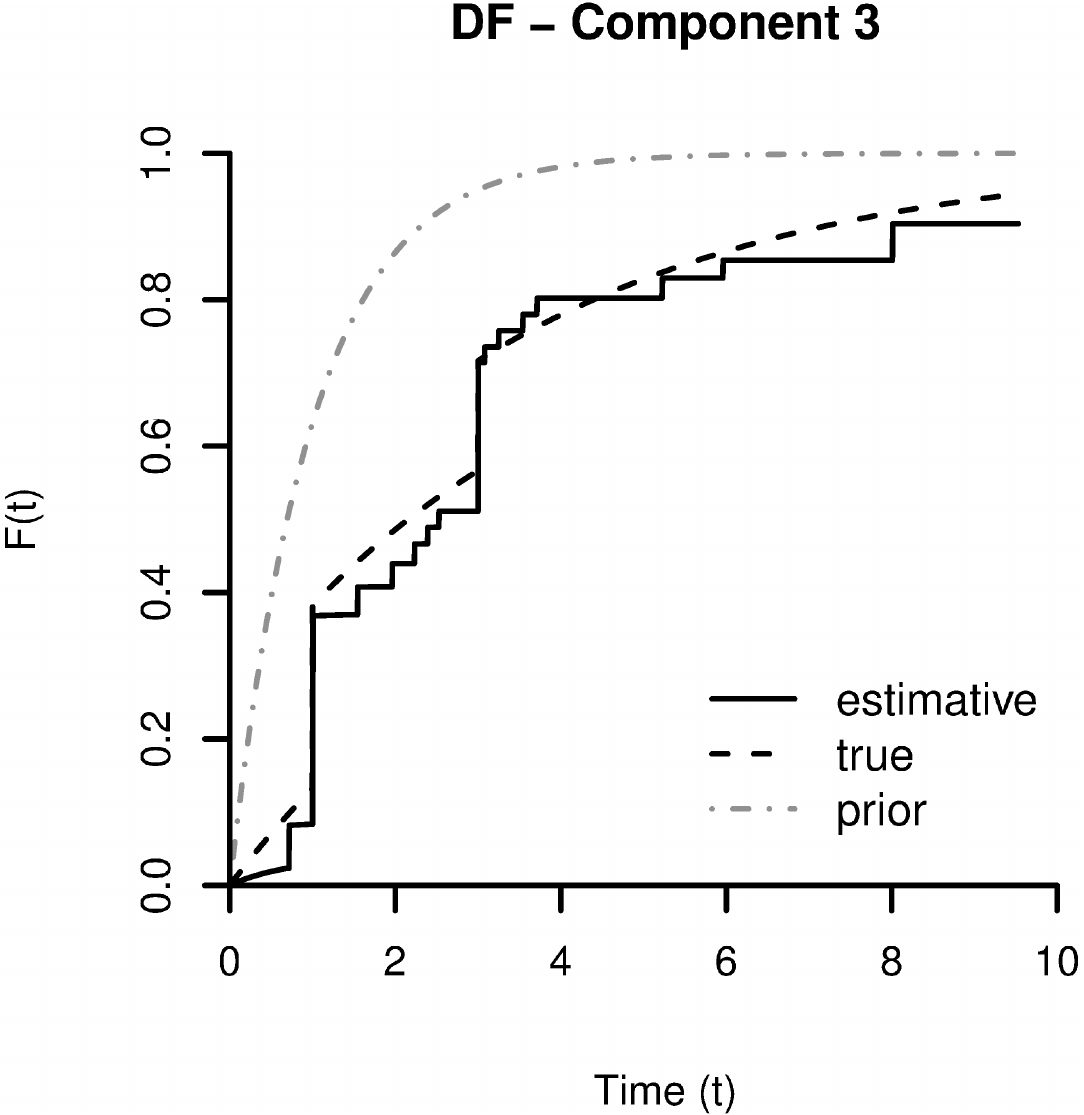}
		\subcaption{ }
	\end{minipage}	
	\begin{minipage}[b]{0.32\linewidth}
		\includegraphics[width=\linewidth]{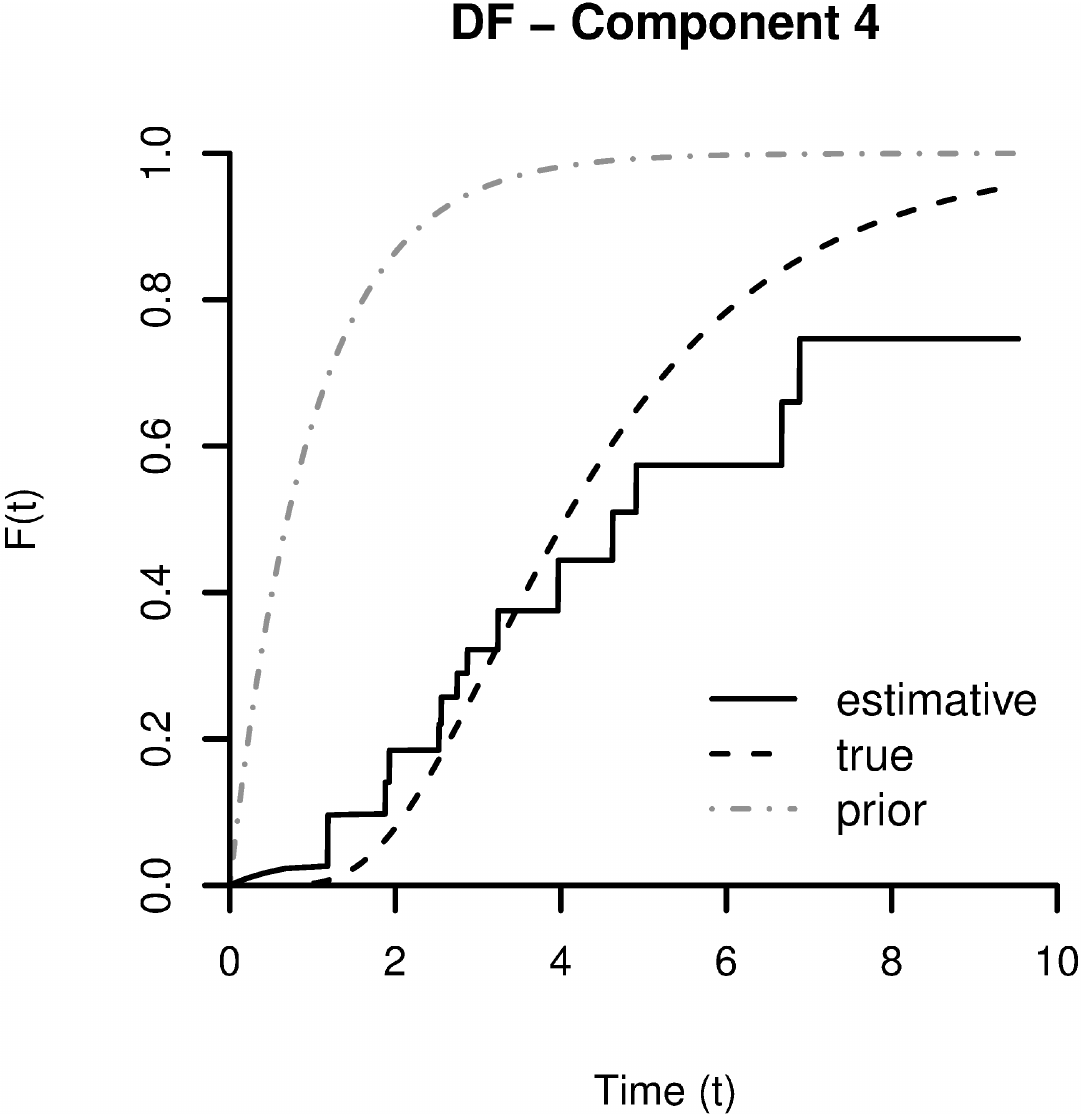}
		\subcaption{ }
	\end{minipage}
	\begin{minipage}[b]{0.32\linewidth}
		\includegraphics[width=\linewidth]{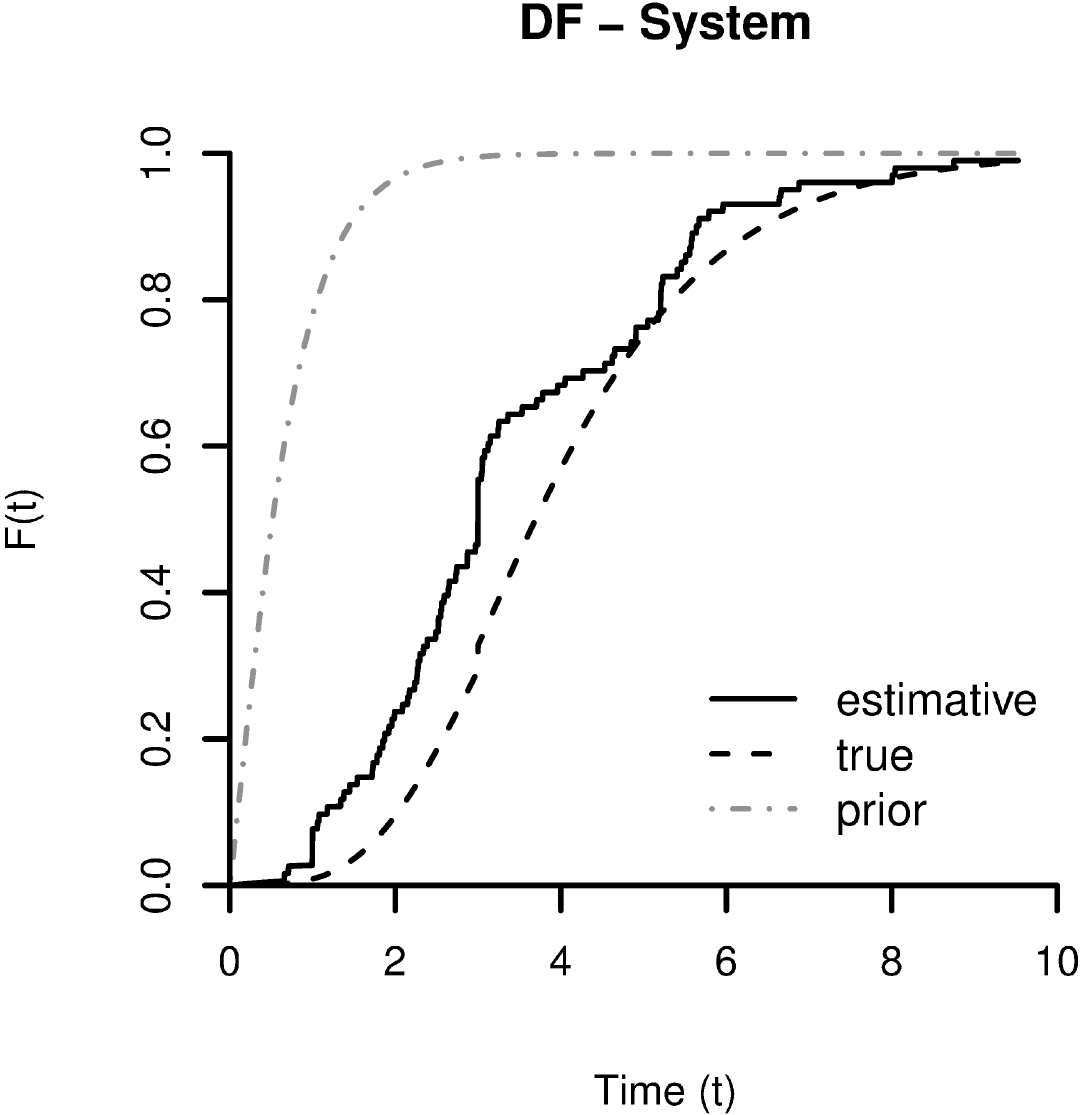}
		\subcaption{ }
	\end{minipage}
	\caption{Estimates for the Example \ref{ex-pss4}.}
	\label{fig_ex2-est:sub:1}
\end{figure}

	Figure  \ref{fig_ex2-est:sub:1} presents the estimates of the five distribution functions: components 1 through 4 and the system. In all the plots, the {\it true} distribution functions (dashed lines) and the prior mean (dashed-dot line) are also illustrated. The conditional reliabilities of the components relative to the system are $\widehat{\rho}_1 \cong 0.2887$, $\widehat{\rho}_2 \cong 0.2887$, $\widehat{\rho}_3 \cong 0.2887$, and $\widehat{\rho}_4 \cong 0.1303$.

\chapter{Weibull model} 
\label{weibull_model}

We present a Bayesian parametric approach for inferences about components' reliabilities in coherent systems when system failure time and status of each component in moment of system failure are available.  
For that, the method does not need the assumption of identically distributed components lifetimes. The main assumption is that components' lifetime are mutually independent.  

We assume the three-parameter Weibull distribution as the component failure time distribution,  presented in Section \ref{Weib_dist}, a very general distribution that can approximate most of the lifetimes distributions. The Weibull model is presented in Section \ref{Wei_model}. In Section \ref{sec_falha_conh_ex} simulated datasets are considered  and in Section \ref{DeviceG_ex} a real dataset is considered in order to show the applicability of presented model. 

\section{Three-parameter Weibull distribution}
\label{Weib_dist}
The Weibull distribution was thus named after Waloddi Weibull presented it to the world scientific community in 1951 \citep{Weibull1951}, but previously considered this distribution in data of resistance of materials \citep{Weibull1939}.

The Weibull distribution is undoubtedly one of the most popular models in statistics because of its ability to fit data from a variety of areas, from survival data to weather data or observations made in economics, hydrology, biology or engineering \citep{Rinne}.

The three-parameter Weibull reliability function is given by
\begin{eqnarray}
R(t \mid {\bm \theta}_j) = \exp\left[-\left(\frac{t-\mu_j}{\eta_j}\right)^{\beta_j}\right], \label{confia_weibull}
\end{eqnarray}
for $t > 0$, where ${\bm \theta}_j= (\beta_j, \eta_j, \mu_j)$ and $\beta_j > 0$ (shape), $\eta_j > 0$ (scale) and $0<\mu_j<t$ (location). The three-parameter Weibull density function is 
\begin{eqnarray}
f(t \mid {\bm \theta}_j)=\frac{\beta_j}{\eta_j}\Bigg(\frac{t-\mu_j}{\eta_j}\Bigg)^{\beta_j-1}\exp
	\left[-\Bigg(\frac{t-\mu_j}{\eta_j}\Bigg)^{\beta_j} \right]. \label{dens_weibull}
\end{eqnarray}

The Weibull distribution has characteristics that make this distribution a great candidate to model components lifetimes. One of them is that variation of parameter values implies changes in both distribution shape and hazard rates. We can have increasing, decreasing and constant failure rates in this family of Weibull distributions \cite{Rinne}. Besides, the three-parameter Weibull family is a very rich family since most real situations will have random aspects that can be represented by an element of the family.  

The Weibull distribution with two parameters ($\mu_j=0$) is the most celebrated case in the literature.  However, the location parameter $\mu_j$ that represents the baseline lifetime has an important meaning in reliability and survival analysis. In reliability, a component under test may be used at an earlier moment and the beginning of the test is not necessarily the beginning of the component's life. 
In medicine for instance, a patient may have the disease before the onset medical appoitment. Not taking account this initial time can underestimates the other parameters. Clearly, in a situation of new component testing $\mu_j$ may be $0$. 

\section{Weibull model}
\label{Wei_model}

We assume that $X_1, \ldots, X_m$ are mutually independent. This assumption is necessary under the considered approach, as it is proved in a Theorem at \citet{rodrigues2018bayesian}.

Considering the reliability and density functions in (\ref{confia_weibull}) and (\ref{dens_weibull}) in the likelihood function presented in (\ref{vero}), the likelihood function for the Weibull parameters ${\bm \theta}_j$ is given by
\begin{eqnarray}
&&{\rm L}({\bm \theta}_j \mid {\bf l}_{j},{\bf u}_{j})  =  \prod_{i=1}^n  {\Bigg\{\frac{\beta_j}{\eta_j}\Bigg(\frac{l_{ji}-\mu_j}{\eta_j}\Bigg)^{\beta_j-1}\exp
	\Bigg[-\Bigg(\frac{l_{ji}-\mu_j}{\eta_j}\Bigg)^{\beta_j} \Bigg]\Bigg\}}^{{\rm I}_{\{l_{ji}=u_{ji}\}}} \nonumber \\
   && {\Bigg\{\exp
   	\Bigg[-\Bigg(\frac{l_{ji}-\mu_j}{\eta_j}\Bigg)^{\beta_j} \Bigg]-
	\exp
	\Bigg[-\Bigg(\frac{u_{ji}-\mu_j}{\eta_j}\Bigg)^{\beta_j} \Bigg]\Bigg\}}^{1-{\rm I}_{\{l_{ji}=u_{ji}\}}} , \label{vero_weibull}
\end{eqnarray}
where ${\rm I}_{\{TRUE\}}=1$ or ${\rm I}_{\{FALSE\}}=0$, ${\bf l_{j}}=(l_{j1},\ldots,l_{jn})$ and ${\bf u_{j}}=(u_{j1},\ldots,u_{jn})$. 

The posterior density of ${\bm \theta}_j= (\beta_j, \eta_j, \mu_j)$ comes out to be
\begin{eqnarray}
&& \pi(\beta_j, \eta_j, \mu_j \mid {\bf l}_{j},{\bf u}_{j})  \propto  \nonumber \\ 
&& \pi(\beta_j, \eta_j, \mu_j)\prod_{i=1}^n   \left\{  \left( \frac{l_{ji}-\mu_j}{\eta_j} \right)^{\beta_j-1}\frac{\beta_j}{\eta_j} \exp \left[-\left(\frac{l_{ji}-\mu_j}{\eta_j} \right)^{\beta_j} \right] \right\}^{{\rm I}_{\{l_{ji}=u_{ji}\}}} \nonumber \\
&&\times  \left\{ \exp \left[-\left(\frac{l_{ji}-\mu_j}{\eta_j}\right)^{\beta_j}\right] - \exp\left[-\left(\frac{u_{ji}-\mu_j}{\eta_j}\right)^{\beta_j}\right]\right\}^{1-{\rm I}_{\{l_{ji}=u_{ji}\}}}, \label{posteriori}
\end{eqnarray}
for which $\pi(\beta_j, \eta_j, \mu_j)$ is the prior density of ${\bm \theta}_j=(\beta_j, \eta_j, \mu_j)$ that we consider to be:
\begin{equation}
\pi(\beta_j, \eta_j, \mu_j) =\frac{1}{\eta_j}\frac{1}{\beta_j}. \label{priori}
\end{equation}	


Even (\ref{priori}) being not a proper prior $-$ its integral is not finite $-$  the posterior density in Equation (\ref{posteriori}) is proper, as stated by the following result \citep{rodrigues2018bayesian}.   
	
\begin{Theorem} \label{theo}
	Let a class of non-informative prior given by
	\begin{eqnarray}
	\pi(\beta_j, \eta_j, \mu_j)=\frac{1}{\eta_j \beta_j^b},~~ b\geq 0. \nonumber
	\end{eqnarray}
	Although, for $b \geq 0$, $n = 1$ and the existence of a failure, the posterior in (\ref{posteriori}) is not proper, for $n > 1$, the posterior in (\ref{posteriori}) is proper.
\end{Theorem}

The proof of this result can be seen at \citet{rodrigues2018bayesian}.
The importance of the above result is that one can perform Bayesian inferences even with low prior information. 
   	
Because the posterior density (\ref{posteriori}) has not a closed form, statistical inferences about the parameters can rely on Markov-Chain Monte-Carlo (MCMC) simulations. Here we consider an adaptive-Metropolis-Hasting algorithm with a multivariate distribution \citep{Haario}.

Discarding burn-in (first generated values discarded to eliminate the effect of the assigned initial values for parameters) and jump samples (spacing among generated values to avoid correlation problems) a sample of size $n_p$ from the joint posterior distribution of ${\bm \theta}_j$ is obtained. For the $j$th component, the sample from the posterior can be expressed as $(\beta_{j}^{(1)},\beta_{j}^{(2)},\ldots,\beta_{j}^{(n_p)})$, $(\eta_{j}^{(1)},\eta_{j}^{(2)},\ldots,\eta_{j}^{(n_p)})$ and $(\mu_{j}^{(1)},\mu_{j}^{(2)},\ldots,\mu_{j}^{(n_p)})$. Consequently, posterior quantities of reliability function $R(t\mid {\bm \theta}_j)$ can be easily obtained \citep{RobertCasella}. For instance, the posterior mean of the reliability function is
\begin{eqnarray}
{\rm E}\Big[R(t\mid {\bm \theta}_j) \mid {\bf l}_{j},{\bf u}_{j}\Big] = \frac{1}{n_p}\sum_{k=1}^{n_p}{R\Big(t \mid \bm{\theta}_{j}^{(k)}\Big)},~~ \mbox{for each} ~ t > 0. \label{relia_bayes}
\end{eqnarray}

\section{Simulated datasets}
\label{sec_falha_conh_ex}

\subsection{Parallel system simulated data}
A sample of $n=30$ parallel systems with $m=3$ components was generated. In the generation process, $X_1$ was generated from a log-normal distribution with mean $5.5$ and variance $7$, $X_2$ from a Weibull distribution with mean $4$ and variance $12$ and $X_3$ was generated from three-parameter Weibull distribution with mean $5$ and variance $3$. In a parallel structure, the system works when at least one component works, so the system lifetime is given by $T=\max\{X_{1},X_{2},X_{3}\}$. The generated data are presented in Table \ref{apeB.paralelo}, Appendix \ref{apenB}. In this structure, when a given component is not the last to fail, it is censored to the left. In observed sample, component 1 presents $60\% $ of left-censored data and components 2 and 3 are $70\%$ left-censored data each.

\begin{figure}[h]\centering
	\begin{minipage}[b]{0.32\linewidth}
		\includegraphics[width=\linewidth]{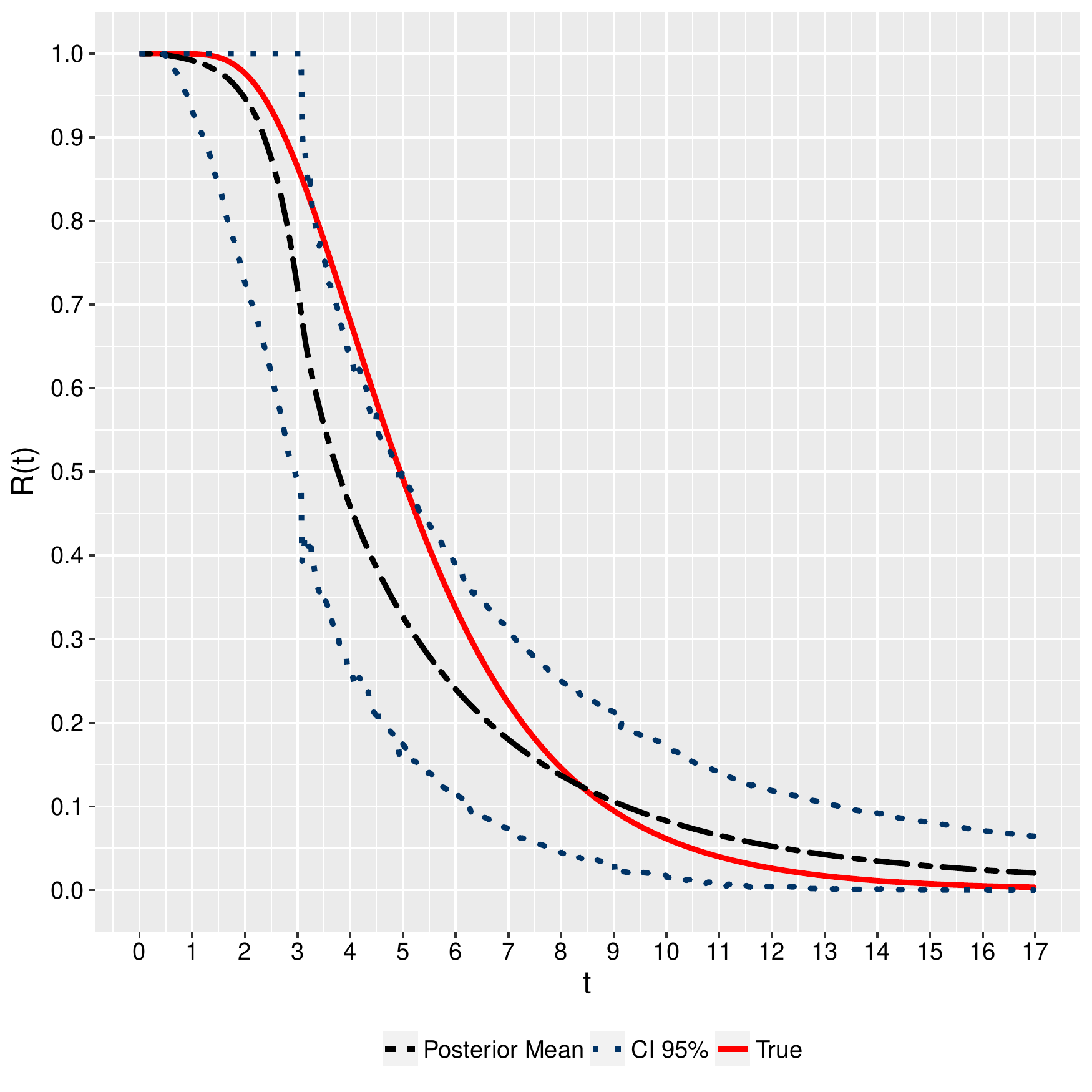}
		\subcaption{Component 1}
	\end{minipage} 
	\begin{minipage}[b]{0.32\linewidth}
		\includegraphics[width=\linewidth]{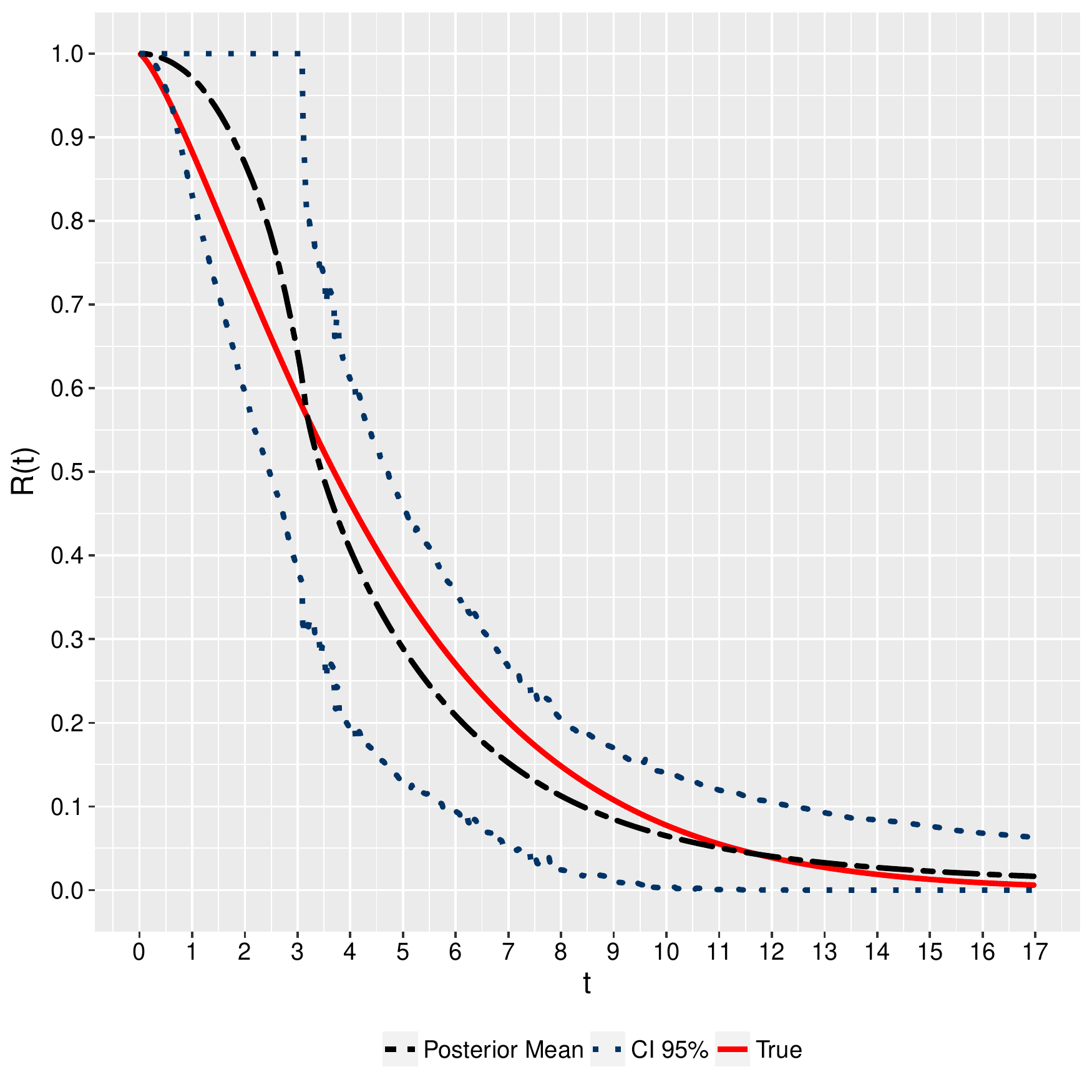}
		\subcaption{Component 2}
	\end{minipage}
	\begin{minipage}[b]{0.32\linewidth}
		\includegraphics[width=\linewidth]{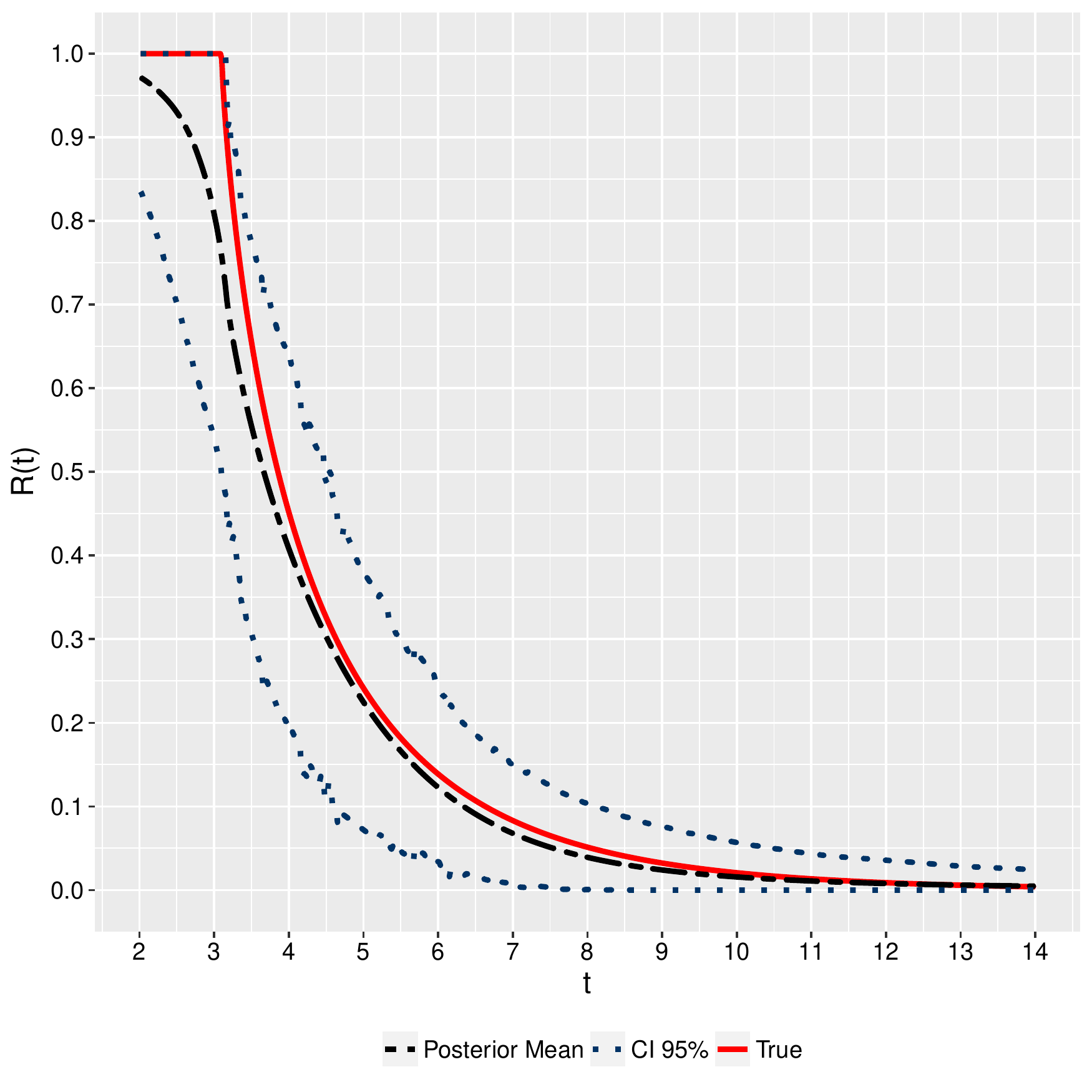}
		\subcaption{Component 3}
	\end{minipage}	
	\caption{Generating reliability functions, posterior means and 95\% HPD intervals (CI 95\%) for the components involved in parallel structure with three components.}
	\label{ex_falhaConhecida_paralelo}
\end{figure}

To obtain posterior quantities based on posterior density (\ref{posteriori}) through MCMC simulations, $20{,}000$ samples were generated, in which the first $10{,}000$ of them were discarded as burn-in samples and jump of size $10$ was chosen to avoid correlation problems. Consequently, samples of size $n_p = 1{,}000$ were obtained of each posterior quantitites. The chains convergence was monitored for good convergence results to be obtained \citep{RobertCasella}. Posterior measures of Weibull parameters $\beta_j$, $\eta_j$ and $\mu_j$, for $j=1,2,3$, are presented in Table \ref{est_exeSimu_paralelo} and posterior measures of $R(t\mid {\bm \theta}_j)$ for some values of $t$ are shown in Table \ref{medidas_confia_paralelo}. The posterior means of reliability functions can be visualized in Figure \ref{ex_falhaConhecida_paralelo} besides the empirical 95\% HPD intervals (Highest Posterior Density). The reliability posterior means are close to the true reliability curves, in general. Even for values of $t$ that the posterior mean is more distant from the true values, the lower or upper limit of HPD interval is very close to the true curve. 

\begin{table}[htbp]
	\centering
	\scriptsize
	\caption{Posterior measures of Weibull parameters for components involved in parallel system.}
	\begin{tabular}{ccccccccl}
		\hline
		\multicolumn{9}{c}{Component 1} \\
		\hline
		& Min   & 1Qt   & Median & Mean & 3Qt   & Max   & SD   & \multicolumn{1}{c}{CI 95\%} \\
		$\beta_1$ & 0.297 & 0.636 & 0.780 & 0.817 & 0.959 & 2.053 & 0.258 & 0.360 - 1.359 \\
		$\eta_1$ & 0.258 & 1.619 & 2.297 & 2.495 & 3.144 & 6.574 & 1.215 & 0.454 - 5.062 \\
	    $\mu_1$   & 0.014 & 1.794 & 2.386 & 2.184 & 2.783 & 3.131 & 0.764 & 0.465 - 3.131 \\
		\hline
		\multicolumn{9}{c}{Component 2} \\
		\hline
		& Min   & 1Qt   & Median & Mean & 3Qt   & Max   & SD    & \multicolumn{1}{c}{CI 95\%} \\
		$\beta_2$  & 0.132 & 0.612 & 0.874 & 0.935 & 1.182 & 2.557 & 0.422 & 0.272 - 1.774 \\
		$\eta_2$ & 0.016 & 1.522 & 2.598 & 2.656 & 3.713 & 7.318 & 1.423 & 0.156 - 5.078 \\
		 $\mu_2$     & 0.005 & 0.709 & 1.656 & 1.642 & 2.581 & 3.156 & 0.998 & 0.109 - 3.137 \\
		\hline
		\multicolumn{9}{c}{Component 3} \\
		\hline
		& Min   & 1Qt   & Median & Mean & 3Qt   & Max   & SD    & \multicolumn{1}{c}{CI 95\%} \\
		$\beta_3$  & 0.164 & 0.734 & 1.005 & 1.157 & 1.449 & 3.635 & 0.581 & 0.340 - 2.421 \\
		$\eta_3$ & 0.124 & 1.008 & 1.557 & 1.883 & 2.528 & 5.405 & 1.142 & 0.250 - 4.328 \\
		 $\mu_3$      & 0.003 & 1.856 & 2.630 & 2.295 & 2.971 & 3.156 & 0.862 & 0.365 - 3.156 \\
		\hline
	\end{tabular}%
	\label{est_exeSimu_paralelo}%
\end{table}%

\begin{table}[htbp]
	\centering
	\scriptsize
	\caption{Posterior measures of reliability functions for some values of $t$ in parallel components.}
	\begin{tabular}{ccccccccc}
		\hline
		\multicolumn{8}{c}{Component 1}          
		                    &  \\
		\hline 
		t     & Min   & 1Qt   & Median & Mean  & 3Qt   & Max   & SD    & CI 95\% \\
		0.50  & 0.899 & 1.000 & 1.000 & 0.998 & 1.000 & 1.000 & 0.009 & 0.997 - 1.000 \\
		10.00 & 0.010 & 0.051 & 0.073 & 0.083 & 0.106 & 0.271 & 0.043 & 0.015 - 0.170 \\
		16.50 & 0.000 & 0.006 & 0.014 & 0.022 & 0.030 & 0.154 & 0.022 & 0 - 0.068 \\
		\hline
		\multicolumn{8}{c}{Component 2}                               &  \\
		\hline
		t     & Min   & 1Qt   & Median & Mean  & 3Qt   & Max   & SD    & CI 95\% \\
		1.00  & 0.565 & 0.975 & 1.000 & 0.971 & 1.000 & 1.000 & 0.061 & 0.831 - 1.000 \\
		11.50 & 0.000 & 0.019 & 0.037 & 0.045 & 0.063 & 0.244 & 0.035 & 0 - 0.112 \\
		17.00 & 0.000 & 0.002 & 0.008 & 0.016 & 0.023 & 0.146 & 0.022 & 0 - 0.063 \\
		\hline
		\multicolumn{8}{c}{Component 3}                               &  \\
		\hline
		t     & Min   & 1Qt   & Median & Mean  & 3Qt   & Max   & SD    & CI 95\% \\
		0.30  & 0.958 & 1.000 & 1.000 & 1.000 & 1.000 & 1.000 & 0.002 & 0.998 - 1.000 \\
		9.00  & 0.000 & 0.007 & 0.017 & 0.024 & 0.034 & 0.182 & 0.025 & 0 - 0.076 \\
		12.50 & 0.000 & 0.000 & 0.002 & 0.007 & 0.007 & 0.140 & 0.013 & 0 - 0.032 \\
		\hline
	\end{tabular}%
	\label{medidas_confia_paralelo}%
\end{table}%

\newpage
\subsection{$2$-out-of-$3$ system simulated data}

A sample of $n=50$ $2$-out-of-$3$ systems was generated. In the generation process, $X_1$ was generated from a Weibull distribution with mean $15$ and variance $8$, $X_2$ from a gamma distribution with mean $18$ and variance $12$ and $X_3$ from a lognormal distribution with mean $20$ and variance $10$. In a $2$-out-of-$3$ structure, the system stops working when the second component fails, so the system lifetime is given by $T=\max\{\min\{X_{1},X_{2}\},$ $\min\{X_{1},X_{3}\},\min\{X_{2},X_{3}\}\}$. The generated data are presented in Table \ref{apeB.2de3}, Appendix \ref{apenB}.

In this structure, when a given component is not the second component to fail, it is censored either to the right or to the left side. In observed sample, component 1 presents $76\% $ of censored data ($ 70 \%$ to the left and 
$6 \%$ to the right), component 2 presents $48\%$ of censored data ($22\%$ to the left and $26\%$ to the right) and component 3 presents $76\%$ of censored data ($8\%$ to the left and $68\% $ to the right).

\begin{figure}[h]\centering
	\begin{minipage}[b]{0.32\linewidth}
		\includegraphics[width=\linewidth]{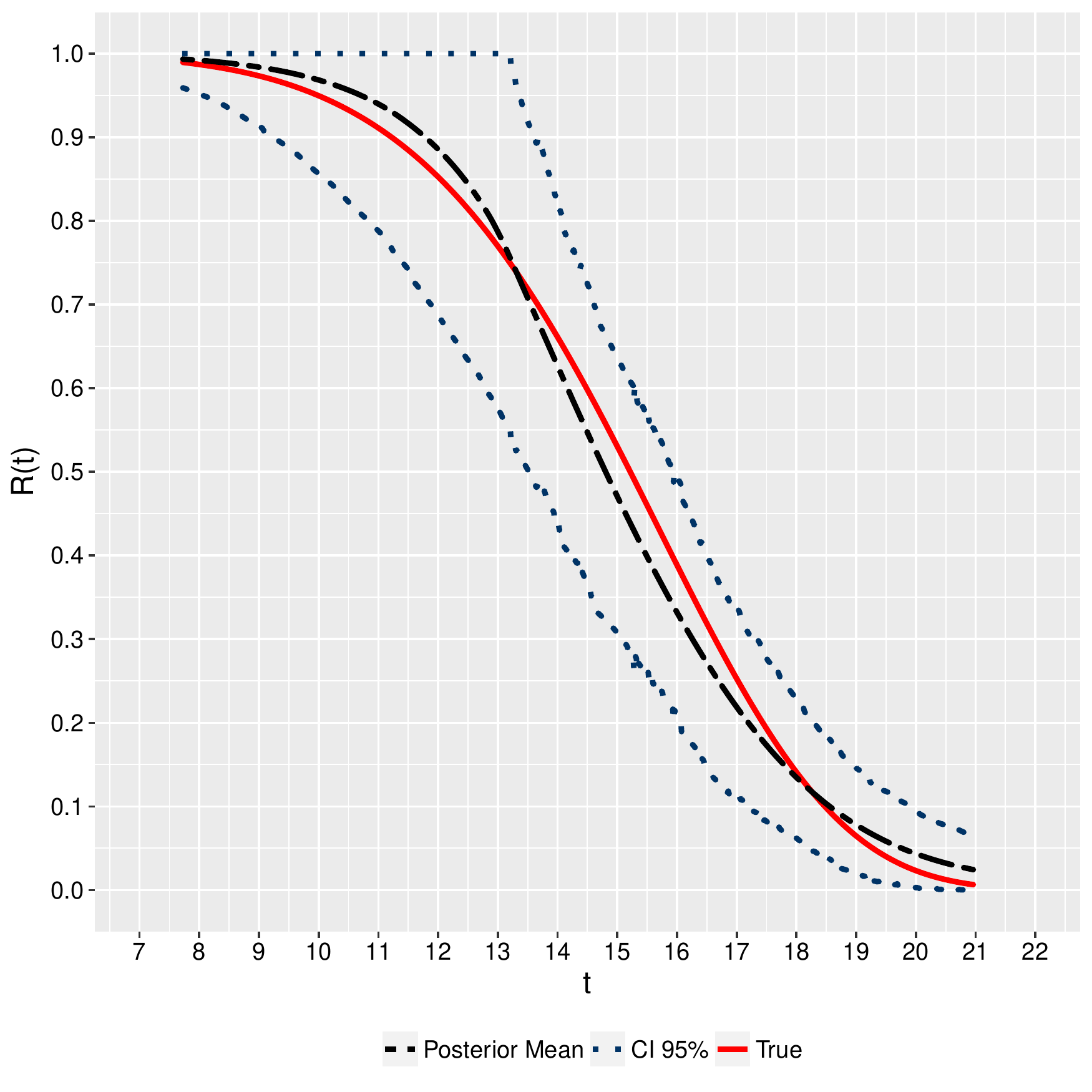}
		\subcaption{Component 1}
	\end{minipage} 
	\begin{minipage}[b]{0.32\linewidth}
		\includegraphics[width=\linewidth]{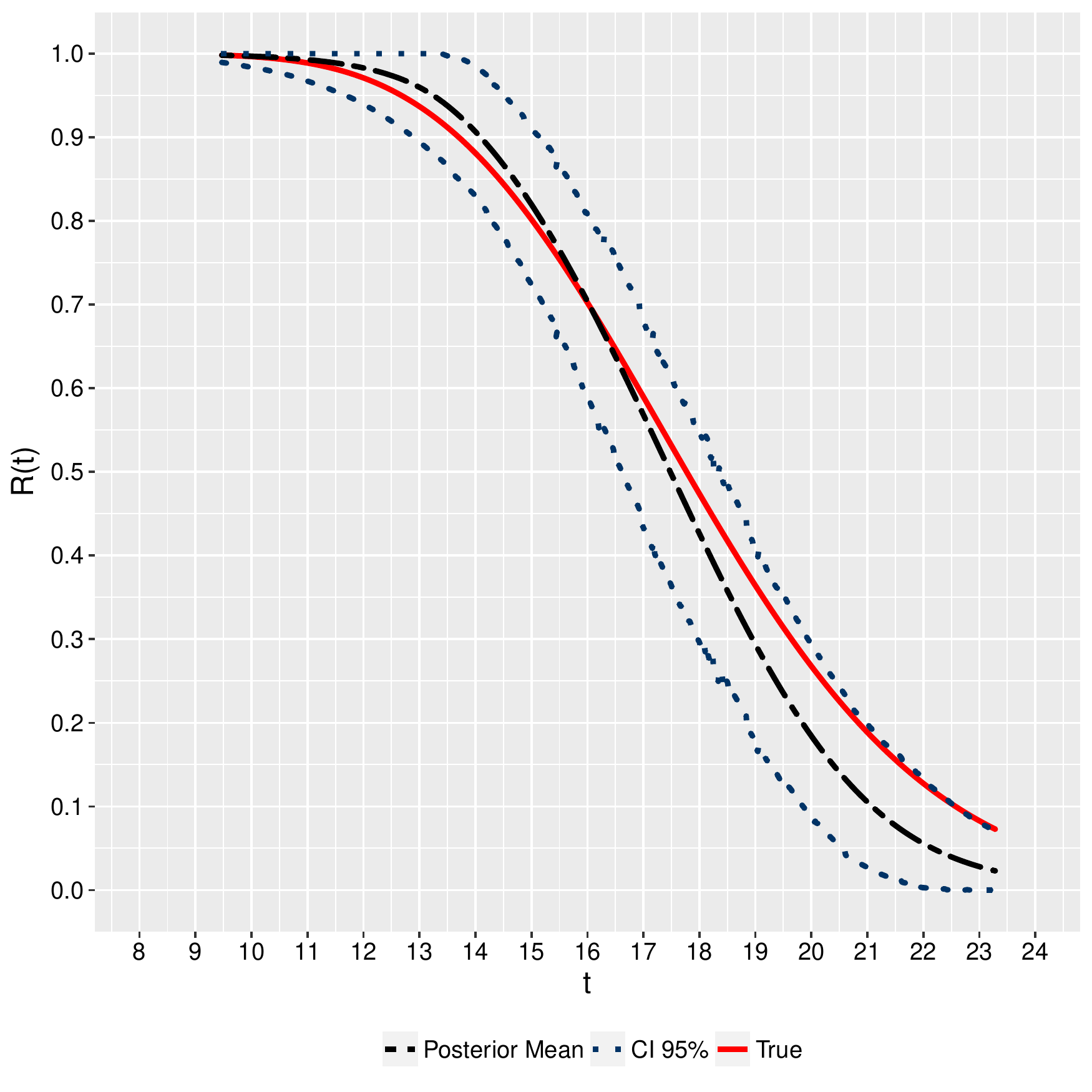}
		\subcaption{Component 2}
	\end{minipage}
	\begin{minipage}[b]{0.32\linewidth}
		\includegraphics[width=\linewidth]{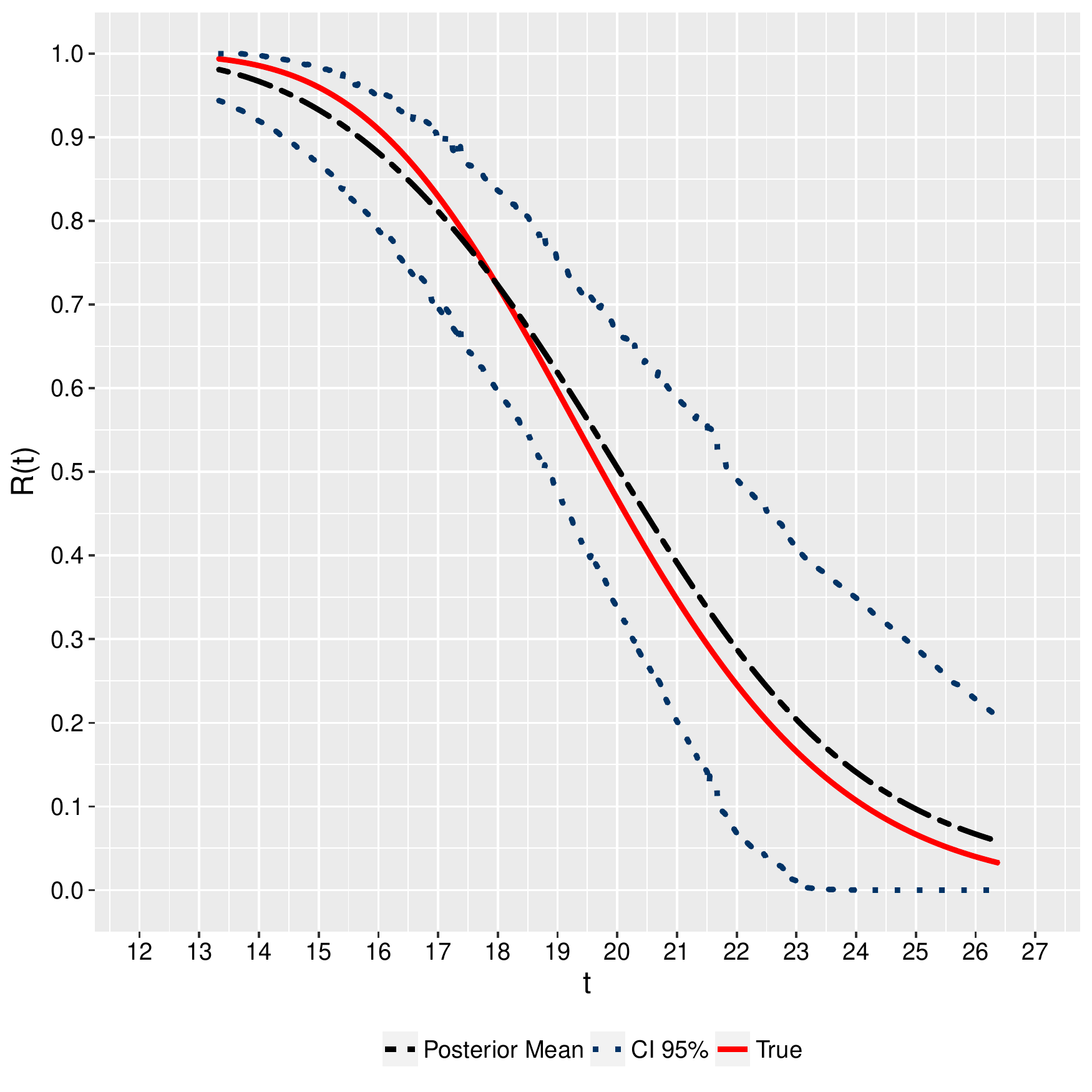}
		\subcaption{Component 3}
	\end{minipage}	
	\caption{Generating reliability functions, posterior means and 95\% HPD intervals (CI 95\%) for the components involved in $2$-out-of-$3$ structure.}
	\label{ex_falhaConhecida_2de3}
\end{figure}

To obtain posterior quantities based on posterior density (\ref{posteriori}) through MCMC simulations, $20{,}000$ samples were generated, in which the first $10{,}000$ of them were discarded as burn-in samples and jump of size $10$ was chosen to avoid correlation problems. Consequently, samples of size $n_p = 1{,}000$ were obtained of each posterior quantitites. The chains convergence was monitored for good convergence results to be obtained \citep{RobertCasella}. Posterior measures of Weibull parameters $\beta_j$, $\eta_j$ and $\mu_j$, for $j=1,2,3$, are presented in Table \ref{est_exeSimu_2de3} and posterior measures of $R(t\mid {\bm \theta}_j)$ for some values of $t$ are shown in Table \ref{medidas_confia}. The posterior means of reliability functions can be visualized in Figure \ref{ex_falhaConhecida_2de3} besides the empirical 95\% HPD intervals. The reliability posterior means are close to the true reliability curves mainly for components 1 and 3. For component 2, even for values of $t$ that the posterior mean is more distant from the true values of the reliability, the upper limit of HPD interval is very close to the true curve.

\begin{table}[htbp]
    \scriptsize
	\centering
	\caption{Posterior measures of Weibull parameters for $2$-out-of-$3$ components.}
	\begin{tabular}{ccccccccc}
		\hline
		\multicolumn{9}{c}{Component 1} \\
		\hline
		& Min & 1Qt & Median & Mean & 3Qt & Max & SD  &  CI 95\%   \\
		$\beta_1$ & 0{.}552 & 1{.}480 & 2{.}281 & 2{.}619 & 3{.}562 & 7{.}297 & 1{.}394 & 0{.}601 - 5{.}376 \\
		$\eta_1$ & 1{.}232 & 3{.}819 & 6{.}262 & 7{.}260 & 10{.}240 & 16{.}540 & 4{.}088 & 1{.}651 -  15{.}101 \\
		$\mu_1$  & 0{.}001 & 5{.}443 & 9{.}328 & 8{.}416 & 11{.}900 & 13{.}310 & 3{.}950  & 0{.}708 - 13{.}212 \\
		\hline
		\multicolumn{9}{c}{Component 2} \\
		\hline
		& Min & 1Qt & Median & Mean & 3Qt & Max & SD  & CI 95\%   \\
		$\beta_2$ & 1{.}006 & 2{.}268 & 3{.}169 & 3{.}622 & 4{.}606 & 9{.}703 & 1{.}693 &  1{.}442 - 7{.}192 \\
		$\eta_2$ & 4{.}107 & 6{.}409 & 8{.}445 & 9{.}612 & 12{.}190 & 19{.}520 & 3{.}968 & 4{.}364 -  17{.}503 \\
		$\mu_2$ & 0{.}035 & 6{.}255 & 9{.}974 & 8{.}816 & 12{.}020 & 13{.}350 & 3{.}813 & 
		1{.}077 - 13{.}347 \\
		\hline
		\multicolumn{9}{c}{Component 3}  \\
		\hline
		& Min & 1Qt & Median & Mean & 3Qt & Max & SD  & CI 95\%   \\
		$\beta_3$ & 0{.}922 & 2{.}381 & 3{.}396 & 3{.}992 & 5{.}243 & 12{.}540 & 2{.}028 & 1{.}318 - 8{.}005 \\
		$\eta_3$ & 6{.}135 & 9{.}210 & 11{.}810 & 12{.}750 & 15{.}740 & 24{.}440 & 4{.}107 & 7{.}122 - 20{.}820 \\
		$\mu_3$ & 0{.}001 & 5{.}656 & 9{.}752 & 8{.}666 & 12{.}290 & 13{.}350 & 4{.}035 & 0{.}837 -  13{.}353 \\
		\hline
	\end{tabular}%
	\label{est_exeSimu_2de3}%
\end{table}%

\begin{table}[htbp]
   \scriptsize
	\centering
	\caption{Posterior measures of reliability functions for some values of $t$ in $2$-out-of-$3$ components.}
	\begin{tabular}{cccccccccc}
		\hline
		\multicolumn{10}{c}{Component 1}  \\
		\hline
		t   & Min & 1Qt & Median & Mean & 3Qt & Max & SD  & CI 95\% & \\
		7{.}70 & 0{.}811 & 0{.}997 & 1{.}000 & 0{.}994 & 1{.}000 & 1{.}000 & 0{.}017  & 0{.}959 - 1{.}000 \\
		15{.}00 & 0{.}204 & 0{.}405 & 0{.}471 & 0{.}471 & 0{.}532 & 0{.}725 & 0{.}090 & 0{.}307 - 0{.}637 \\
		20{.}00 & 0{.}002 & 0{.}023 & 0{.}038 & 0{.}043 & 0{.}056 & 0{.}191 & 0{.}027  & 0{.}003 - 0{.}094 \\
		\hline
		\multicolumn{10}{c}{Component 2} \\
		\hline
		t   & Min & 1Qt & Median & Mean & 3Qt & Max & SD   & CI 95\% &  \\
		9{.}45 & 0{.}955 & 0{.}999 & 1{.}000 & 0{.}998 & 1{.}000 & 1{.}000 & 0{.}004 &  0{.}990 - 1{.}000 \\
		16{.}00 & 0{.}520 & 0{.}665 & 0{.}705 & 0{.}703 & 0{.}745 & 0{.}858 & 0{.}058 & 0{.}588 - 0{.}807 \\
		22{.}00 & 0{.}001 & 0{.}028 & 0{.}048 & 0{.}055 & 0{.}074 & 0{.}259 & 0{.}037 & 0{.}003 - 0{.}131 \\
		\hline
		\multicolumn{10}{c}{Component 3} \\
		\hline
		t   & Min & 1Qt & Median & Mean & 3Qt & Max & SD  & HPD 95\% &  \\
		14{.}00 & 0{.}830 & 0{.}955 & 0{.}973 & 0{.}967 & 0{.}985 & 0{.}999 & 0{.}024 & 0{.}919 - 0{.}997 \\
		22{.}00 & 0{.}038 & 0{.}205 & 0{.}277 & 0{.}288 & 0{.}368 & 0{.}654 & 0{.}113 & 0{.}067 - 0{.}492 \\
		26{.}00 & 0{.}000 & 0{.}007 & 0{.}035 & 0{.}067 & 0{.}100 & 0{.}482 & 0{.}079 & 0{.}000 - 0{.}227 \\
		\hline
	\end{tabular}%
	\label{medidas_confia}%
\end{table}%

\newpage

\subsection{Bridge system simulated data}

A sample of $n=100$ bridge systems as representation in Figure \ref{fig:bridgenew} was simulated. In the generation process, $X_1$ was generated from a Weibull distribution with mean $17$ and variance $8$, $X_2$ from a log-normal distribution with mean $16$ and variance $22$, $X_3$ from a log-normal distribution with mean $15$ and variance $15$, $X_4$ from a gamma distribution with mean $15$ and variance $6$ and $X_5$ from a gamma distribution with mean $20$ and variance $12$. In a bridge structure, the system stops depending on a combination of components failures, so the system lifetime is given by $T = \max \{\min\{X_{1},X_{4}\}, \min\{X_{2},X_{5}\},\min\{X_{1},X_{3},X_{5}\},\min\{X_{2},X_{3},X_{4}\}\}$. The generated data are presented in Table \ref{apeB.bridge}, in Appendix \ref{apenB}. In observed sample, component 1 presents $83\% $ of censored data ($31\%$ to the left and 
$52\%$ to the right), component 2 presents $71\%$ of censored data ($49\%$ to the left and $22\%$ to the right), component 3 presents $89\%$ of censored data ($64\%$ to the left and $25\% $ to the right), component 4 presents $74\%$ of censored data ($56\%$ to the left and $18\% $ to the right) and component 5 presents $83\%$ of censored data ($11\%$ to the left and $72\% $ to the right).

\begin{figure}[h!]\centering
	\begin{minipage}[b]{0.32\linewidth}
		\includegraphics[width=\linewidth]{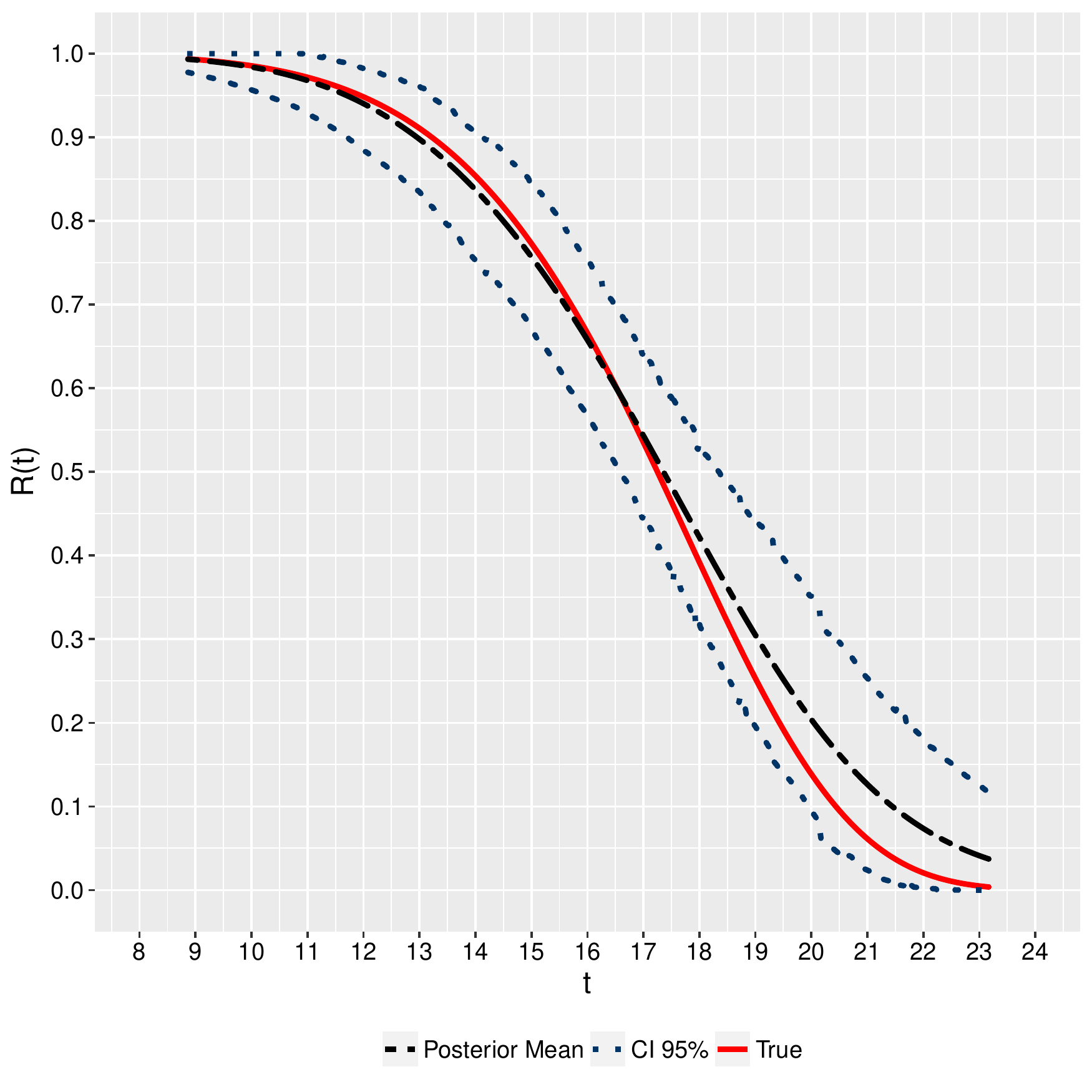}
		\subcaption{Component 1}  
	\end{minipage} 
	\begin{minipage}[b]{0.32\linewidth}
		\includegraphics[width=\linewidth]{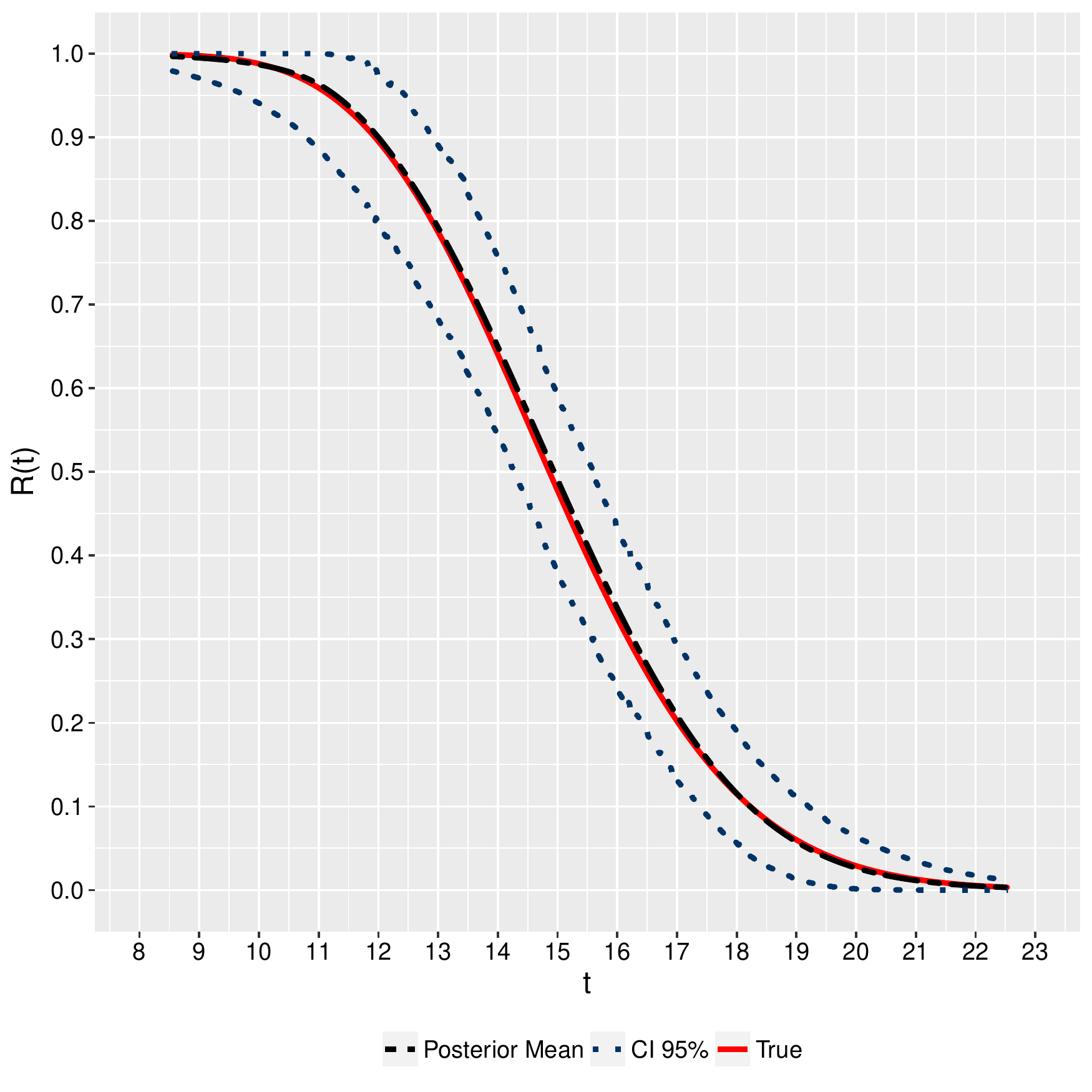}	
		\subcaption{Component 2}  
	\end{minipage}
	\begin{minipage}[b]{0.32\linewidth}
		\includegraphics[width=\linewidth]{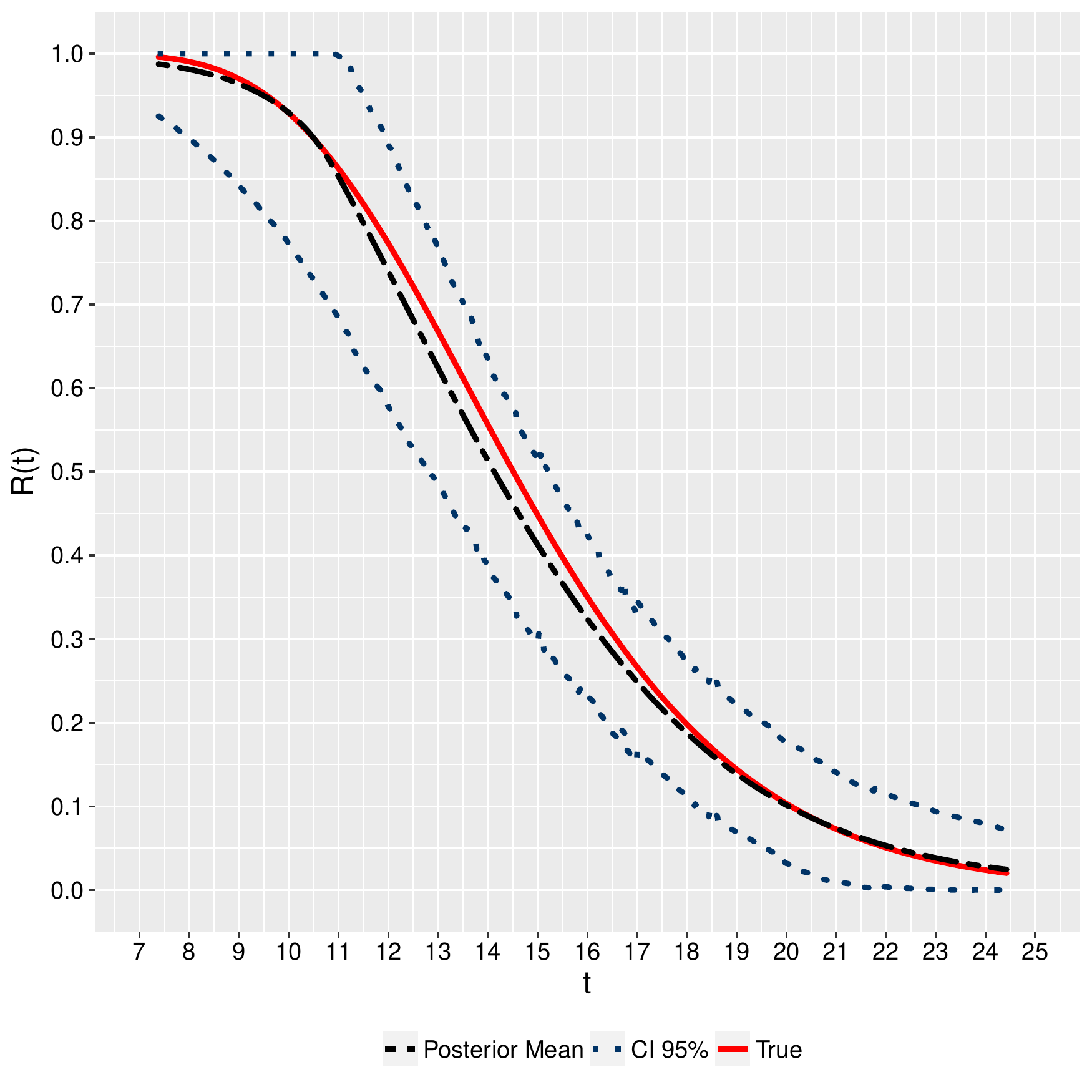}
		\subcaption{Component 3} \label{ex_simu_chin_c}
	\end{minipage}	
	\begin{minipage}[b]{0.32\linewidth}
		\includegraphics[width=\linewidth]{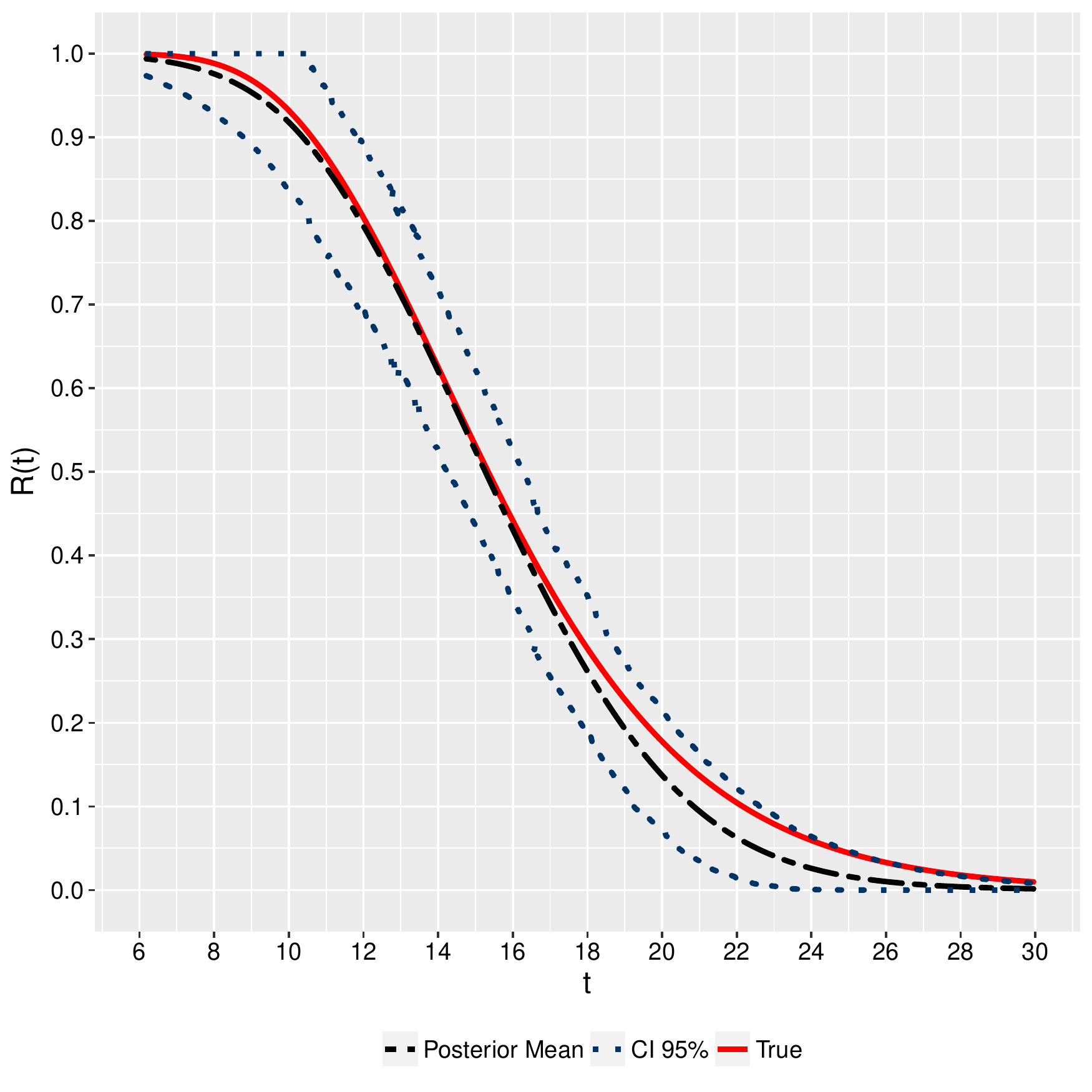}
		\subcaption{Component 4}   
	\end{minipage}	
	\begin{minipage}[b]{0.32\linewidth}
		\includegraphics[width=\linewidth]{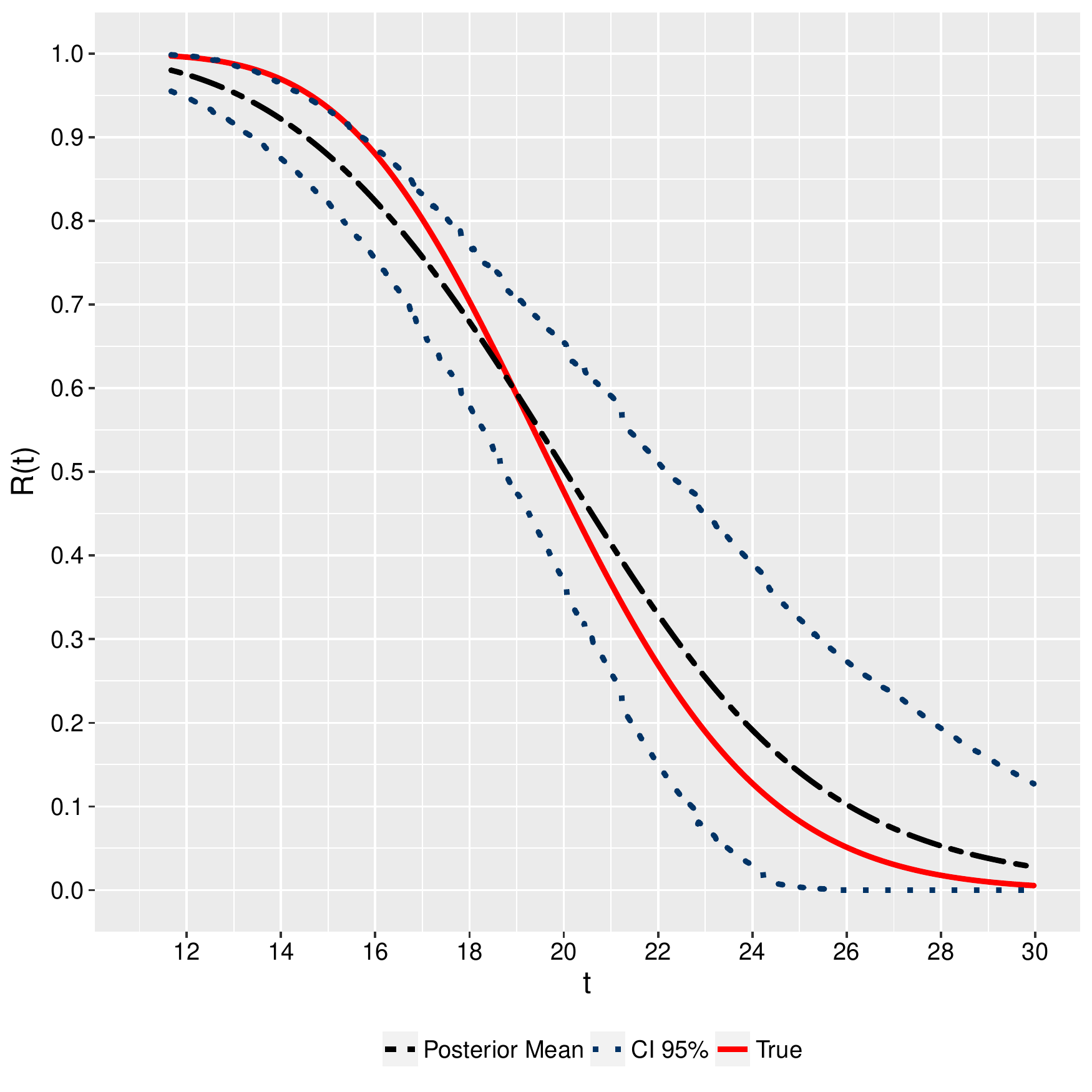}
		\subcaption{Component 5}  
	\end{minipage}	
	\caption{Generating reliability functions, posterior means and 95\% HPD intervals (CI 95\%) for the components involved in the bridge structure.}
	\label{ex_simu_bridge}
\end{figure}

To obtain posterior quantities based on posterior density (\ref{posteriori}) through MCMC simulations, $30{,}000$ samples were generated, in which the first $10{,}000$ of them were discarded as burn-in samples and jump of size $20$ was chosen to avoid correlation problems. Consequently, samples of size $n_p = 1{,}000$ were obtained of each posterior quantitites. The chains convergence was monitored for good convergence results to be obtained \citep{RobertCasella}. Posterior measures of Weibull parameters $\beta_j$, $\eta_j$ and $\mu_j$, for $j=1,\ldots,5$, are presented in Table \ref{est_exeSimu_bridge} and posterior measures of $R(t\mid {\bm \theta}_j)$ for some values of $t$ are shown in Table \ref{medidas_confia_bridge}. The posterior means of reliability functions can be visualized in Figure \ref{ex_simu_bridge} besides the empirical 95\% HPD intervals. The reliability posterior means are close to the true reliability curves for all components, mainly for component 2. Even for values of $t$ that the posterior mean is more distant from the true values, the lower or upper limit of HPD interval is very close to the true curve. 

\begin{table}[htbp]
	\centering
	\scriptsize
	\caption{Posterior measures of Weibull parameters for the components in the bridge structure.}
	\begin{tabular}{ccccccccc}
		\hline
		\multicolumn{9}{c}{Component 1} \\
		\hline
		& Min   & 1Qt   & Median & Mean  & 3Qt   & Max   & SD    & \multicolumn{1}{c}{CI 95\%} \\
		$\beta_1$ & 0.993 & 3.561 & 4.523 & 4.611 & 5.560 & 8.665 & 1.409 & 2.061 - 7.335 \\
		$\eta_1$  & 7.236 & 11.766 & 14.245 & 14.089 & 16.596 & 19.790 & 2.971 & 8.861 - 18.917 \\
		$\mu_1$    & 0.010 & 1.909 & 4.145 & 4.408 & 6.708 & 10.801 & 2.913 & 0.010 - 9.436 \\
		\hline
		\multicolumn{9}{c}{Component 2} \\
         \hline
          & Min   & 1Qt   & Median & Mean  & 3Qt   & Max   & SD    & \multicolumn{1}{c}{CI 95\%} \\
      $\beta_2$ & 1.163 & 2.219 & 2.785 & 2.829 & 3.408 & 5.509 & 0.822 & 1.333 - 4.343 \\
      $\eta_2$  & 5.009 & 9.398 & 11.644 & 11.742 & 14.328 & 17.710 & 3.066 & 6.035 - 16.791 \\
      $\mu_2$   & 0.001 & 2.563 & 5.043 & 4.955 & 7.264 & 10.684 & 2.924 & 0.008 - 9.78 \\
       \hline
		\multicolumn{9}{c}{Component 3} \\
		\hline
		& Min   & 1Qt   & Median & Mean  & 3Qt   & Max   & SD    & \multicolumn{1}{c}{CI 95\%} \\
		$\beta_3$ & 0.568 & 1.272 & 1.608 & 1.830 & 2.174 & 4.951 & 0.776 & 0.772 - 3.534 \\
		$\eta_3$  & 2.607 & 5.124 & 6.389 & 7.558 & 9.088 & 16.909 & 3.241 & 3.534 - 14.742 \\
		$\mu_3$     & 0.018 & 6.399 & 9.036 & 7.907 & 10.273 & 10.899 & 3.029 & 1.282 - 10.899 \\
		\hline
				\multicolumn{9}{c}{Component 4} \\
		\hline
		& Min   & 1Qt   & Median & Mean  & 3Qt   & Max   & SD    & \multicolumn{1}{c}{CI 95\%} \\
		$\beta_4$ & 1.203 & 2.167 & 2.641 & 3.039 & 3.605 & 8.107 & 1.197 & 1.403 - 5.491 \\
		$\eta_4$  & 4.137 & 5.280 & 6.341 & 7.464 & 8.811 & 16.534 & 2.916 & 4.284 - 14.346 \\
		$\mu_4$    & 0.008 & 7.096 & 9.495 & 8.320 & 10.468 & 10.897 & 2.813 & 1.674 - 10.897 \\
		\hline
		\multicolumn{9}{c}{Component 5} \\
		\hline
		& Min   & 1Qt   & Median & Mean  & 3Qt   & Max   & SD    & \multicolumn{1}{c}{CI 95\%} \\
		$\beta_5$ & 1.034 & 2.570 & 3.342 & 3.597 & 4.450 & 7.778 & 1.289 & 1.562 - 6.098 \\
		$\eta_5$  & 9.044 & 12.236 & 14.542 & 15.060 & 17.716 & 24.581 & 3.373 & 9.823 - 21.195 \\
		$\mu_5$    & 0.028 & 4.185 & 7.301 & 6.662 & 9.555 & 10.894 & 3.231 & 0.777 - 10.894 \\
		\hline
	\end{tabular}%
	\label{est_exeSimu_bridge}%
\end{table}%
	
\begin{table}[htbp]
	\centering
	\scriptsize
	\caption{Posterior measures of components' reliability functions for some values of $t$ in the bridge structure.}
	\begin{tabular}{ccccccccc}
		\hline
		\multicolumn{9}{c}{Component 1} \\
		\hline
		t     & Min   & 1Qt   & Median & Mean  & 3Qt   & Max   & SD    & CI 95\% \\
		8.95  & 0.922 & 0.990 & 0.995 & 0.993 & 0.999 & 1.000 & 0.008 & 0.976 - 1.000 \\
		15.00 & 0.598 & 0.727 & 0.758 & 0.757 & 0.790 & 0.869 & 0.046 & 0.671 - 0.845 \\
		22.00 & 0.001 & 0.031 & 0.059 & 0.070 & 0.098 & 0.346 & 0.052 & 0.001 - 0.174 \\
		\hline
        \multicolumn{9}{c}{Component 2} \\
        \hline
        t     & Min   & 1Qt   & Median & Mean  & 3Qt   & Max   & SD    & CI 95\% \\
        6.50  & 0.927 & 0.988 & 0.998 & 0.992 & 1.000 & 1.000 & 0.012 & 0.967 - 1.00 \\
        18.00 & 0.142 & 0.233 & 0.259 & 0.261 & 0.286 & 0.404 & 0.041 & 0.190 - 0.351 \\
        32.00 & 0.000 & 0.000 & 0.000 & 0.001 & 0.000 & 0.012 & 0.002 & 0 - 0.004 \\
        \hline
		\multicolumn{9}{c}{Component 3} \\
		\hline
		t     & Min   & 1Qt   & Median & Mean  & 3Qt   & Max   & SD    & CI 95\% \\
		7.50  & 0.767 & 0.992 & 1.000 & 0.987 & 1.000 & 1.000 & 0.030 & 0.92 - 1.000 \\
		13.00 & 0.348 & 0.576 & 0.626 & 0.624 & 0.677 & 0.834 & 0.074 & 0.485 - 0.769 \\
		24.00 & 0.000 & 0.010 & 0.021 & 0.028 & 0.036 & 0.174 & 0.025 & 0 - 0.079 \\
		\hline
		\multicolumn{9}{c}{Component 4} \\
		\hline
		t     & Min   & 1Qt   & Median & Mean  & 3Qt   & Max   & SD    & CI 95\% \\
		9.00  & 0.933 & 0.996 & 1.000 & 0.995 & 1.000 & 1.000 & 0.011 & 0.971 - 1.000 \\
		14.00 & 0.427 & 0.610 & 0.651 & 0.649 & 0.689 & 0.801 & 0.057 & 0.542 - 0.756 \\
		20.00 & 0.001 & 0.013 & 0.022 & 0.026 & 0.036 & 0.145 & 0.019 & 0.001 - 0.063 \\
		\hline
		\multicolumn{9}{c}{Component 5} \\
		\hline
		t     & Min   & 1Qt   & Median & Mean  & 3Qt   & Max   & SD    & CI 95\% \\
		12.00 & 0.896 & 0.967 & 0.978 & 0.975 & 0.986 & 0.998 & 0.015 & 0.947 - 0.997 \\
		20.00 & 0.295 & 0.455 & 0.503 & 0.503 & 0.550 & 0.697 & 0.073 & 0.371 - 0.654 \\
		29.50 & 0.000 & 0.002 & 0.012 & 0.032 & 0.041 & 0.319 & 0.048 & 0 - 0.142 \\
		\hline
	\end{tabular}%
	\label{medidas_confia_bridge}%
\end{table}%

\newpage	
	
\section{Device-G dataset}
\label{DeviceG_ex}
A dataset presented by \cite{MeekerEscobar} is considered. A sample of $n=30$ units were installed in typical service environments. Cause of failure information was determined for each unit that failed. Mode S failures were caused by an accumulation of randomly occurring damage from power-line voltage spikes during electric storms. 
Mode W failures were caused by normal product wear that began to appear after many cycles of use. The causes W and S competed to be responsible for device-G failure, i.e., it is a system with $2$ components in series. The observed dataset is presented in Table \ref{deviceG_data} (Appendix \ref{apenB}), in which $23.33\%$ of observed system failure because of component W (component S is right-censored), $50\%$ of systems failed because of component S failure (component W is right-censored) and the remaining $26.67\%$ were censored systems and for them, both components are right-censored observations. 

To obtain posterior quantities related to the posterior distribution of $\mbox{\boldmath{$\theta$}}=(\beta,\eta,\mu)$ from posterior distribution in (\ref{posteriori}) through MCMC simulations, we discarded the first $10{,}000$ as burn-in samples and jump of size $10$ to avoid correlation problems, obtaining a sample of size $n_p=1{,}000$. The chains convergence was monitored for all simulation scenarios and good convergence results were obtained.

\begin{figure}[h]\centering
	\begin{minipage}[b]{0.43\linewidth}
		\includegraphics[width=\linewidth]{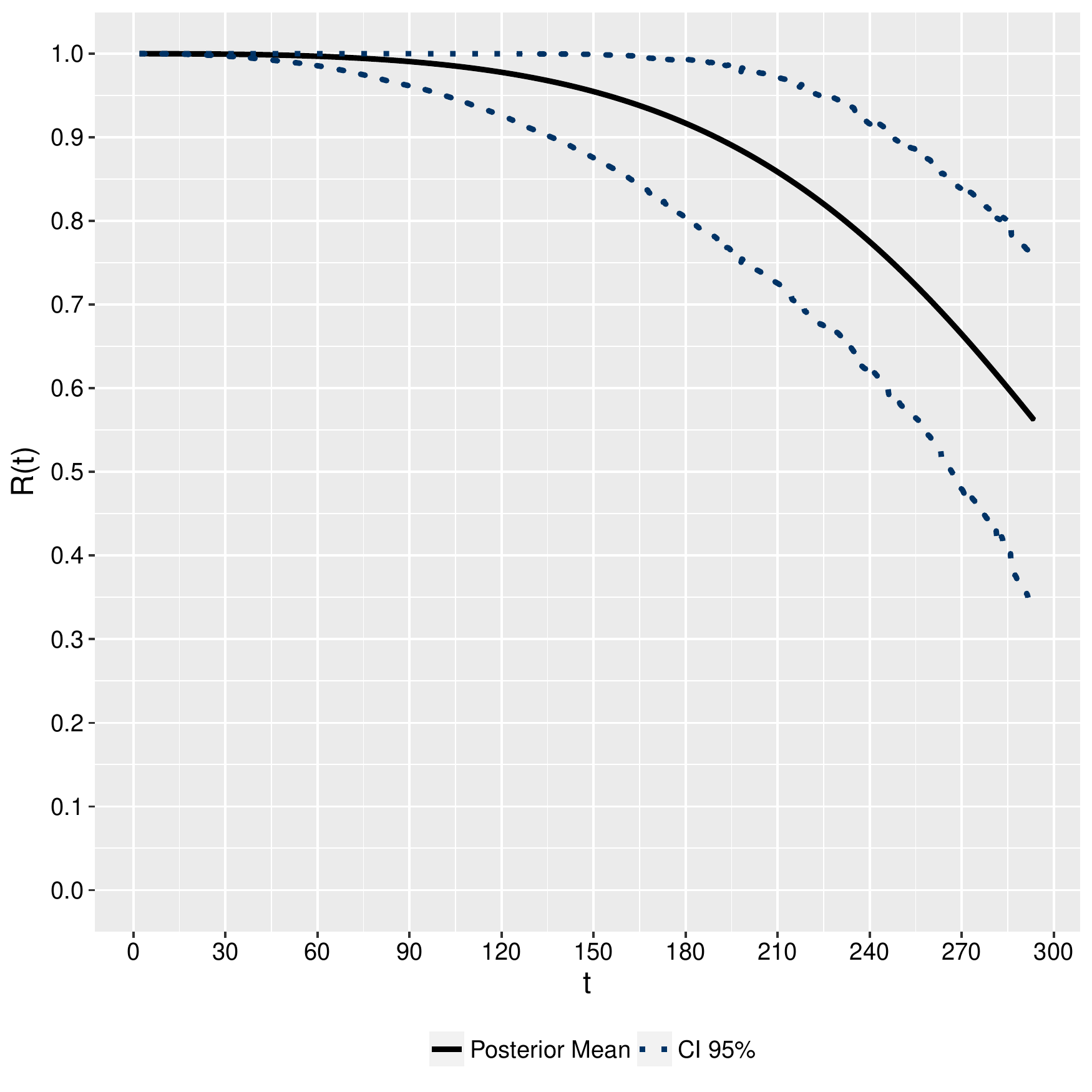}
		\subcaption{Component W}
	\end{minipage} 
	\begin{minipage}[b]{0.43\linewidth}
		\includegraphics[width=\linewidth]{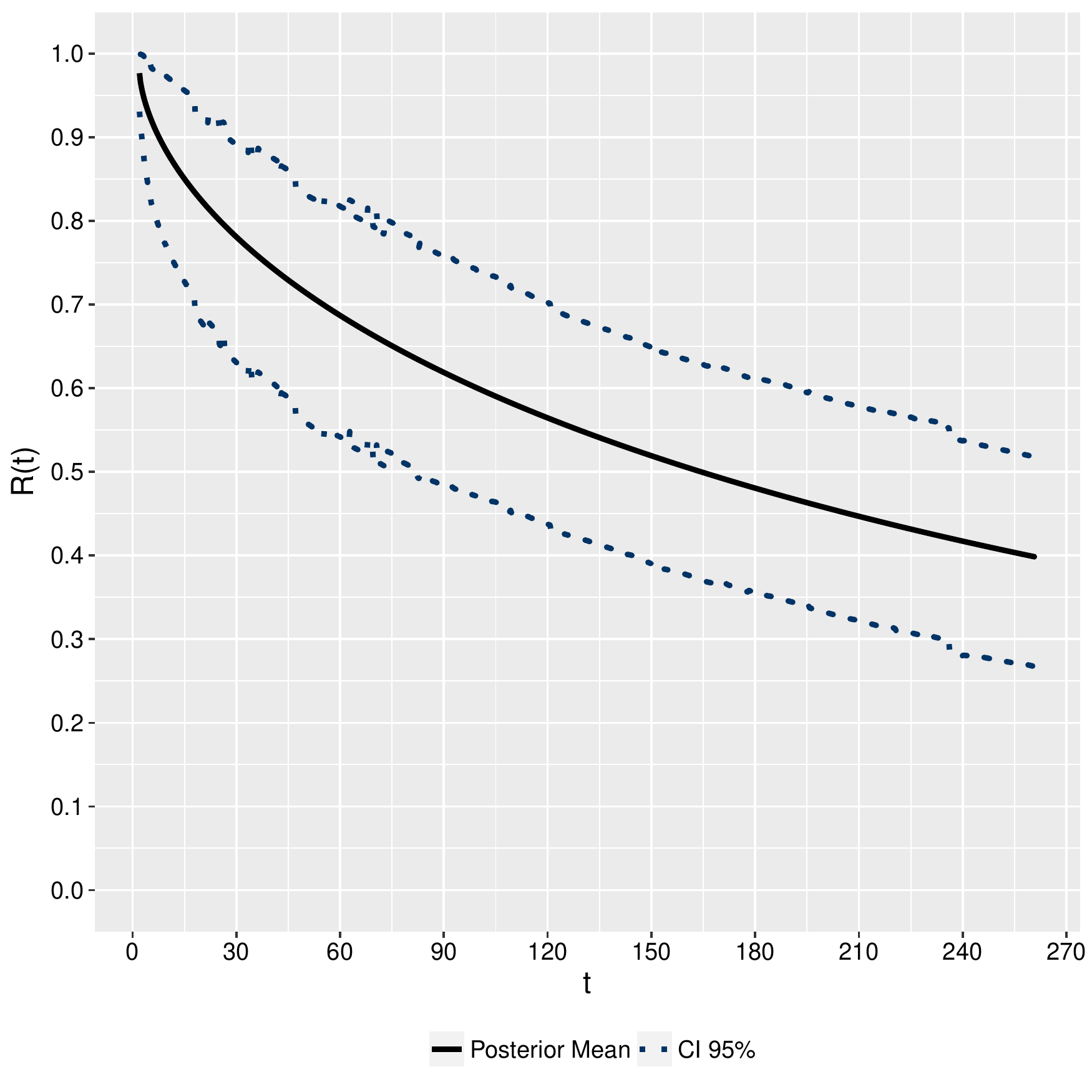}
		\subcaption{Component S}
	\end{minipage}
	\caption{Generating reliability functions, posterior means and 95\% HPD intervals (CI 95\%) for the components involved in Device-G dataset.}
	\label{DeviceG_dataset_curvas}
\end{figure}

Table \ref{est_param_DeviceG} lists the posterior quantities for the parameters of shape ($\beta$), scale ($\eta$) and location ($\mu$). The posterior measures of $R(t\mid {\bm \theta}_j)$ for some values of $t$ are shown in Table \ref{medidas_confia_DeviceG} and posterior means of reliability functions can be visualized in Figure \ref{DeviceG_dataset_curvas} besides the empirical 95\% HPD intervals.

\begin{table}[htbp]
	\centering
	\scriptsize
	\caption{Posterior measures of Weibull parameters for Device-G dataset.}
	\begin{tabular}{ccccccccl}
		\hline
		\multicolumn{9}{c}{Componente W ($j=1$)} \\
		\hline
		& Min   & 1Qt   & Median & Mean  & 3Qt   & Max   & SD    & \multicolumn{1}{c}{CI 95\%} \\
		$\beta_1$ & 1.440 & 3.329 & 4.117 & 4.246 & 4.999 & 9.931 & 1.314 & 1.895 - 6.875 \\
		$\eta_1$ & 265.166 & 315.879 & 335.632 & 341.645 & 360.237 & 581.516 & 37.428 & 283.810 - 418.451 \\
		$\mu_1$  & 0.001 & 0.479 & 1.004 & 0.994 & 1.518 & 1.999 & 0.587 & 0.004 - 1.894 \\
		\hline
		\multicolumn{9}{c}{Componente S ($j=2$)} \\
		\hline
		& Min   & 1Qt   & Median & Mean  & 3Qt   & Max   & SD    & \multicolumn{1}{c}{CI 95\%} \\
		$\beta_2$ & 0.271 & 0.521 & 0.613 & 0.620 & 0.709 & 1.296 & 0.148 & 0.372 - 0.913 \\
		$\eta_2$ & 102.848 & 238.267 & 294.820 & 311.932 & 367.790 & 722.754 & 98.473 & 163.956 - 535.053 \\
		$\mu_2$  & 0.001 & 0.881 & 1.558 & 1.357 & 1.871 & 2.000 & 0.593 & 0.204 - 2.000 \\
		\hline
	\end{tabular}%
	\label{est_param_DeviceG}%
\end{table}%

\begin{table}[htbp]
	\centering
	\scriptsize
	\caption{Posterior measures of reliability functions for some values of $t$ in Device-G dataset.}
	\begin{tabular}{ccccccccc}
		\hline
		\multicolumn{9}{c}{Componente W} \\
		\hline
		t     & Min   & 1Qt   & Median & Mean  & 3Qt   & Max   & SD    & CI 95\% \\
		11.046 & 0.996 & 1.000 & 1.000 & 1.000 & 1.000 & 1.000 & 0.000 & 0.999 - 1.000 \\
		150.230 & 0.761 & 0.938 & 0.966 & 0.955 & 0.982 & 0.999 & 0.039 & 0.875 - 0.999 \\
		250.023 & 0.423 & 0.687 & 0.748 & 0.741 & 0.804 & 0.940 & 0.086 & 0.58 - 0.896 \\
		\hline
		\multicolumn{9}{c}{Componente S} \\
		\hline
		t     & Min   & 1Qt   & Median & Mean  & 3Qt   & Max   & SD    & CI 95\% \\
		14.988 & 0.573 & 0.811 & 0.861 & 0.850 & 0.898 & 0.984 & 0.063 & 0.728 - 0.956 \\
		120.191 & 0.353 & 0.512 & 0.564 & 0.564 & 0.614 & 0.778 & 0.071 & 0.437 - 0.702 \\
		250.071 & 0.187 & 0.359 & 0.404 & 0.408 & 0.456 & 0.600 & 0.067 & 0.275 - 0.528 \\
		\hline
	\end{tabular}%
	\label{medidas_confia_DeviceG}%
\end{table}%

\clearpage

\chapter{Masked data} 
\label{masked_failure}

Chapters \ref{nonparametric} and \ref{weibull_model} considered the problem of component estimation in coherent systems under the Bayesian paradigm in nonparametric and parametric approaches, respectively. In both cases the status of each component at the time of system failure is considered to be known. In some situation, however, the component that causes the system failure is not identified exactly and can only be narrowed down to a smaller set of components, known as masked data failure cause.

 \cite{Flehinger2002} presents a masked data problem in which consists of computer hard-drives failure times monitored over a period of $4$ years. There were three possible causes of failure competing to be computer hard-drives failures: eletronic hard, head flyability and head/disc magnetics. For some units, the real cause of hard-drive failure is not identified only that it belongs to a subset of components candidates to be system failure cause.  
 
The masked data problem formulation is developed and a Bayesian three-parameter Weibull model for marginal components' reliabilities in masked data scenario is presented. This chapter is based on the work of \citet{rodriguesMasked2017} and we suggest seeing this article for more details.

In Section \ref{masked_explicacao} masked data problem is discussed. Weibull model for masked data scenario is presented in \ref{masked_weibull}. Section \ref{secao_exemplos} presents simulated datasets for some system structures and Section \ref{computer_data_sec} presents computer hard-drives dataset solution. 
 
\section{Masked data scenario} \label{masked_explicacao}

Consider again that a random sample of $n=10$ $2$-out-of-$3$ systems (Figure \ref{system_2de3_SPS_PSS}) is observed. Unlike the previous situations, the component whose failure causes the system failure is not identified for all units. A set of components index, say $s$, which includes the component responsible for the system failure is specified. The data are presented in Table \ref{2de3_ex_masked}. For instance, system ID=4 failed at time $2.55$ but the component that cause the system failure is not identified, it is just known that the failure of component 2 or the failure of component 3 caused system failure, once $s=\{2,3\}$. If your interest is inference about component 2, por example, it is not known whether the failure of component 2 produced the failure of the system or if component 2 is left-censored at time $2.55$. 

Note that for system ID=5 there is only one component in set $s$, $s=\{1\}$. Thus, it is not a case of masked data, in which component 1 failed at time $1.89$ and its failure caused the system to fail. 

\begin{table}[htbp]
	\centering
	\caption{Observed data of $n=10$ $2$-out-of-$3$ systems in masked data scenario.}
	\begin{tabular}{c|c|c}
		\hline
		System ID & t  & s \\
		\hline
		1     & 1.95  & \{1,3\} \\
		2     & 2.09  & \{1,2\} \\
		3     & 3.56  & \{2\} \\
		4     & 2.55  & \{2,3\} \\
		5     & 1.89  & \{1\} \\
		6     & 3.01  & \{2,3\} \\
		7     & 2.43  & \{1,2\} \\
		8     & 1.51  & \{2\} \\
		9     & 3.55  & \{1\} \\
		10    & 2.35  & \{2,3\} \\
		\hline
	\end{tabular}%
	\label{2de3_ex_masked}%
\end{table}%

\newpage

\section{Weibull model} \label{masked_weibull}


Consider again a system of $m$ components and $X_j$ denoting the failure time of the $j$th component, $j=1,\ldots,m$. We assume that $X_1,X_2,\ldots,X_m$ are mutually independent. Let $T$ the random variable that represents the system failure time and $t$ a observation of $T$. Associated to each component $j$, let $\delta_j$ an indicator of censor.

The observation of $X_j$ can be: $X_j = t$, the failure time of $X_j$ it is not censored ($\delta_j = 1$); $X_j > t$, the failure time is right censored ($\delta_j = 2$); and $X_j \leq t$, the failure time is left censored ($\delta_j = 3$). Also, $j$th component can belong to the masked set or not.

Let $s$ be the set of index indicating possible components that produced the failure of the system, that is, components that have their failure time masked, $s$ is a subset of $\{1, \ldots, m\}$.

 Let $t_1, \ldots, t_n$ a sample of system failure time of size $n$, and $s_i$ is the set of masked components in the $i$th sample, $i = 1, \ldots, n$. Also, $\upsilon_{ji} = 1$ if the $j$th component has the failure time masked ($j \in s_i$ and $s_i$ is not unitary), and $\upsilon_{ji} = 0$ otherwise ($j \notin s_i$), for $j = 1, \ldots, m$. The observation of $j$th component will be one of the following:

\begin{description}
	\item[uncensored; not masked:] $\delta_{ji} = 1$ and $\upsilon_{ji} = 0$;
	\item[right censored; not masked:] $\delta_{ji} = 2$ and $\upsilon_{ji} = 0$;
	\item[left censored; not masked:] $\delta_{ji} = 3$ and $\upsilon_{ji} = 0$;
	\item[masked:]  $\upsilon_{ji} = 1$.
\end{description}
If a component has the failure time masked, the component can be the one that produced the failure of the system (uncensored), right censored or left censored. Consider that
\begin{eqnarray}
{\lambda_1}_j(t) &= \Pr(\upsilon_j = 1 \mid t, \delta_j = 1), \nonumber \\ 
{\lambda_2}_j(t) &= \Pr(\upsilon_j = 1 \mid t, \delta_j = 2),  \nonumber \\ 
{\lambda_3}_j(t) &= \Pr(\upsilon_j = 1 \mid t, \delta_j = 3), \nonumber
\end{eqnarray}
where ${\lambda_1}_j(t)$ is the conditional probability of the $j$th component be masked, given the failure time of the system $t$, and the censor type $\delta_j = 1$. ${\lambda_2}_j(t)$ and ${\lambda_3}_j(t)$ are analogous to ${\lambda_1}_j(t)$. Here, we consider that ${\lambda_1}_j(t) = {\lambda_1}_j$, ${\lambda_2}_j(t) = {\lambda_2}_j$, and ${\lambda_3}_j(t) = {\lambda_3}_j$, that is, the probability of a component be masked does not depend on the failure time $t$.

For each component, we can observe the triple $(t_i, \delta_{ji}, \upsilon_{ji})$, $i, \ldots, n$. Our interest consists in the estimation of the distribution function, $F$, of the $j$-h component. 
We consider a parametric family model for $F$ with parameter $\bm{\theta}_j$, then the estimation of the parameter $\bm{\theta}_j$ induces the distribution function $F$. The available information from the data is one of the following types:
\begin{enumerate}
	\item $\Pr(X_{ji} \in (t_i, t_i], \upsilon_i = 0 \mid \bm{\theta}_j) = f(t_i \mid \bm{\theta}_j) (1- {\lambda_1}_j)$, if the $i$th observation is uncensored and not masked;
	\item $\Pr(X_{ji} \in (t_i, \infty), \upsilon_i = 0 \mid \bm{\theta}_j) = [1-F(t_i \mid \bm{\theta}_j)](1- {\lambda_2}_j)$, if the $i$th observation is right censored and not masked;
	\item $\Pr(X_{ji} \in (0, t_i], \upsilon_i = 0 \mid \bm{\theta}_j) = F(t_i \mid \bm{\theta}_j) (1- {\lambda_3}_j)$, if the $i$th observation is left censored and not masked;
	\item $\Pr(X_{ji} \in (t_i, t_i], \upsilon_i = 1 \mid \bm{\theta}_j) = f(t_i \mid \bm{\theta}_j) {\lambda_1}_j$, if the $i$th observation is uncensored and masked;
	\item $\Pr(X_{ji} \in (t_i, \infty), \upsilon_i = 1 \mid \bm{\theta}_j) = [1-F(t_i \mid \bm{\theta}_j)] {\lambda_2}_j$, if the $i$th observation is right censored and masked; and
	\item $\Pr(X_{ji} \in (0, t_i], \upsilon_i = 1 \mid \bm{\theta}_j) = F(t_i \mid \bm{\theta}_j) {\lambda_3}_j$, if the $i$th observation is left censored and masked.
\end{enumerate}

However, we do not have information about the cases 4 to 6, since when the data is masked, we do not know if that component was censored or not. 
Considering an augmented data procedure (latent variable), define $d_{1ji} = 1$ if the masked observation is not censored or $d_{1ji} = 0$ otherwise, $d_{2ji} = 1$ if the masked observation is right censored or $d_{2ji} = 0$ otherwise, and $d_{3ji} = 1$ if the masked observation is left censored or $d_{3ji} = 0$ otherwise. Besides, ${\bm d}_{ji}=(d_{1ji},d_{2ji},d_{3ji})$ and $\sum_{l=1}^3d_{lji}=1$.

Let $R(t_i \mid \bm{\theta}_j) = 1-F(t_i \mid \bm{\theta}_j)$ the reliability function. The likelihood function of the $j$th component can be writen as a part of non-masked data and a part for masked data (augmented data), that is,
\begin{eqnarray}
& L(\bm{\theta}_j, {\lambda_1}_j, {\lambda_2}_j, {\lambda_3}_j, \bm{d}_j \mid \bm{t}, \bm{\delta}_j, \bm{\upsilon}_j)  = \prod\limits_{i: ~ \upsilon_{ji} = 0} \Big\{ \big[f(t_i \mid \bm{\theta}_j) ~ (1- {\lambda_1}_j)\big]^{I(\delta_{ji} = 1)} \nonumber \\
&\times \big[R(t_i \mid \bm{\theta}_j) ~ (1- {\lambda_2}_j)\big]^{I(\delta_{ji} = 2)}  \big[F(t_i \mid \bm{\theta}_j) ~ (1- {\lambda_3}_j)\big]^{I(\delta_{ji} = 3)} \Big\} \nonumber \\
&\times\prod\limits_{i: ~ \upsilon_{ji} = 1} \Big\{ \big[f(t_i \mid \bm{\theta}_j) ~ {\lambda_1}_j \big]^{d_{1ji}}  \big[R(t_i \mid \bm{\theta}_j) ~ {\lambda_2}_j \big]^{d_{2ji}}  \big[F(t_i \mid \bm{\theta}_j) ~ {\lambda_3}_j \big]^{d_{3ji}} \Big\}, \label{vero_masked}
\end{eqnarray}
where $I(A) = 1$, if $A$ is true, and $I(A) = 0$ otherwise, $\bm{t} =(t_1, \ldots, t_n)$, $\bm{\upsilon}_j =(\upsilon_{j1}, \ldots, \upsilon_{jn})$, $\bm{d}_j = ({\bm d}_{ji}:i \in \{\upsilon_{ji} = 1\})$ and $\bm{\delta}_j = (\delta_{ji}:i \in \{\upsilon_{ji} = 0\})$.

The likelihood function in (\ref{vero_masked}) is defined generically and it is straightforward to any probability distribution. We propose the three-parameter Weibull distribution. Thus, the reliability and density functions are given in (\ref{confia_weibull}) and (\ref{dens_weibull}), respectively.

The estimation work is performed under a Bayesian perspective of inference, and thus, the priori distribution for $(\bm{\theta}_j, {\lambda_1}_j, {\lambda_2}_j, {\lambda_3}_j, \bm{d}_j)$ needs to be defined. The prior distributions of all parameters are considered independent with gamma distribution with mean $1$ and variance $1000$ for $\beta_j$, $\eta_j$, $\mu_j$ and uniform distribution over $(0,1)$ for ${\lambda_1}_j$, ${\lambda_2}_j$ and ${\lambda_3}_j$. Besides, $\Pr(d_{lji}=1)=\Pr(d_{lji}=0)=0.5$, for $l=1,2,3$.

Here, no prior information about component's operation is known and noninformative prior is considered. However, it is possible to express a prior information about the component functioning in the system through the opinion of an expert and/or through past experiences. 

The posterior density of $(\bm{\theta}_j, {\lambda_1}_j, {\lambda_2}_j, {\lambda_3}_j, \bm{d}_j)$ comes out to be
\begin{eqnarray} 
&&\pi(\bm{\theta}_j, {\lambda_1}_j, {\lambda_2}_j, {\lambda_3}_j, \bm{d}_j  \mid \bm{t},\bm{\delta}_j, \bm{\upsilon}_j)  \propto   \nonumber \\
&& \pi(\bm{\theta}_j, {\lambda_1}_j, {\lambda_2}_j, {\lambda_3}_j, \bm{d}_j) L(\bm{\theta}_j, {\lambda_1}_j, {\lambda_2}_j, {\lambda_3}_j, \bm{d}_j \mid \bm{t}, \bm{\delta}_j, \bm{\upsilon}_j),    \label{posteriori_masked}
\end{eqnarray}
where $\pi(\bm{\theta}_j, {\lambda_1}_j, {\lambda_2}_j, {\lambda_3}_j, \bm{d}_j)$ is the prior distribution of $(\bm{\theta}_j, {\lambda_1}_j, {\lambda_2}_j, {\lambda_3}_j, \bm{d}_j)$. 

The posterior density in Equation (\ref{posteriori_masked}) has not close form. An alternative is to rely on Markov-Chain Monte-Carlo (MCMC) simulations. Here we consider Metropolis within Gibbs algorithm. This algorithm is suitable in this situation because it is possible direct sampling from conditional distribution for some parameters but for others this is not possible \citep{Tierney}. The algorithm works in the following steps: 

\vspace{0.2cm}
{\it
	\begin{enumerate}
		\item Attribute initial values $\bm{\theta}_j^{(0)}$, $\lambda_{1j}^{(0)}$, $\lambda_{2j}^{(0)}$ and $\lambda_{3j}^{(0)}$ for   $\bm{\theta}_j=(\beta_j,\eta_j,\mu_j)$, $\lambda_{1j}$, $\lambda_{2j}$ and $\lambda_{3j}$, respectively, and set $b=1$;
		\item For $i \in \{\bm{\upsilon}_j=1\}$, draw $\bm{d}_{ji}^{(b)}$ from $\pi(\bm{d}_{ji} \mid \bm{t}, \bm{\delta}_j, \bm{\upsilon}_j, \bm{\theta}_j^{(b-1)},\lambda_{1j}^{(b-1)},\lambda_{2j}^{(b-1)},\lambda_{3j}^{(b-1)})$, in which:
		\begin{eqnarray} 
		\pi(\bm{d}_{ji} \mid \bm{t}, \bm{\delta}_j, \bm{\upsilon}_j, \bm{\theta}_j,\lambda_{1j} ,\lambda_{2j},\lambda_{3j} )&\propto & \Big\{\lambda_{1j} f(t_{i}|\bm{\theta}_j) \Big\}^{d_{1ji}}
		\Big\{\lambda_{2j}R(t_{i}|\bm{\theta}_j)\Big\}^{d_{2ji}}\nonumber \\
		&&\times \Big\{\lambda_{3j}F(t_{i}|\bm{\theta}_j)\Big\}^{d_{3ji}}, \nonumber                 
		\end{eqnarray}
		that is, $\bm{d}_{ji} \mid \bm{t},\bm{\delta}_j,\bm{\upsilon}_j,\bm{\theta}_j,\lambda_{1j} ,\lambda_{2j},\lambda_{3j} \sim \mbox{Multinomial}(1;p_{1ji},p_{2ji},p_{3ji})$ in which $p_{1ji}=\lambda_{1j}f(t_{i}|\bm{\theta}_j)/C$, $p_{2ji}=\lambda_{2j} R(t_{i}|\bm{\theta}_j)/C$ and $p_{3ji}=\lambda_{3j}F(t_{i}|\bm{\theta}_j)/C$, where $C=\lambda_{1j}f(t_{i}|\bm{\theta}_j)+\lambda_{2j} R(t_{i}|\bm{\theta}_j)+\lambda_{3j}F(t_{i}|\bm{\theta}_j)$;
		\item Draw $\bm{\theta}_j^{(b)}$ from $\pi(\bm{\theta}_j\mid \bm{t}, \bm{\delta}_j, \bm{\upsilon}_j,\bm{d}_j^{(b)},\lambda_{1j}^{(b-1)} ,\lambda_{2j}^{(b-1)} ,\lambda_{3j}^{(b-1)})$ through Metropolis-Hastings algorithm \citep{RobertCasella}, where 
		\begin{eqnarray} 
		& \pi(\bm{\theta}_j\mid \bm{t}, \bm{\delta}_j, \bm{\upsilon}_j,\bm{d}_j,\lambda_{1j} ,\lambda_{2j} ,\lambda_{3j})    \propto  ~~ \pi(\bm{\theta}_j) \prod\limits_{i: ~ \upsilon_{ji} = 0} \Big\{ \big[f(t_i \mid \bm{\theta}_j) ~ (1- {\lambda_1}_j)\big]^{I(\delta_{ji} = 1)} \nonumber \\
		&\times  \big[R(t_i \mid \bm{\theta}_j) ~ (1- {\lambda_2}_j)\big]^{I(\delta_{ji} = 2)} \big[F(t_i \mid \bm{\theta}_j) ~ (1- {\lambda_3}_j)\big]^{I(\delta_{ji} = 3)} \Big\} \nonumber \\
		&\times  \prod\limits_{i: ~ \upsilon_{ji} = 1} \Big\{ \big[f(t_i \mid \bm{\theta}_j) ~ {\lambda_1}_j \big]^{d_{1ji}}  \big[R(t_i \mid \bm{\theta}_j) ~ {\lambda_2}_j \big]^{d_{2ji}}  \big[F(t_i \mid \bm{\theta}_j) ~ {\lambda_3}_j \big]^{d_{3ji}} \Big\}.     \nonumber
		\end{eqnarray}
		
		\item Simulate $\lambda_{1j}$ from $\pi(\lambda_{1j}\mid  \bm{t}, \bm{\delta}_j, \bm{\upsilon}_j,\bm{d}_j^{(b)},\bm{\theta}_j^{(b)} ,\lambda_{2j}^{(b-1)} ,\lambda_{3j}^{(b-1)})$ in which
		\begin{eqnarray} 
		\pi(\lambda_{1j}\mid \bm{t}, \bm{\delta}_j, \bm{\upsilon}_j,\bm{d}_j,\bm{\theta}_j ,\lambda_{2j},\lambda_{3j}) & ~~ \propto &  ~~   \lambda_{1j}^{\sum\limits_{i: ~ \upsilon_{ji} = 1}d_{1ji}}(1-\lambda_{1j})^{n_f},    \nonumber                 
		\end{eqnarray}
		that is, $\lambda_{1j}\mid (\bm{t}, \bm{\delta}_j, \bm{\upsilon}_j,\bm{d}_j,\bm{\theta}_j ,\lambda_{2j} ,\lambda_{3j}) \sim \mbox{Beta}(\sum\limits_{i: ~ \upsilon_{ji} = 1}d_{1ji}+1,n_f+1)$, in which $n_f$ is the number of systems in which component $j$ is  known to be the responsible of system failure.

		\item Simulate $\lambda_{2j}$ from $\pi(\lambda_{2j}\mid  \bm{t}, \bm{\delta}_j, \bm{\upsilon}_j,\bm{d}_j^{(b)},\bm{\theta}_j^{(b)} ,\lambda_{1j}^{(b-1)} ,\lambda_{3j}^{(b-1)})$ in which
		\begin{eqnarray} 
		\pi(\lambda_{2j}\mid \bm{t}, \bm{\delta}_j, \bm{\upsilon}_j,\bm{d}_j,\bm{\theta}_j ,\lambda_{1j},\lambda_{3j}) & ~~ \propto &  ~~   \lambda_{2j}^{\sum\limits_{i: ~ \upsilon_{ji} = 1}d_{2ji}}(1-\lambda_{2j})^{n_r},    \nonumber                 
		\end{eqnarray}
		that is, $\lambda_{2j}\mid (\bm{t}, \bm{\delta}_j, \bm{\upsilon}_j,\bm{d}_j,\bm{\theta}_j ,\lambda_{1j} ,\lambda_{3j}) \sim \mbox{Beta}(\sum\limits_{i: ~ \upsilon_{ji} = 1}d_{2ji}+1,n_r+1)$, in which $n_r$ is the number of systems in which component $j$ is observed to be right-censored.		
		\item Simulate $\lambda_{3j}$ from $\pi(\lambda_{3j}\mid  \bm{t}, \bm{\delta}_j, \bm{\upsilon}_j,\bm{d}_j^{(b)},\bm{\theta}_j^{(b)} ,\lambda_{1j}^{(b-1)} ,\lambda_{2j}^{(b-1)})$ in which
		\begin{eqnarray} 
		\pi(\lambda_{3j}\mid \bm{t}, \bm{\delta}_j, \bm{\upsilon}_j,\bm{d}_j,\bm{\theta}_j ,\lambda_{1j},\lambda_{2j}) & ~~ \propto &  ~~   \lambda_{3j}^{\sum\limits_{i: ~ \upsilon_{ji} = 1}d_{3ji}}(1-\lambda_{3j})^{n_l},    \nonumber                 
		\end{eqnarray}
		that is, $\lambda_{3j}\mid (\bm{t}, \bm{\delta}_j, \bm{\upsilon}_j,\bm{d}_j,\bm{\theta}_j ,\lambda_{1j} ,\lambda_{2j}) \sim \mbox{Beta}(\sum\limits_{i: ~ \upsilon_{ji} = 1}d_{3ji}+1,n_l+1)$, in which $n_l$ is the number of systems in which component $j$ is observed to be left-censored.
		\item Let $b=b+1$ and repeat steps $2)$ to $7)$ until $b=B$, where $B$ is pre-set number of simulated samples of $(\bm{\theta}_j,\lambda_{1j},\lambda_{2j},\lambda_{3j},\bm{d}_j)$.
	\end{enumerate}
}

\vspace{0.2cm}

Discarding burn-in (first generated values discarded to eliminate the effect of the assigned initial values for parameters) and jump samples (spacing among generated values to avoid correlation problems), a sample of size $n_p$ from the joint posterior distribution of $(\bm{\theta}_j,\lambda_{1j},\lambda_{2j},\lambda_{3j},\bm{d}_j)$ is obtained. For the $j$th component, the sample from the posterior can be expressed as $(\bm{\theta}_{j}^{(1)},\bm{\theta}_{j}^{(2)},\ldots,\bm{\theta}_{j}^{(n_p)})$, $(\lambda_{1j}^{(1)},\lambda_{1j}^{(2)},\ldots,\lambda_{1j}^{(n_p)})$, $(\lambda_{2j}^{(1)},\lambda_{2j}^{(2)},\ldots,\lambda_{2j}^{(n_p)})$, $(\lambda_{3j}^{(1)},\lambda_{3j}^{(2)},\ldots,\lambda_{3j}^{(n_p)})$ and $(\bm{d}_j^{(1)},\bm{d}_j^{(2)},\ldots,\bm{d}_j^{(n_p)})$. Consequently, posterior quantities of reliability function $R(t\mid\bm{\theta}_{j})$ can be easily obtained \citep{RobertCasella}. For example, the posterior mean of the reliability function is given by
\begin{eqnarray}
{\rm E}[R(t\mid \bm{\theta}_{j}) \mid Data] = \frac{1}{n_p}\sum_{k=1}^{n_p}{R\Big(t \mid \bm{\theta}_{j}^{(k)}\Big)},~~ \mbox{for each} ~ t > 0. \nonumber
\end{eqnarray}

\section{Simulated datasets in masked scenario} \label{secao_exemplos}

\begin{figure}[h]
	\centering
	\includegraphics[width=0.4\linewidth]{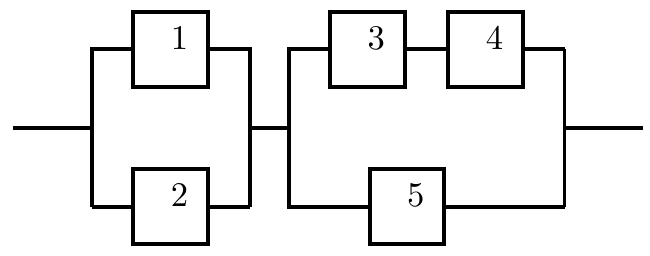}
	\caption{System 2 representation.}
	\label{fig:sistemachines}
\end{figure}

We consider two simulated examples of complex system structure. The system 1, the $2$-out-of-$3$ system is presented in Figure \ref{system_2de3_SPS_PSS}. System in Figure \ref{fig:sistemachines}, system 2, can be denoted as a series system of components 1 and 2 with components 3, 4 and 5. 

For these simulated examples, we consider that $s_i$ consists of components that no longer work in the system failure, that is, there is no right censored for components in $s_i$, which leads to  $\lambda_{2j}=0$. Besides, for $s_i$ not unitary, $j \in s_i$ only if $j$ belongs to the minimal cut set that caused the $i$th system failure. A cut set is a set of components which by failing causes the system to fail. A cut set is said to be minimal if it can not be reduced without loosing its status as a cut set. For the bridge system, for instance, there are four minimal cut sets, they are: $\{1,2\}$, $\{4,5\}$, $\{1,3,5\}$ and $\{2,3,4\}$. 

Imagine a situation that this bridge system fail and the components 1, 2 and 3 do not work at the moment of system failure. Then, only components 1 and 2 belong to set $s$, once the component 3 does not belong to the minimal cut that caused the system failure and, in fact, the component 3 is observed to be left censored failure time. In this approach, we fit the proposed model under symmetric assumption (that is, $\lambda_{1j}=\lambda_{3j}$).

%

The simulated systems have the following characteristics:
\begin{itemize}
	\item System 1 ($2$-out-of-$3$): $m=3$ and $X_{1}$ was generated from a Weibull distribution with mean $15$ and variance $8$, $X_{2}$ from a gamma distribution with mean $18$ and variance $12$, $X_{3}$ from a lognormal distribution with mean $20$ and variance $10$ and the system failure time is $T=\max \{\min\{X_{1},X_{2}\}, \min\{X_{1},X_{3}\}, \min\{X_{2},X_{3}\}\}$ Besides, $n=300$ and $p=0.4$, where $p$ is the proportion of masked data system. 
	\item System 2: $m=5$ and $X_{1}$ was generated from a Weibull distribution with mean $12$ and variance $15$, $X_{2}$ from a gamma distribution with mean $11$ and variance $11$, $X_{3}$ from a three-parameter Weibull distribution with mean $12$ and variance $9$, $X_{4}$ from a lognormal distribution with mean $12$ and variance $7$ and $X_{5}$ from a three-parameter Weibull distribution with mean $11$ and variance $14$. In this structure, the system lifetime is given by $T = \min \{\max\{X_{1},X_{2}\}, \max\{\min\{X_{3},X_{4}\},X_{5}\}\}$, $n=100$ and the proportion of masked data systems is $p=0.3$.
\end{itemize}

For all systems, we generated $30{,}000$ values of each parameter, disregarding the first $10{,}000$ iterations as burn-in samples and jump of size $20$ was chosen to avoid correlation problems, obtaining a sample of size $n_p=1{,}000$. The chains convergence was monitored and good convergence results were obtained.



\subsection{System 1} \label{ex_simu_2de3_texto}


The generated data are presented in Table \ref{2de3_masked_ex} in Appendix \ref{apeC}. Table \ref{2de3_masked_porc_censura} presents the percentages of observations types of each component. Component 1, for instance, has $46.67\%$ of censored observations ($40.67\%$ to the left and $6\%$ to the right) and it has  $37.33\%$ of systems in masked scenario.

\begin{figure}[h]\centering
	\begin{minipage}[b]{0.32\linewidth}
		\includegraphics[width=\linewidth]{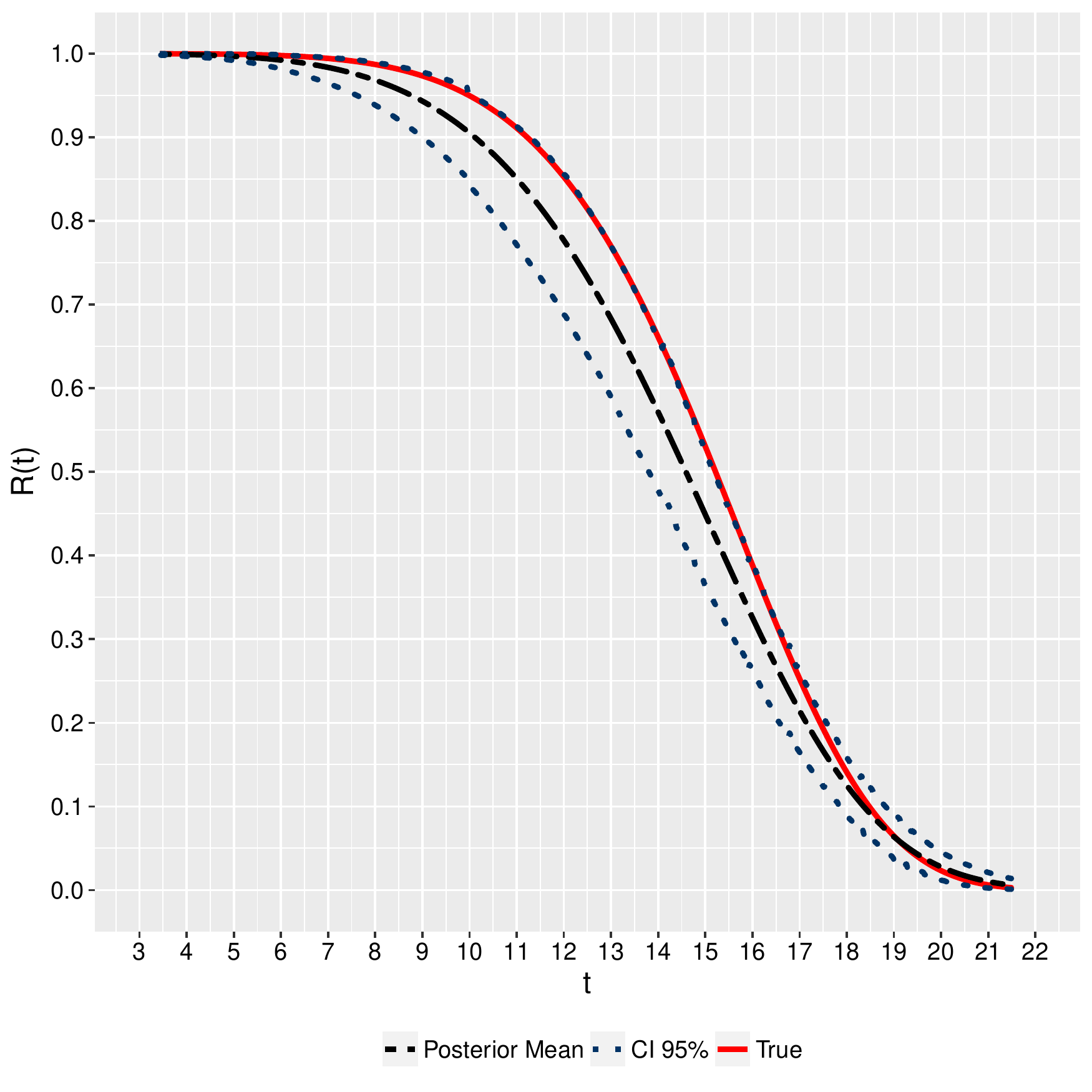}
		\subcaption{Component 1}
	\end{minipage} 
	\begin{minipage}[b]{0.32\linewidth}
		\includegraphics[width=\linewidth]{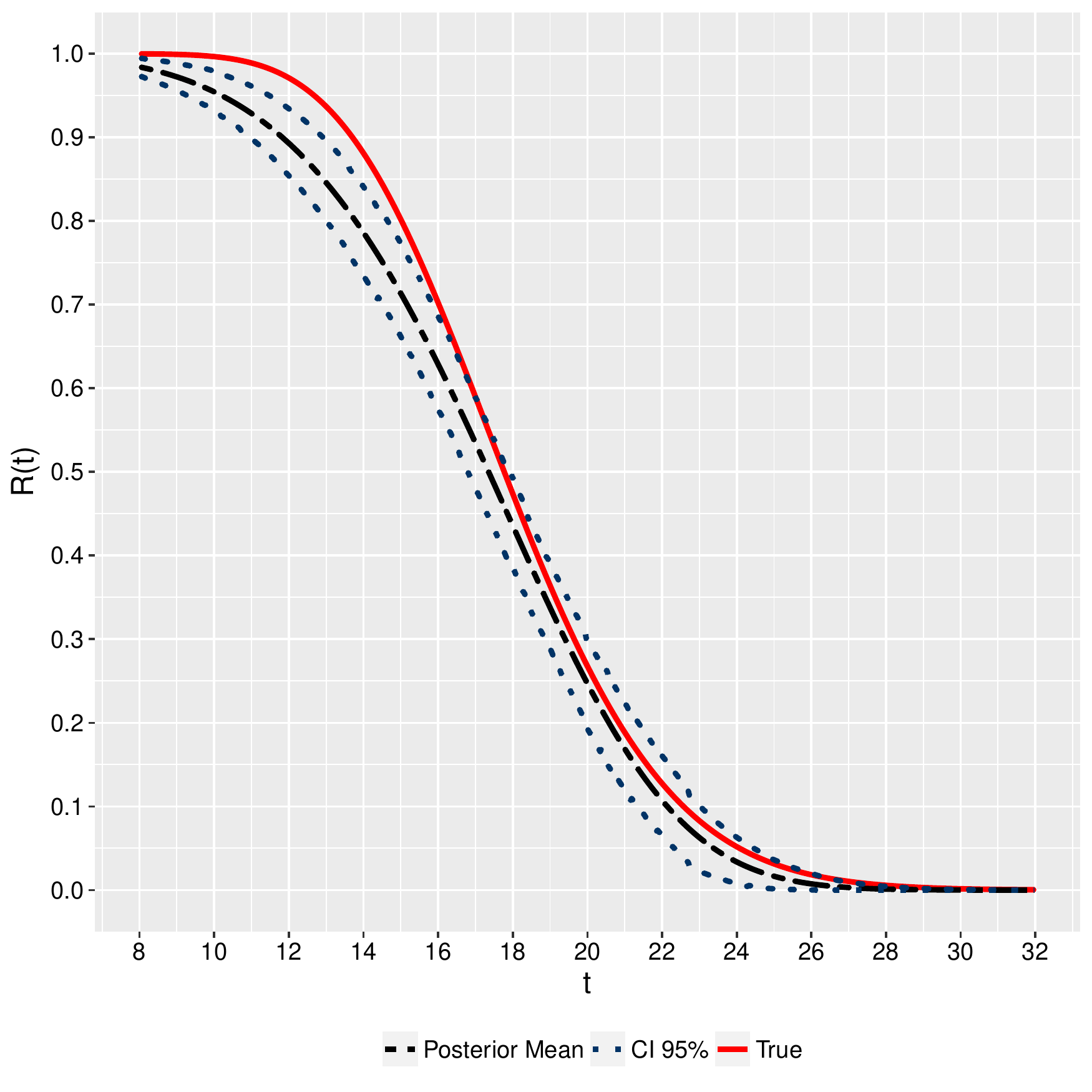}
		\subcaption{Component 2}
	\end{minipage}
	\begin{minipage}[b]{0.32\linewidth}
		\includegraphics[width=\linewidth]{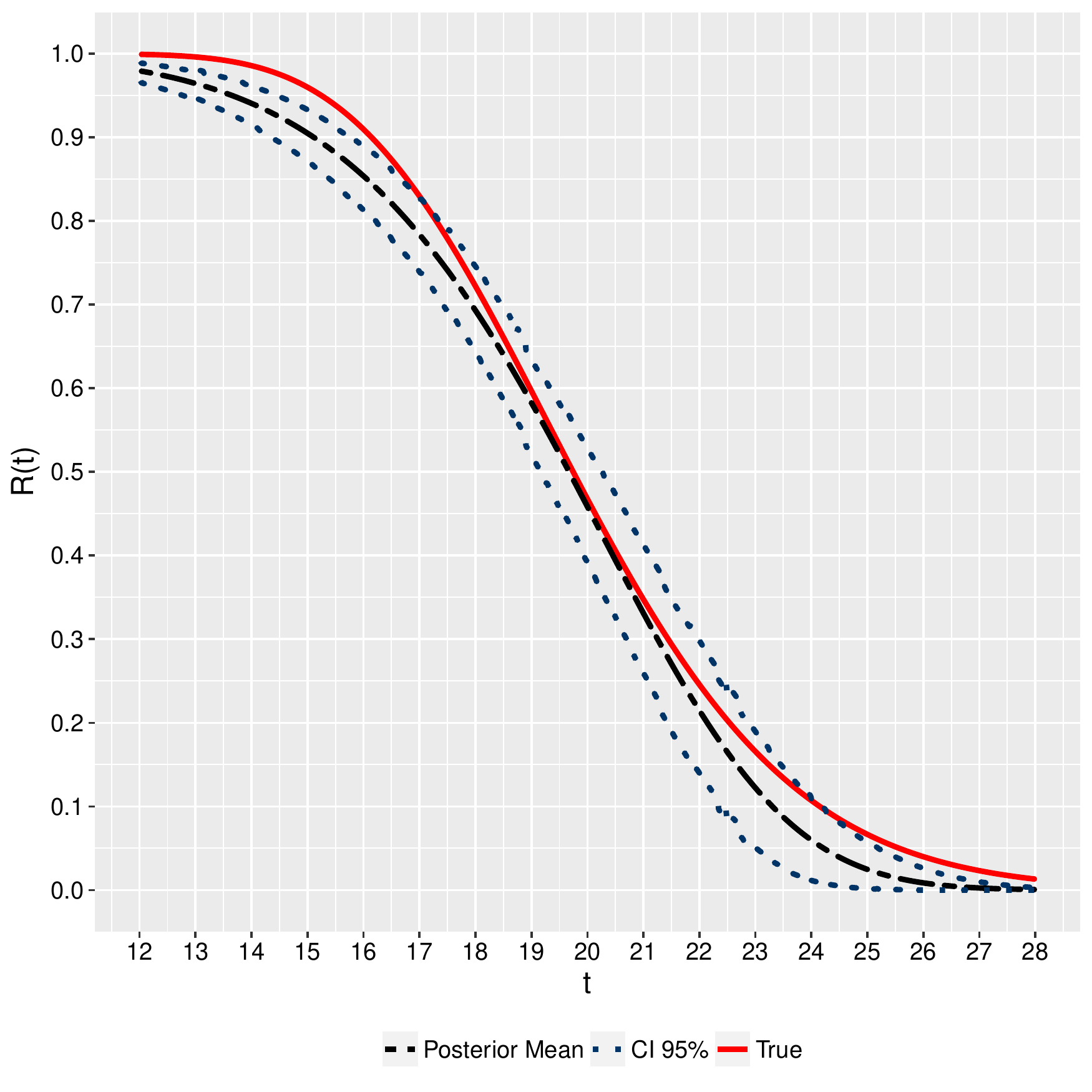}
		\subcaption{Component 3}
	\end{minipage}	
	\caption{Generating reliability functions, posterior means and 95\% HPD intervals (CI 95\%) for the components involved in $2$-out-of-$3$ structure in masked scenario.}
	\label{ex_2de3_masked}
\end{figure}

 Posterior measures of Weibull parameters $\beta_j$, $\eta_j$ and $\mu_j$, for $j=1,2,3$, are presented in Table \ref{est_param_2de3_mask}. The posterior quantities of $R(t\mid{\bm \theta}_j)$ for some values of $t$ are shown in Table \ref{medidas_posterior_2de3_mask}.  The posterior means of reliability functions can be visualized in Figure \ref{ex_2de3_masked} besides the empirical 95\% HPD intervals. The reliability posterior means are close to the true reliability curves, in general. Even for values of $t$ that the posterior mean is more distant from the true values, the lower or upper limit of HPD interval is very close to the true curve.

\begin{table}[htbp]
	\centering
	\scriptsize
	\caption{Percentages of type of observations for each component in $2\mbox{-out-of-}3$ structure in masked scenario.}
	\begin{tabular}{cccc}
		\hline
		Components & Type  & Number of systems & \% \\
		\hline
		1     & Known failure & 48    & 16.00\% \\
		& Left censored & 122   & 40.67\% \\
		& Right censored & 18    & 6.00\% \\
		& Masked  & 112   & 37.33\% \\
		\hline
		2     & Known failure & 75    & 25.00\% \\
		& Left censored & 49    & 16.33\% \\
		& Right censored & 99    & 33.00\% \\
		& Masked  & 77    & 25.67\% \\
		\hline
		3     & Known failure & 57    & 19.00\% \\
		& Left censored & 9     & 3.00\% \\
		& Right censored & 183   & 61.00\% \\
		& Masked  & 51    & 17.00\% \\
		\hline
	\end{tabular}%
	\label{2de3_masked_porc_censura}%
\end{table}%

\begin{table}[htbp]
	\centering
	\scriptsize
	\caption{Posterior measures of Weibull parameters for $2$-out-of-$3$ components in masked scenario.}
	\begin{tabular}{ccccccccc}
		\hline
		\multicolumn{9}{c}{Componente 1} \\
		\hline
		& Min   & 1Qt   & Median & Mean & 3Qt   & Max   & SD    & IC 95\% \\
		$\beta_1$ & 3.722 & 4.894 & 5.242 & 5.250 & 5.645 & 7.090 & 0.532 & 4.209 - 6.244 \\
		$\eta_1$ & 14.420 & 15.440 & 15.650 & 15.630 & 15.850 & 16.610 & 0.305 & 15.029 - 16.207 \\
		$\mu_1$   & 1.64E-09 & 4.16E-08 & 1.46E-07 & 4.42E-07 & 4.68E-07 & 3.30E-06 & 6.58E-07 & 1.63E-09 - 2.04E-06 \\
		\hline
		\multicolumn{9}{c}{Componente 2} \\
		\hline
		& Min   & 1Qt   & Median & Mean & 3Qt   & Max   & SD    & IC 95\% \\
		$\beta_2$ & 3.648 & 4.690 & 4.964 & 4.962 & 5.230 & 6.448 & 0.439 & 4.101 - 5.833 \\
		$\eta_2$ & 17.850 & 18.500 & 18.690 & 18.690 & 18.870 & 19.780 & 0.288 & 18.169 - 19.274 \\
		$\mu_2$    & 1.08E-07 & 5.18E-07 & 9.55E-06 & 1.92E-05 & 2.70E-05 & 0.000221 & 3.05E-05 & 1.08E-07 - 6.13E-05 \\
		\hline
		\multicolumn{9}{c}{Componente 3} \\
		\hline
		& Min   & 1Qt   & Median & Mean & 3Qt   & Max   & SD    & IC 95\% \\
		$\beta_3$ & 5.308 & 6.695 & 7.118 & 7.143 & 7.557 & 9.216 & 0.610 & 6.008 - 8.296 \\
		$\eta_3$ & 19.640 & 20.380 & 20.620 & 20.610 & 20.820 & 21.830 & 0.339 & 19.987 - 21.333 \\
		$\mu_3$  & 0.000 & 0.002 & 0.023 & 0.113 & 0.225 & 0.864 & 0.144 & 1.08E-04 - 0.362 \\
		\hline
	\end{tabular}%
	\label{est_param_2de3_mask}%
\end{table}%

\begin{table}[htbp]
	\centering
	\scriptsize
	\caption{Posterior measures of components' reliability functions involved in $2$-out-of-$3$ masked system for some values of $t$.}
	\begin{tabular}{ccccccccc}
		\hline
		\multicolumn{9}{c}{Component 1} \\
		\hline
		t     & Min   & 1Qt   & Median & Mean  & 3Qt   & Max   & SD    & CI 95\% \\
		3.5 & 0.995 & 0.999 & 1.000 & 0.999 & 1.000 & 1.000 & 0.001 & 0.998 -  1.000 \\
		10.0 & 0.788 & 0.889 & 0.908 & 0.905 & 0.927 & 0.966 & 0.029 & 0.842 -   0.951 \\
		20.0 & 0.007 & 0.021 & 0.026 & 0.028 & 0.033 & 0.059 & 0.009 & 0.012 -  0.046 \\
		\hline
		\multicolumn{9}{c}{Component 2} \\
		\hline
         t     & Min   & 1Qt   & Median & Mean  & 3Qt   & Max   & SD    & CI 95\% \\
		8.5 & 0.938 & 0.975 & 0.980 & 0.979 & 0.984 & 0.994 & 0.007 & 0.965 - 0.992 \\
		15.0 & 0.602 & 0.692 & 0.712 & 0.712 & 0.732 & 0.793 & 0.030 & 0.660 -   0.772 \\
		26.0 & 0.001 & 0.003 & 0.006 & 0.007 & 0.009 & 0.056 & 0.006 & 0.001 -  0.020 \\
		\hline
		\multicolumn{9}{c}{Component 3} \\
		\hline
		t     & Min   & 1Qt   & Median & Mean  & 3Qt   & Max   & SD    & CI 95\% \\
		12.5 & 0.936 & 0.968 & 0.974 & 0.973 & 0.978 & 0.989 & 0.008 &  0.958 -  0.986 \\
		20.0 & 0.347 & 0.437 & 0.460 & 0.458 & 0.480 & 0.555 & 0.034 &  0.392 -  0.528 \\
		25.0 & 0.001 & 0.012 & 0.021 & 0.024 & 0.033 & 0.119 & 0.016 &  0.002 - 0.056 \\
		\hline
	\end{tabular}%
	\label{medidas_posterior_2de3_mask}%
\end{table}%

\newpage

\subsection{System 2}

The generated data are presented in Table \ref{chin_masked_ex} in Appendix \ref{apeC}. Table \ref{chin_masked_porc_censura} presents the percentages of observations types of each component. Component 3, for instance, has $81\%$ of censored observations ($26\%$ to the left and $55\%$ to the right) and it has  $9\%$ of systems in masked scenario.

\begin{figure}[h!]\centering
	\begin{minipage}[b]{0.32\linewidth}
		\includegraphics[width=\linewidth]{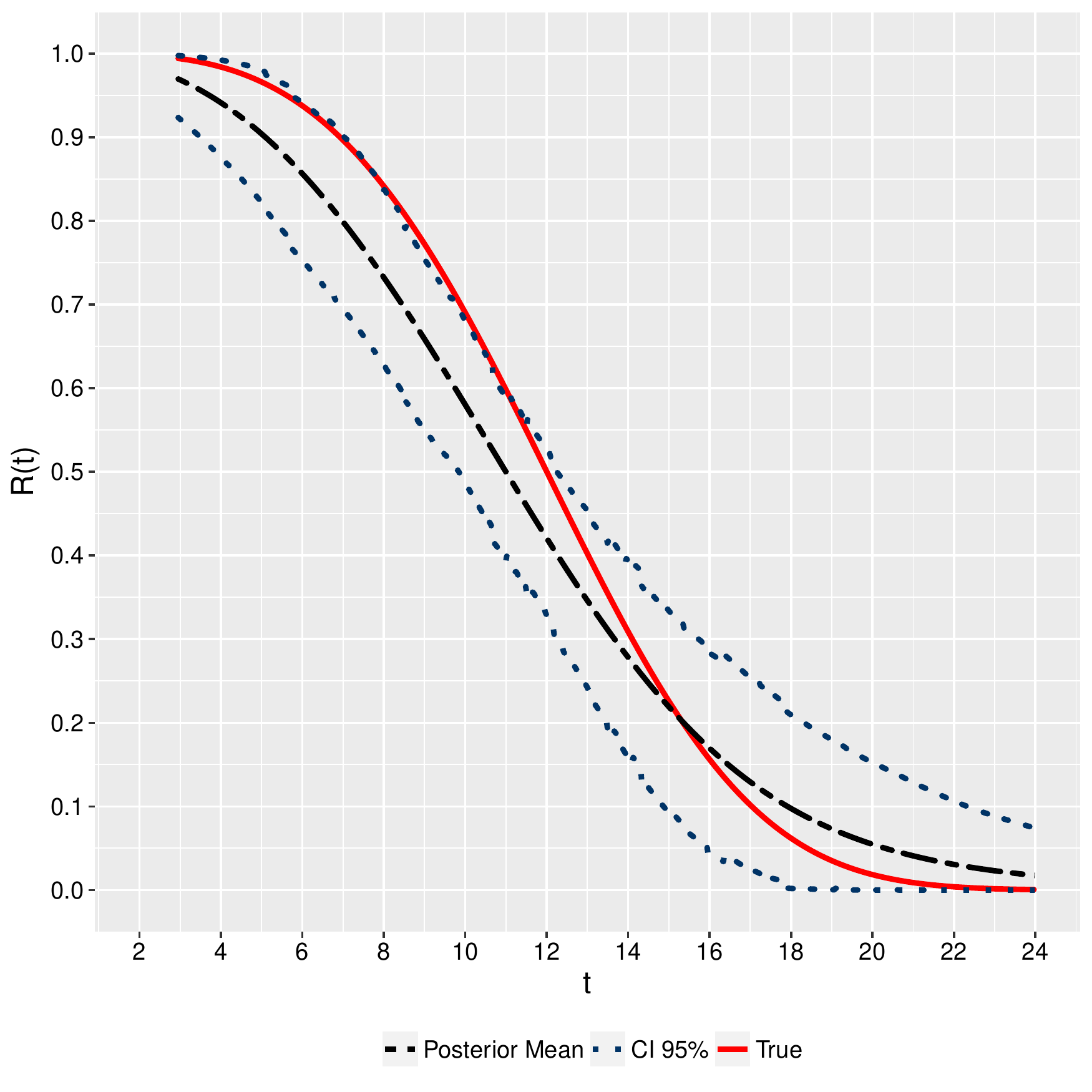}
		\subcaption{Component 1}
	\end{minipage} 
	\begin{minipage}[b]{0.32\linewidth}
		\includegraphics[width=\linewidth]{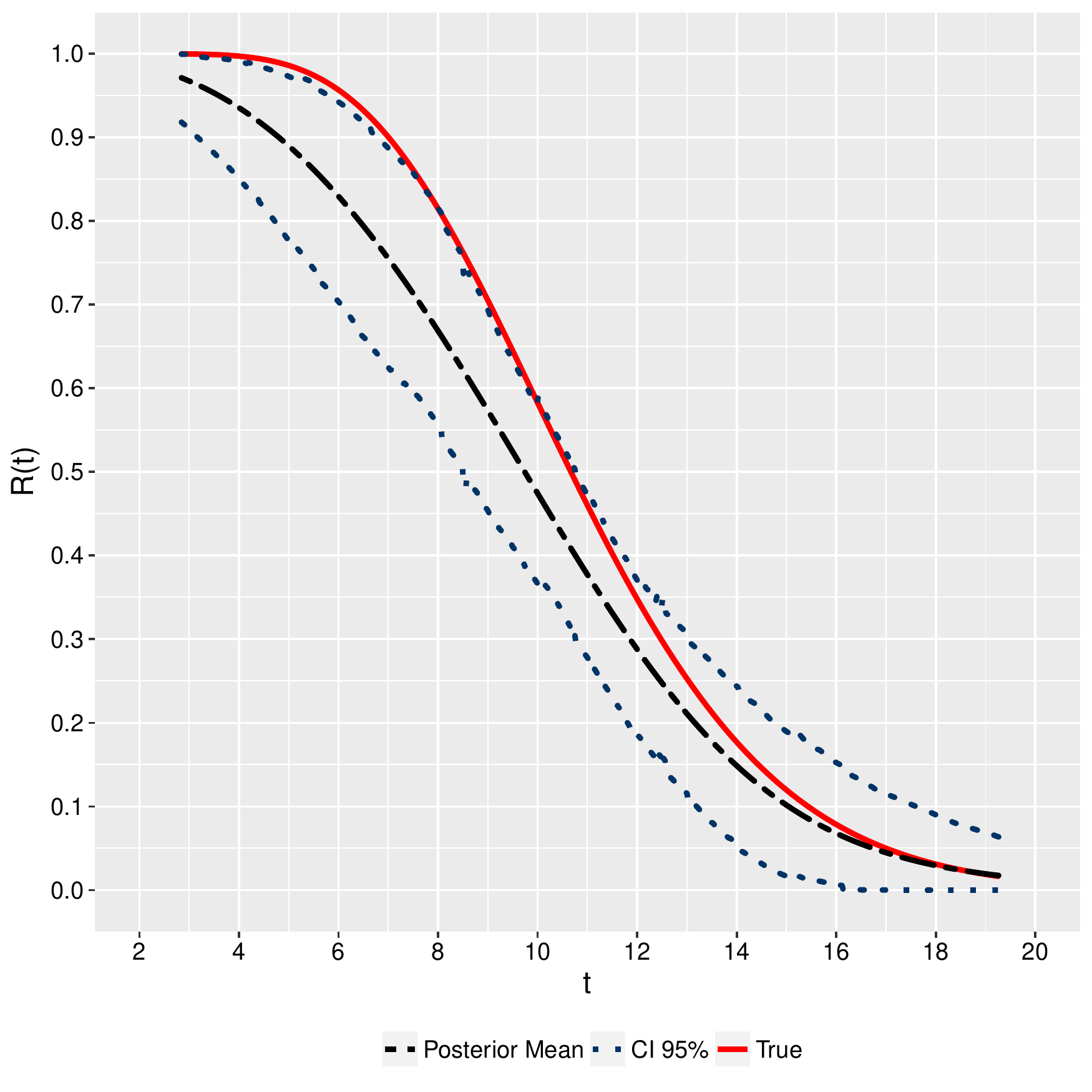}
		\subcaption{Component 2}
	\end{minipage}
	\begin{minipage}[b]{0.32\linewidth}
		\includegraphics[width=\linewidth]{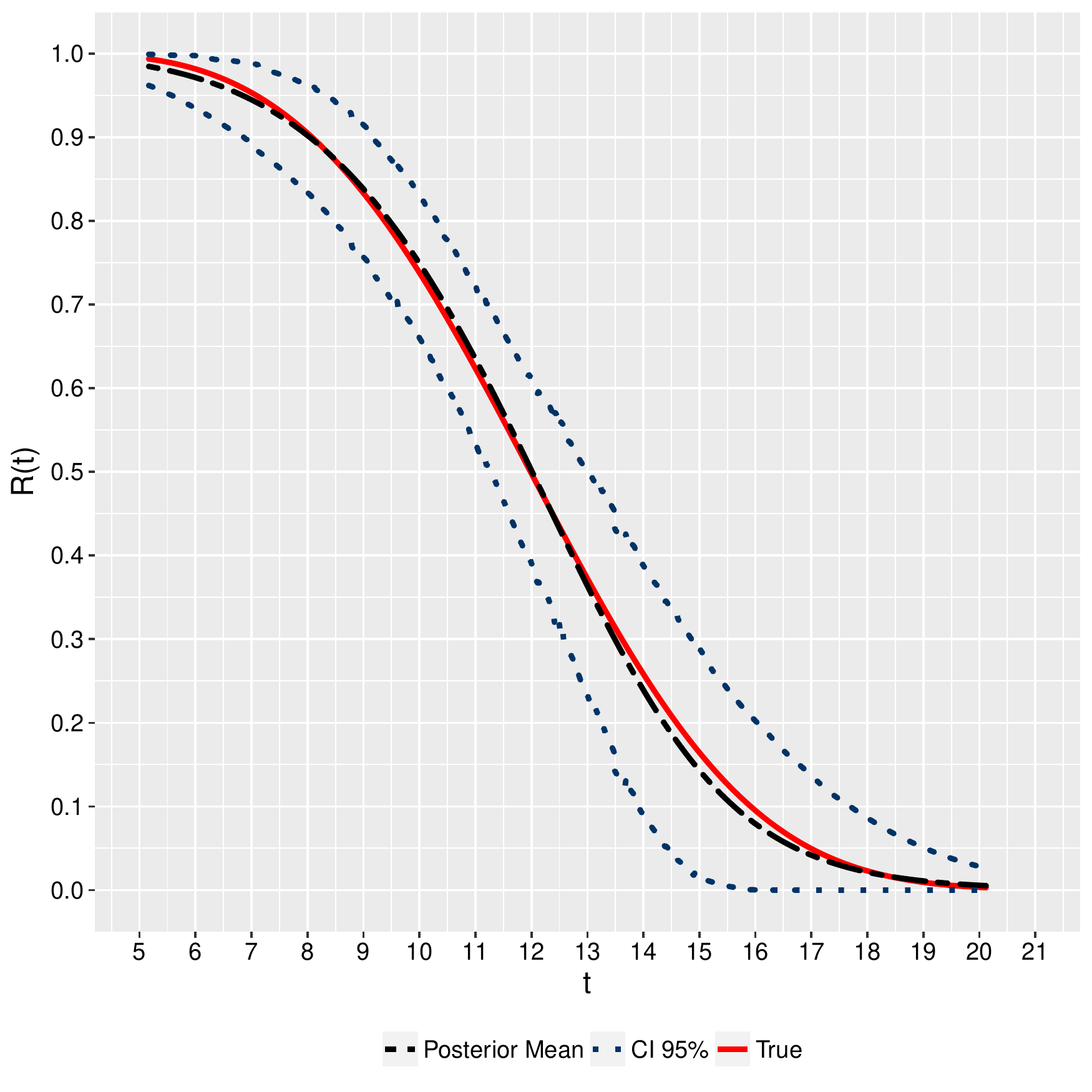}
		\subcaption{Component 3}
	\end{minipage}	
	\begin{minipage}[b]{0.32\linewidth}
	\includegraphics[width=\linewidth]{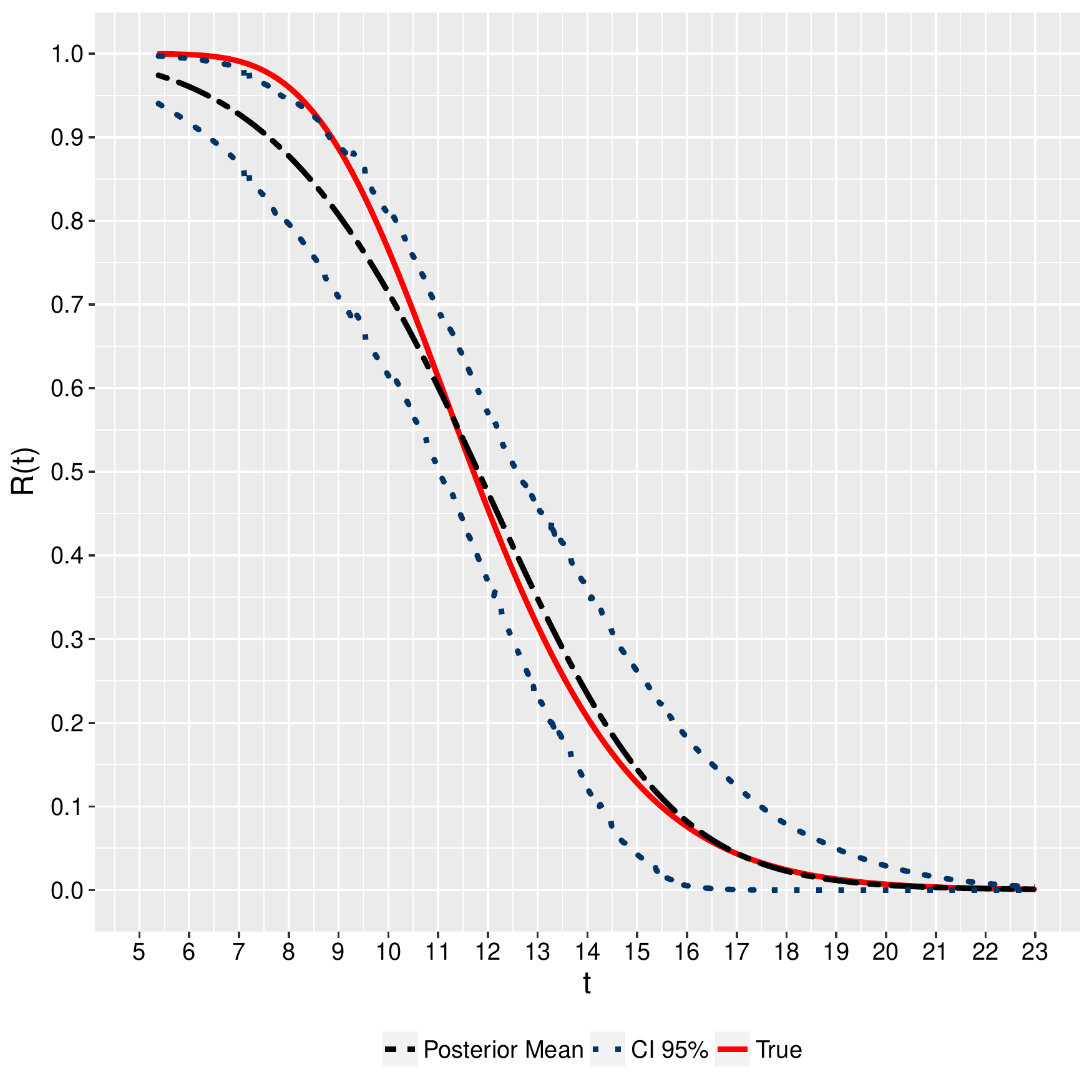}
	\subcaption{Component 4}
    \end{minipage}
	\begin{minipage}[b]{0.32\linewidth}
	\includegraphics[width=\linewidth]{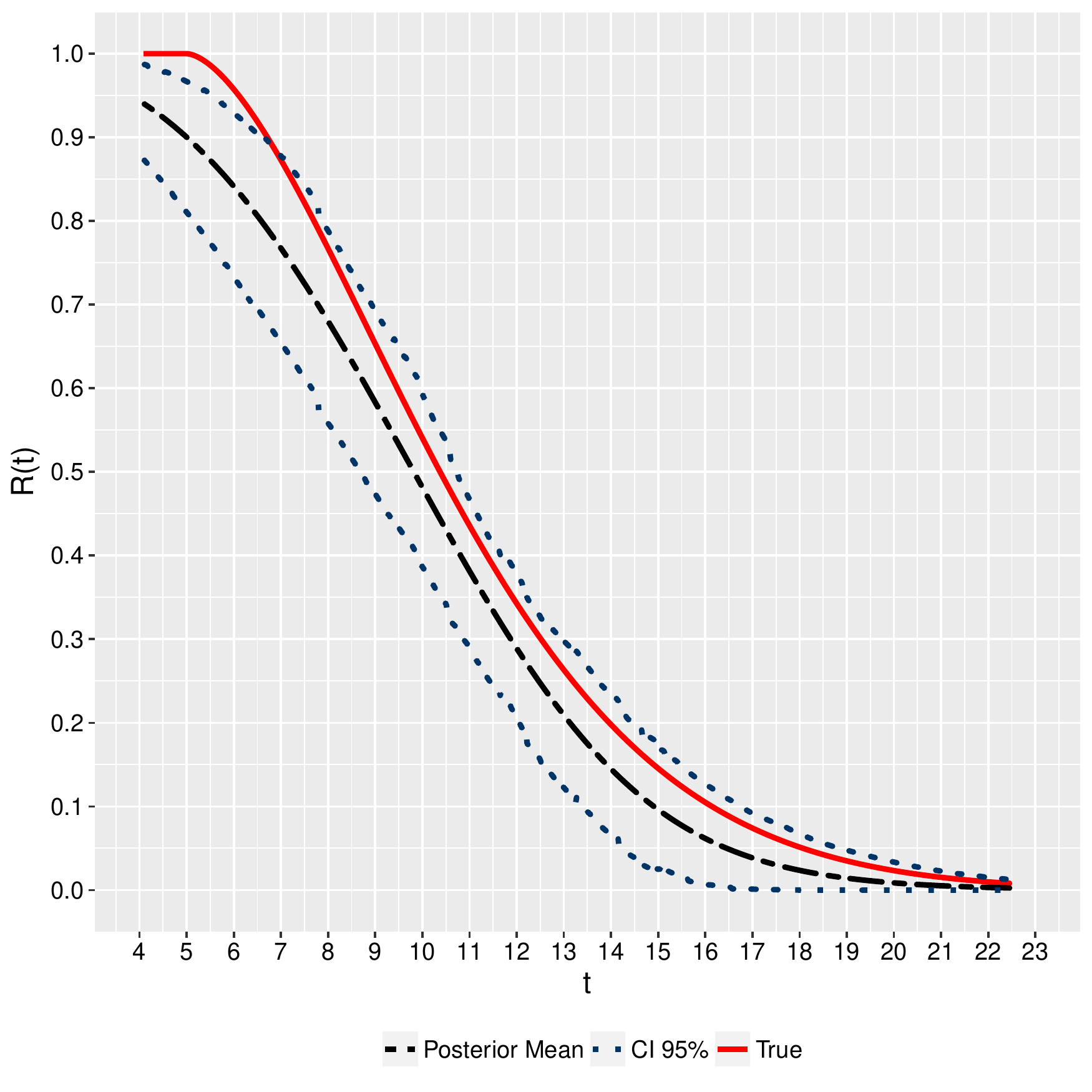}
	\subcaption{Component 5}
    \end{minipage}
	\caption{Generating reliability functions, posterior means and 95\% HPD intervals (CI 95\%) for components involved in \ref{fig:sistemachines} structure in masked scenario.}
	\label{ex_chin_masked}
\end{figure}

 Posterior measures of Weibull parameters $\beta_j$, $\eta_j$ and $\mu_j$, for $j=1,\ldots,5$, are presented in Table \ref{est_param_chin_mask}. The posterior quantities of $R(t\mid{\bm \theta}_j)$ for some values of $t$ are shown in Table \ref{medidas_posterior_chin_mask}.  The posterior means of reliability functions can be visualized in Figure \ref{ex_chin_masked} besides the empirical 95\% HPD intervals. The reliability posterior means are close to the true reliability curves, mainly for component 3. Even for values of $t$ that the posterior mean is more distant from the true values, the lower or upper limit of HDP interval is very close to the true curve. 
 
 \begin{table}[htbp]
 	\centering
 	\scriptsize
 	\caption{Percentages of type of observations for each component in \ref{fig:sistemachines} structure.}
 	\begin{tabular}{cccc}
 		\hline
 		Components & Type  & Number of systems & \% \\
 		\hline
 		1     & Known failure & 13    & 13.00\% \\
 		& Left censored & 37    & 37.00\% \\
 		& Right censored & 37    & 37.00\% \\
 		& Masked  & 13    & 13.00\% \\
 		\hline
 		2     & Known failure & 12    & 12.00\% \\
 		& Left censored & 48    & 48.00\% \\
 		& Right censored & 27    & 27.00\% \\
 		& Masked  & 13    & 13.00\% \\
 		\hline
 		3     & Known failure & 10    & 10.00\% \\
 		& Left censored & 26    & 26.00\% \\
 		& Right censored & 55    & 55.00\% \\
 		& Masked  & 9     & 9.00\% \\
 		\hline
 		4     & Known failure & 12    & 12.00\% \\
 		& Left censored & 30    & 30.00\% \\
 		& Right censored & 47    & 47.00\% \\
 		& Masked  & 11    & 11.00\% \\
 		\hline
 		5     & Known failure & 16    & 16.00\% \\
 		& Left censored & 54    & 54.00\% \\
 		& Right censored & 22    & 22.00\% \\
 		& Masked  & 8     & 8.00\% \\
 		\hline
 	\end{tabular}%
 	\label{chin_masked_porc_censura}%
 \end{table}%

\begin{table}[htbp]
	\centering
	\scriptsize
	\caption{Posterior measures of Weibull parameters for components in \ref{fig:sistemachines} structure in masked scenario.}
	\begin{tabular}{ccccccccc}
		\hline
		\multicolumn{9}{c}{Component 1} \\
		\hline
		& Min   & 1Qt   & Median & Mean  & 3Qt   & Max   & SD    & IC 95\% \\
		$\beta_1$  & 1.017 & 2.188 & 2.568 & 2.572 & 2.922 & 4.449 & 0.558 & 1.561 - 3.677 \\
		$\eta_1$   & 10.650 & 12.190 & 12.700 & 12.780 & 13.290 & 16.590 & 0.815 & 11.287 - 14.242 \\
		$\mu_1$   & 1.49E-33 & 6.03E-27 & 3.44E-24 & 8.25E-13 & 8.34E-19 & 2.48E-11 & 3.15E-12 & 0 - 0 \\
		\hline
		\multicolumn{9}{c}{Component 2} \\
		\hline
		& Min   & 1Qt   & Median & Mean  & 3Qt   & Max   & SD    & IC 95\% \\
		$\beta_2$ & 1.340 & 2.415 & 2.812 & 2.842 & 3.240 & 5.844 & 0.624 & 1.634 - 4.016 \\
		$\eta_2$  & 9.101 & 10.730 & 11.120 & 11.100 & 11.440 & 13.110 & 0.579 & 9.982 - 12.162 \\
		$\mu_2$    & 9.79E-38 & 5.69E-35 & 2.86E-34 & 7.33E-23 & 6.66E-30 & 3.90E-21 & 4.50E-22 & 0 - 0 \\
		\hline
		\multicolumn{9}{c}{Component 3} \\
		\hline
		& Min   & 1Qt   & Median & Mean  & 3Qt   & Max   & SD    & IC 95\% \\
		$\beta_3$  & 2.283 & 4.195 & 4.789 & 4.844 & 5.454 & 8.149 & 0.944 & 3.130 - 6.838 \\
		$\eta_3$  & 11.690 & 12.660 & 12.960 & 13.040 & 13.360 & 15.530 & 0.566 & 12.023 - 14.212 \\
		$\mu_3$   & 5.45E-19 & 1.49E-17 & 7.45E-16 & 7.03E-13 & 4.91E-13 & 1.52E-11 & 1.90E-12 & 0 - 0 \\
		\hline
		\multicolumn{9}{c}{Component 4} \\
		\hline
		& Min   & 1Qt   & Median & Mean  & 3Qt   & Max   & SD    & IC 95\% \\
		$\beta_4$  & 1.858 & 3.867 & 4.391 & 4.412 & 4.925 & 6.707 & 0.774 & 2.95 - 5.943 \\
		$\eta_4$   & 11.330 & 12.530 & 12.830 & 12.880 & 13.180 & 15.800 & 0.505 & 12.009 - 13.887 \\
		$\mu_4$   & 4.09E-20 & 2.53E-18 & 2.13E-17 & 3.33E-12 & 1.20E-13 & 2.77E-10 & 2.40E-11 & 0 - 0 \\
		\hline
		\multicolumn{9}{c}{Component 5} \\
		\hline
		& Min   & 1Qt   & Median & Mean  & 3Qt   & Max   & SD    & IC 95\% \\
		$\beta_5$  & 1.559 & 2.562 & 2.914 & 2.932 & 3.300 & 4.524 & 0.534 & 1.947 - 3.96 \\
		$\eta_5$  & 9.526 & 10.820 & 11.120 & 11.130 & 11.450 & 13.190 & 0.513 & 10.224 - 12.307 \\
		$\mu_5$   & 1.99E-23 & 2.11E-22 & 1.48E-19 & 4.95E-12 & 4.26E-16 & 1.92E-10 & 1.81E-11 & 0 - 0 \\
		\hline
	\end{tabular}%
	\label{est_param_chin_mask}%
\end{table}%

\begin{table}[htbp]
	\centering
	\scriptsize
	\caption{Posterior measures of components reliability functions involved in structure \ref{fig:sistemachines} in masked scenario for some values of $t$.}
	\begin{tabular}{ccccccccc}
		\hline
		\multicolumn{9}{c}{Component 1} \\
		\hline
		t     & Min   & 1Qt   & Median & Mean  & 3Qt   & Max   & SD    & IC 95\% \\
		3.5   & 0.779 & 0.944 & 0.963 & 0.956 & 0.977 & 0.997 & 0.029 & 0.900 - 0.996 \\
		11.0  & 0.340 & 0.463 & 0.499 & 0.500 & 0.535 & 0.651 & 0.051 & 0.394 - 0.591 \\
		18.5  & 0.001 & 0.041 & 0.072 & 0.085 & 0.121 & 0.297 & 0.057 & 0.001 - 0.194 \\
		\hline
		\multicolumn{9}{c}{Component 2} \\
		\hline
		t     & Min   & 1Qt   & Median & Mean  & 3Qt   & Max   & SD    & IC 95\% \\
		4.00  & 0.760 & 0.914 & 0.945 & 0.935 & 0.964 & 0.998 & 0.040 & 0.851 - 0.990 \\
		12.00 & 0.162 & 0.252 & 0.287 & 0.288 & 0.319 & 0.446 & 0.049 & 0.181 - 0.371 \\
		17.00 & 0.001 & 0.017 & 0.035 & 0.044 & 0.061 & 0.226  & 0.037  & 0.001 - 0.115 \\
		\hline
		\multicolumn{9}{c}{Component 3} \\
		\hline
		t     & Min   & 1Qt   & Median & Mean  & 3Qt   & Max   & SD    & IC 95\% \\
		6.00  & 0.874 & 0.963 & 0.975 & 0.971 & 0.984 & 0.998 & 0.018  & 0.937 - 0.998 \\
		11.00 & 0.461 & 0.604 & 0.638 & 0.635 & 0.670 & 0.767 & 0.049  & 0.535 - 0.722 \\
		15.00 & 0.008 & 0.083 & 0.134 & 0.143 & 0.191 & 0.408 & 0.075  & 0.020 - 0.289 \\
		\hline
		\multicolumn{9}{c}{Component 4} \\
		\hline
		t     & Min   & 1Qt   & Median & Mean  & 3Qt   & Max   & SD    & IC 95\% \\
		6.00  & 0.772 & 0.951 & 0.965 & 0.961 & 0.976 & 0.994 & 0.022  & 0.917 - 0.994 \\
		12.00 & 0.273 & 0.440 & 0.477 & 0.476 & 0.516 & 0.643 & 0.053  & 0.369 - 0.571 \\
		16.00 & 0.005 & 0.041 & 0.069 & 0.082 & 0.115 & 0.357 & 0.053  & 0.005 - 0.182 \\
		\hline
		\multicolumn{9}{c}{Component 5} \\
		\hline
		t     & Min   & 1Qt   & Median & Mean  & 3Qt   & Max   & SD    & IC 95\% \\
		4.50  & 0.755 & 0.902 & 0.930 & 0.923 & 0.953 & 0.986 & 0.038 & 0.845 - 0.979 \\
		12.00 & 0.165 & 0.260 & 0.286 & 0.289 & 0.317 & 0.470 & 0.044  & 0.203 - 0.376 \\
		16.00 & 0.004 & 0.037 & 0.057 & 0.062 & 0.081 & 0.210 & 0.033  & 0.007 - 0.127 \\
		\hline
	\end{tabular}%
	\label{medidas_posterior_chin_mask}%
\end{table}%

\clearpage

\section{Computer hard-drive dataset} \label{computer_data_sec}

The dataset is available in \cite{Flehinger2002} which consists of $172$ observed failure times of computer hard-drives monitored over a period of $4$ years. There were three possible causes of failure: eletronic hard (component $j=1$), head flyability (component $j=2$) and head/disc magnetics (component $j=3$). That is, the hard-drives are system with three components in series. However, for some of them ($38\%$) the cause of hard-drive fail was not identified. For these masked data systems, $s=\{1,3\}$ or $s=\{1,2,3\}$, that is, there is no possible masked set $s=\{1,2\}$ or $s=\{2,3\}$. Note that in our proposal approach the configuration of set $s$ is not a big deal, once the important information for estimation of $j$th component reliability is if $j$ belongs to $s$ or not. More details about the detection of failure causes can be found in \cite{Flehinger} and in \cite{CraiuReiser}. The dataset is presented in Table \ref{computer_mask} in Appendix \ref{apeC}. 

Since the components in $s_i$ are right censored or responsible for system failure, $\lambda_{3j}=0$, for $j=1,2,3$, and no hard-drive is subject of left censored failure time. Table \ref{causa_falha_drive} presents the percentages of observations types of each component. The component 1 caused the failure of $20.35\%$ of systems, $11.05\%$ of systems failured because of component 2 and component 3 was responsible for $30.23\%$ of system failure. Besides, or component 1 or component 3 caused the failure of $18.60\%$ of systems and the remaining $19.77\%$ of systems broken down because of any of three components.

We generated $35{,}000$ values of each parameter, disregarding the first $5{,}000$ iterations as burn-in samples and jump of size $30$ to avoid correlation problems, obtaining a sample of size $n_p=1{,}000$. The chains convergence was monitored and good convergence results were obtained.

\begin{figure}[h!]\centering
	\begin{minipage}[b]{0.32\linewidth}
		\includegraphics[width=\linewidth]{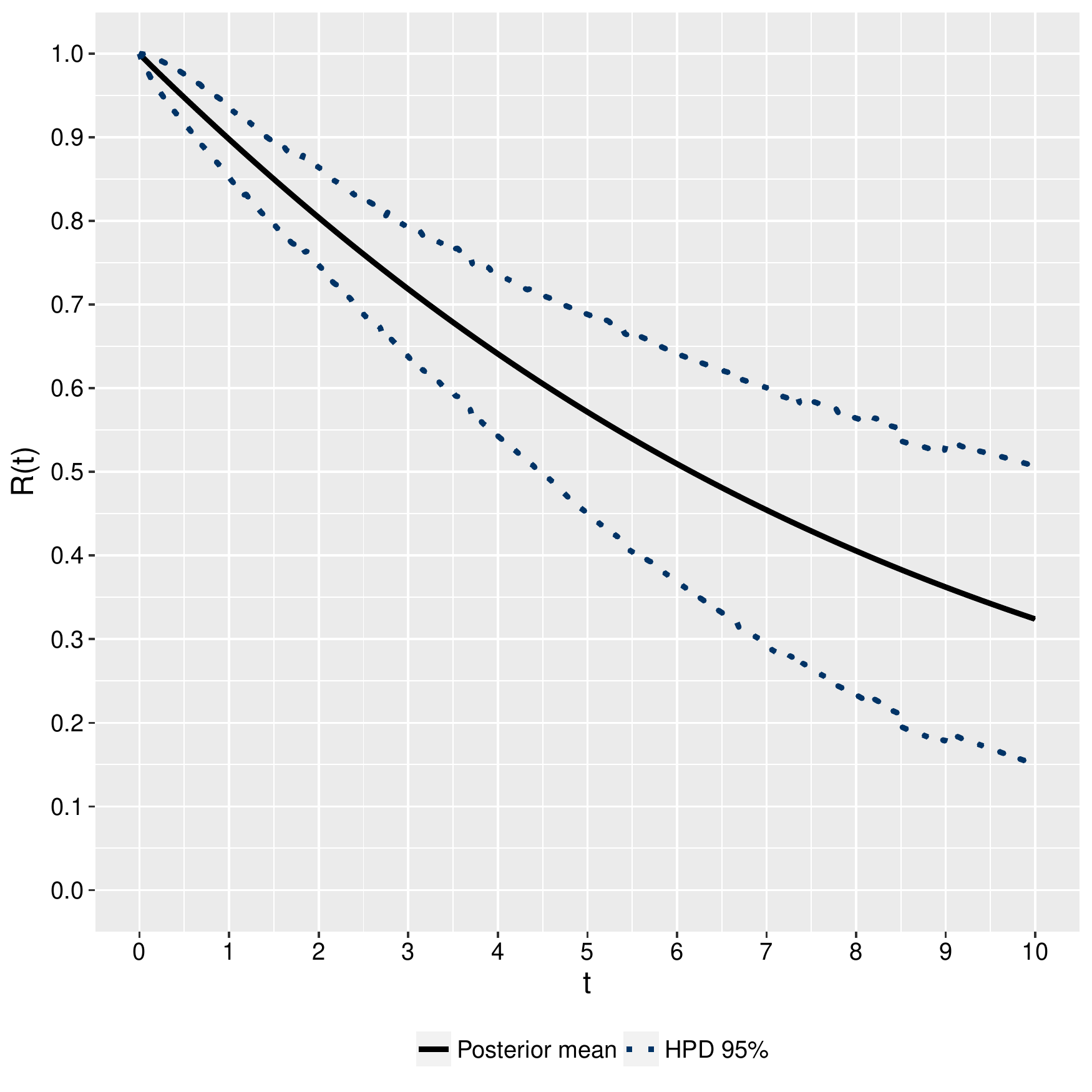}
		\subcaption{Eletronic hard}
	\end{minipage} 
	\begin{minipage}[b]{0.32\linewidth}
		\includegraphics[width=\linewidth]{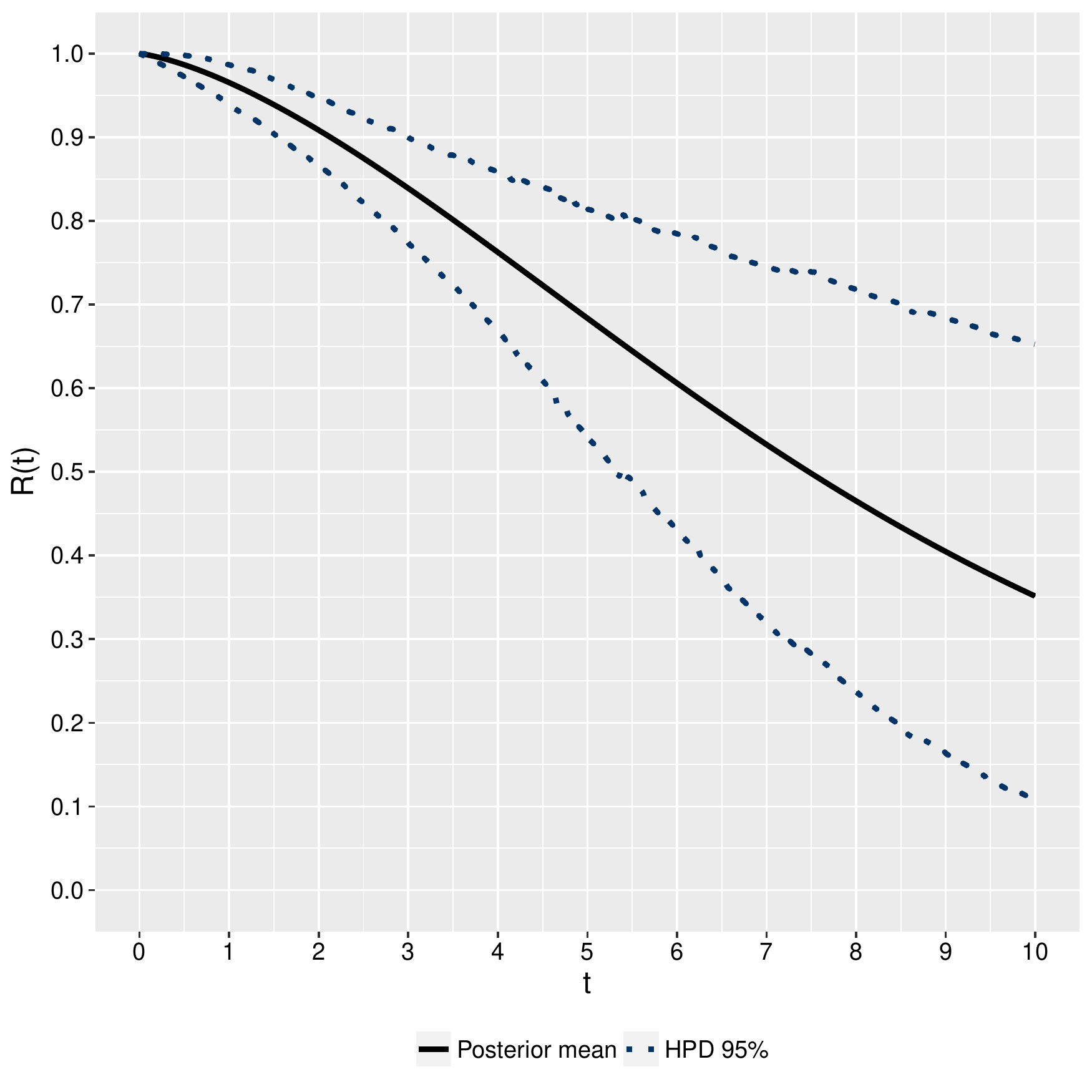}
		\subcaption{Head flyability}
	\end{minipage}
	\begin{minipage}[b]{0.32\linewidth}
		\includegraphics[width=\linewidth]{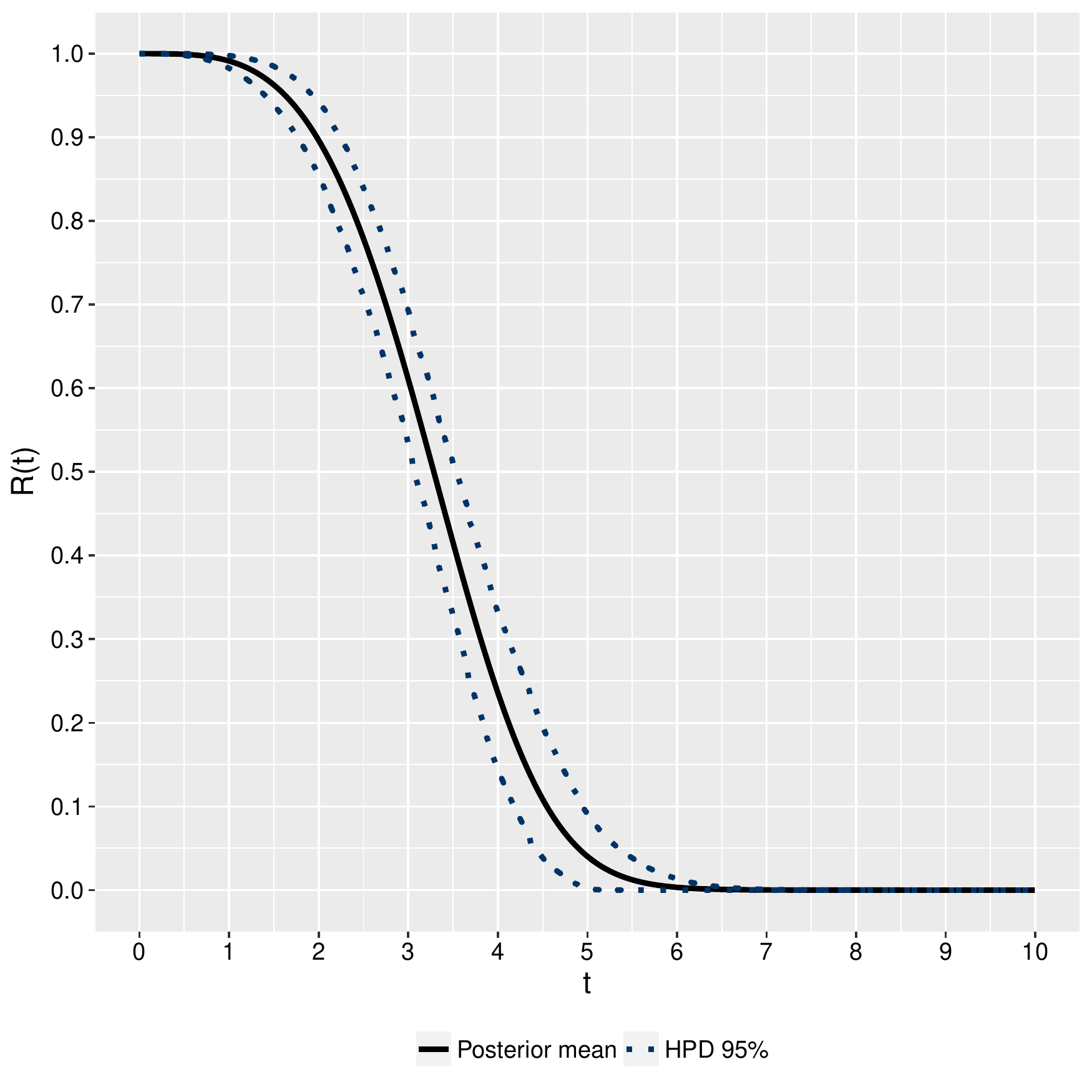}
		\subcaption{Head/disc magnetics}
	\end{minipage}
	\caption{Generating reliability functions, posterior mean and 95\% HPD intervals (CI 95\%) for the components involved in computer hard-drive dataset.}
	\label{aplic_hard}
\end{figure}

Posterior measures of Weibull parameters $\beta_j$, $\eta_j$ and $\mu_j$, for $j=1,2,3$, are presented in Table \ref{est_param_computer_mask}. The posterior quantities of $R(t\mid{\bm \theta}_j)$ for some values of $t$ are shown in Table \ref{medidas_posterior_computer_mask}.  The posterior means of reliability functions can be visualized in Figure \ref{aplic_hard} besides the empirical 95\% HPD intervals. 

\begin{table}[htbp]
	\centering
	\scriptsize
	\caption{Percentages of type of observations for each component in computer hard-drive dataset.}
	\begin{tabular}{cccc}
		\hline
		Components & Type  & Number of systems & \% \\
		\hline
		1     & Known failure & 35    & 20.35\% \\
		& Right censored & 71    & 41.28\% \\
		& Masked  & 66    & 38.37\% \\
		\hline
		2     & Known failure & 19    & 11.05\% \\
		& Right censored & 119   & 69.19\% \\
		& Masked  & 34    & 19.77\% \\
		\hline
		3     & Known failure & 52    & 30.23\% \\
		& Right censored & 54    & 31.40\% \\
		& Masked  & 66    & 38.37\% \\
		\hline
	\end{tabular}%
	\label{causa_falha_drive}%
\end{table}%

\begin{table}[htbp]
	\centering
	\scriptsize
	\caption{Posterior measures of Weibull parameters for components in computer hard drive dataset.}
	\begin{tabular}{ccccccccc}
		\hline
		\multicolumn{9}{c}{Component 1} \\
		\hline
		& Min   & 1Qt   & Median & Mean  & 3Qt    & Max   & SD    & IC95\% \\
		$\beta_1$ & 0.603 & 0.919 & 1.017 & 1.025 & 1.130 & 1.613 & 0.156 & 0.734 - 1.312 \\
		$\eta_1$ & 4.890 & 7.586 & 9.135 & 9.668 & 10.790 & 34.550 & 3.098 & 5.299 - 15.499 \\
		$\mu_1$    & 2.03E-160 & 1.72E-129 & 3.55E-85 & 1.36E-35 & 1.03E-60 & 9.68E-33 & 3.11E-34 & 0 - 0 \\
		\hline
		\multicolumn{9}{c}{Component 2} \\
		\hline
		& Min   & 1Qt   & Median & Mean  & 3Qt    & Max   & SD    & IC95\% \\
		$\beta_2$ & 0.607 & 1.295 & 1.480 & 1.502 & 1.684 & 2.584 & 0.305 & 0.994 - 2.184 \\
		$\eta_2$ & 5.168 & 8.151 & 9.783 & 10.870 & 12.060 & 76.190 & 5.027 & 5.350 - 19.275 \\
		$\mu_2$    & 1.84E-201 & 2.45E-161 & 8.25E-120 & 2.17E-36 & 1.00E-78 & 7.55E-34 & 3.11E-35 & 0 - 0 \\
		\hline
		\multicolumn{9}{c}{Component 3} \\
		\hline
		& Min   & 1Qt   & Median & Mean  & 3Qt    & Max   & SD    & IC95\% \\
		$\beta_3$ & 2.689 & 3.479 & 3.725 & 3.737 & 3.992 & 5.027 & 0.399 & 2.971 - 4.522 \\
		$\eta_3$ & 3.273 & 3.539 & 3.617 & 3.626 & 3.704 & 4.340 & 0.129 & 3.378 - 3.869 \\
		$\mu_3$  & 4.67E-29 & 1.75E-21 & 1.93E-16 & 1.73E-11 & 8.58E-14 & 8.46E-10 & 7.75E-11 & 0 - 0 \\
		\hline
	\end{tabular}%
	\label{est_param_computer_mask}%
\end{table}%

\begin{table}[htbp]
	\centering
	\scriptsize
	\caption{Posterior measures of components' reliability function involved in computer hard-drive dataset for some values of $t$.}
	\begin{tabular}{ccccccccc}
		\hline
		\multicolumn{9}{c}{Component 1} \\
		\hline
		t     & Min   & 1Qt   & Median & Mean  & 3Qt   & Max   & SD    & IC95\% \\
		0.049 & 0.973 & 0.993 & 0.995 & 0.994 & 0.997 & 1.000 & 0.004 & 0.986 - 0.999 \\
		1.502 & 0.739 & 0.835 & 0.852 & 0.850 & 0.869 & 0.932 & 0.026 & 0.797 - 0.899 \\
		4.480 & 0.409 & 0.571 & 0.612 & 0.610 & 0.651 & 0.837 & 0.060 & 0.495 - 0.730 \\
		\hline
		\multicolumn{9}{c}{Component 2} \\
		\hline
		t     & Min   & 1Qt   & Median & Mean  & 3Qt   & Max   & SD    & IC95\% \\
		0.049 & 0.988 & 0.999 & 1.000 & 0.999 & 1.000 & 1.000 & 0.001 & 0.998 - 1.000 \\
		1.502 & 0.845 & 0.927 & 0.939 & 0.938 & 0.952 & 0.985 & 0.018 & 0.903 - 0.971 \\
		4.480 & 0.499 & 0.683 & 0.728 & 0.724 & 0.769 & 0.878 & 0.064 & 0.607 - 0.850 \\
		\hline
		\multicolumn{9}{c}{Component 3} \\
		\hline
		t     & Min   & 1Qt   & Median & Mean  & 3Qt   & Max   & SD    & IC95\% \\
		0.049 & 1.000 & 1.000 & 1.000 & 1.000 & 1.000 & 1.000 & 0.000 & 0.999 - 1.000 \\
		1.502 & 0.907 & 0.954 & 0.963 & 0.961 & 0.971 & 0.987 & 0.013 & 0.933 - 0.982 \\
		4.480 & 0.012 & 0.084 & 0.107 & 0.113 & 0.138 & 0.336 & 0.042 & 0.041 - 0.197 \\
		\hline
	\end{tabular}%
	\label{medidas_posterior_computer_mask}%
\end{table}%

\appendix

\chapter{Datasets - Chapter 2} 
\label{apeD}

\begin{table}[!ht]
	\scriptsize
	\caption {Simulated sample of example \ref{ex-sps4}}
	\label{tabela_data1}
	\centering
\begin{tabular}{cc|cc|cc|cc|cc}
		\hline
		$T$ & $\delta$ & $T$ & $\delta$ & $T$ & $\delta$ & $T$ & $\delta$ & $T$ & $\delta$ \\
		\hline
		~1.25 & 2 & ~1.27 & 4 & ~1.35 & 3 & ~1.38 & 3 & ~1.43 & 3 \\
		~1.47 & 2 & ~1.54 & 4 & ~1.61 & 4 & ~1.89 & 4 & ~1.92 & 1 \\
		~2.22 & 2 & ~2.40 & 3 & ~2.43 & 2 & ~2.58 & 1 & ~2.59 & 1 \\
		~2.60 & 3 & ~2.94 & 4 & ~2.98 & 2 & ~3.01 & 1 & ~3.03 & 1 \\ 
		~3.05 & 1 & ~3.12 & 2 & ~3.14 & 4 & ~3.15 & 2 & ~3.16 & 1 \\
		~3.17 & 2 & ~3.40 & 1 & ~3.41 & 1 & ~3.42 & 1 & ~3.46 & 2 \\ 
		~3.55 & 3 & ~3.67 & 3 & ~3.68 & 2 & ~3.72 & 3 & ~3.83 & 4 \\
		~3.88 & 1 & ~4.14 & 3 & ~4.17 & 2 & ~4.24 & 2 & ~4.40 & 3 \\
		~4.50 & 1 & ~4.66 & 4 & ~4.67 & 4 & ~4.72 & 4 & ~4.80 & 1 \\
		~4.81 & 3 & ~4.85 & 3 & ~4.89 & 1 & ~4.98 & 1 & ~5.00 & 3 \\
		~5.06 & 3 & ~5.12 & 1 & ~5.15 & 2 & ~5.18 & 2 & ~5.22 & 2 \\
		~5.23 & 2 & ~5.38 & 2 & ~5.51 & 2 & ~5.67 & 4 & ~5.68 & 1 \\ 
		~5.78 & 2 & ~5.81 & 3 & ~5.92 & 2 & ~6.01 & 4 & ~6.05 & 1 \\
		~6.10 & 2 & ~6.12 & 3 & ~6.28 & 2 & ~6.41 & 2 & ~6.57 & 3 \\
		~6.67 & 4 & ~6.76 & 3 & ~6.87 & 3 & ~6.95 & 4 & ~7.08 & 2 \\
		~7.09 & 2 & ~7.14 & 1 & ~7.41 & 3 & ~7.48 & 2 & ~7.60 & 1 \\
		~7.62 & 2 & ~7.77 & 1 & ~8.38 & 2 & ~8.44 & 3 & ~8.48 & 2 \\
		~8.69 & 2 & ~8.74 & 2 & ~9.02 & 3 & ~9.36 & 2 & 10.04 & 2 \\ 
		10.07 & 2 & 10.16 & 1 & 11.07 & 2 & 11.27 & 2 & 11.38 & 3 \\
		11.74 & 3 & 11.89 & 3 & 13.36 & 3 & 13.51 & 1 & 15.27 & 3 \\
		\hline
	\end{tabular}
\end{table}

\begin{table}[!ht]
	\caption {Simulated sample of example \ref{ex-pss4}}
	\label{tabela_data2}
	\centering
	\scriptsize
	\begin{tabular}{cc|cc|cc|cc|cc}
		\hline
		$T$ & $\delta$ & $T$ & $\delta$ & $T$ & $\delta$ & $T$ & $\delta$ & $T$ & $\delta$ \\
		\hline
		~0.66 & 2 & ~0.71 & 3 & ~1.00 & 3 & ~1.00 & 3 & ~1.00 & 3 \\
		~1.00 & 3 & ~1.00 & 3 & ~1.06 & 2 & ~1.08 & 2 & ~1.18 & 4 \\
		~1.34 & 2 & ~1.38 & 1 & ~1.45 & 1 & ~1.54 & 3 & ~1.72 & 1 \\
		~1.73 & 1 & ~1.78 & 1 & ~1.81 & 2 & ~1.85 & 1 & ~1.87 & 4 \\
		~1.92 & 4 & ~1.96 & 3 & ~1.99 & 1 & ~2.09 & 1 & ~2.15 & 2 \\
		~2.17 & 2 & ~2.23 & 3 & ~2.26 & 2 & ~2.27 & 2 & ~2.28 & 1 \\
		~2.30 & 1 & ~2.34 & 1 & ~2.39 & 3 & ~2.49 & 1 & ~2.52 & 3 \\
		~2.53 & 4 & ~2.55 & 4 & ~2.56 & 2 & ~2.59 & 1 & ~2.64 & 1 \\
		~2.65 & 1 & ~2.73 & 1 & ~2.74 & 4 & ~2.87 & 1 & ~2.88 & 4 \\
		~2.97 & 2 & ~3.00 & 3 & ~3.00 & 3 & ~3.00 & 3 & ~3.00 & 3 \\
		~3.00 & 3 & ~3.00 & 3 & ~3.00 & 3 & ~3.00 & 3 & ~3.00 & 3 \\
		~3.04 & 1 & ~3.05 & 1 & ~3.06 & 2 & ~3.08 & 3 & ~3.12 & 2 \\
		~3.15 & 2 & ~3.24 & 4 & ~3.25 & 3 & ~3.36 & 2 & ~3.53 & 3 \\
		~3.71 & 3 & ~3.78 & 2 & ~3.96 & 4 & ~4.05 & 1 & ~4.27 & 1 \\
		~4.53 & 2 & ~4.62 & 4 & ~4.65 & 1 & ~4.85 & 1 & ~4.91 & 2 \\
		~4.92 & 4 & ~5.05 & 2 & ~5.18 & 2 & ~5.21 & 2 & ~5.22 & 1 \\
		~5.23 & 1 & ~5.24 & 3 & ~5.25 & 1 & ~5.41 & 2 & ~5.46 & 2 \\
		~5.51 & 1 & ~5.56 & 2 & ~5.58 & 2 & ~5.59 & 1 & ~5.64 & 2 \\
		~5.67 & 1 & ~5.79 & 2 & ~5.96 & 3 & ~6.64 & 2 & ~6.66 & 4 \\
		~6.88 & 4 & ~8.01 & 3 & ~8.04 & 2 & ~8.75 & 3 & ~9.53 & 3 \\
		\hline
	\end{tabular}
\end{table}
 
\chapter{Datasets - Chapter 3} 
\label{apenB}

\begin{table}[htbp]
	\centering
	\caption{Observed data of $n=30$ parallel systems.}
	\begin{scriptsize}
	\begin{tabular}{c|cc|cc|cc}
		\hline
		System ID & \multicolumn{2}{c|}{Component 1} & \multicolumn{2}{c|}{Component 2} & \multicolumn{2}{c}{Component 3} \\
		& l     & u     & l     & u     & l     & u \\
		\hline
		1     & 0.00  & 8.83  & 8.83  & 8.83  & 0.00  & 8.83 \\
		2     & 5.04  & 5.04  & 0.00  & 5.04  & 0.00  & 5.04 \\
		3     & 0.00  & 5.07  & 5.07  & 5.07  & 0.00  & 5.07 \\
		4     & 3.99  & 3.99  & 0.00  & 3.99  & 0.00  & 3.99 \\
		5     & 0.00  & 9.73  & 9.73  & 9.73  & 0.00  & 9.73 \\
		6     & 0.00  & 5.75  & 5.75  & 5.75  & 0.00  & 5.75 \\
		7     & 0.00  & 5.56  & 0.00  & 5.56  & 5.56  & 5.56 \\
		8     & 0.00  & 3.17  & 3.17  & 3.17  & 0.00  & 3.17 \\
		9     & 5.31  & 5.31  & 0.00  & 5.31  & 0.00  & 5.31 \\
		10    & 0.00  & 3.45  & 0.00  & 3.45  & 3.45  & 3.45 \\
		11    & 16.71 & 16.71 & 0.00  & 16.71 & 0.00  & 16.71 \\
		12    & 7.24  & 7.24  & 0.00  & 7.24  & 0.00  & 7.24 \\
		13    & 0.00  & 4.54  & 0.00  & 4.54  & 4.54  & 4.54 \\
		14    & 0.00  & 4.56  & 0.00  & 4.56  & 4.56  & 4.56 \\
		15    & 6.39  & 6.39  & 0.00  & 6.39  & 0.00  & 6.39 \\
		16    & 0.00  & 8.41  & 0.00  & 8.41  & 8.41  & 8.41 \\
		17    & 0.00  & 7.22  & 0.00  & 7.22  & 7.22  & 7.22 \\
		18    & 0.00  & 4.70  & 0.00  & 4.70  & 4.70  & 4.70 \\
		19    & 4.19  & 4.19  & 0.00  & 4.19  & 0.00  & 4.19 \\
		20    & 0.00  & 9.59  & 9.59  & 9.59  & 0.00  & 9.59 \\
		21    & 3.91  & 3.91  & 0.00  & 3.91  & 0.00  & 3.91 \\
		22    & 5.39  & 5.39  & 0.00  & 5.39  & 0.00  & 5.39 \\
		23    & 0.00  & 5.57  & 0.00  & 5.57  & 5.57  & 5.57 \\
		24    & 5.50  & 5.50  & 0.00  & 5.50  & 0.00  & 5.50 \\
		25    & 0.00  & 7.63  & 7.63  & 7.63  & 0.00  & 7.63 \\
		26    & 0.00  & 8.25  & 8.25  & 8.25  & 0.00  & 8.25 \\
		27    & 0.00  & 7.76  & 7.76  & 7.76  & 0.00  & 7.76 \\
		28    & 5.18  & 5.18  & 0.00  & 5.18  & 0.00  & 5.18 \\
		29    & 0.00  & 3.16  & 0.00  & 3.16  & 3.16  & 3.16 \\
		30    & 15.85 & 15.85 & 0.00  & 15.85 & 0.00  & 15.85 \\
		\hline
	\end{tabular}%
  \end{scriptsize}
	\label{apeB.paralelo}%
\end{table}%

\begin{table}[htbp]
	\centering
	\begin{scriptsize}
		\caption{Observed data of $n=50$ $2$-out-of-$3$ systems.}
		\begin{tabular}{c|cc|cc|cc}
			\hline
			System ID & \multicolumn{2}{c|}{Component 1} & \multicolumn{2}{c|}{Component 2} & \multicolumn{2}{c}{Component 3} \\
			& l     & u     & l     & u     & l     & u \\
			\hline
			1     & 0.00  & 20.16 & 20.16 & 20.16 & 20.16 & $\infty$ \\
			2     & 18.95 & 18.95 & 18.95 & $\infty$   & 0.00  & 18.95 \\
			3     & 0.00  & 20.12 & 20.12 & 20.12 & 20.12 & $\infty$ \\
			4     & 18.86 & 18.86 & 0.00  & 18.86 & 18.86 & $\infty$ \\
			5     & 0.00  & 15.80 & 15.80 & 15.80 & 15.80 & $\infty$ \\
			6     & 0.00  & 13.35 & 13.35 & 13.35 & 13.35 & $\infty$ \\
\hline
		\end{tabular}%
		\label{apeB.2de3}%
	\end{scriptsize}
\end{table}%

\begin{table}[htbp]
	\centering
   \begin{scriptsize}
	\caption{Observed data of $n=100$ bridge systems (continues on the next page).}
	\begin{tabular}{c|cc|cc|cc|cc|cc}
		\hline
		System ID  & \multicolumn{2}{c|}{Component 1} & \multicolumn{2}{c|}{Component 2} & \multicolumn{2}{c|}{Component 3} & \multicolumn{2}{c|}{Component 4} & \multicolumn{2}{c}{Component 5} \\
		& l     & u     & l     & u     & l     & u     & l     & u     & l     & u \\
		\hline
		1     & 16.56 & $\infty$   & 0.00  & 16.56 & 0.00  & 16.56 & 16.56 & 16.56 & 16.56 & $\infty$ \\
		2     & 18.98 & 18.98 & 0.00  & 18.98 & 18.98 & $\infty$   & 0.00  & 18.98 & 18.98 & $\infty$ \\
		3     & 0.00  & 17.50 & 17.50 & 17.50 & 0.00  & 17.50 & 0.00  & 17.50 & 17.50 & $\infty$ \\
		4     & 15.21 & $\infty$   & 15.21 & $\infty$   & 0.00  & 15.21 & 15.21 & 15.21 & 0.00  & 15.21 \\
		5     & 0.00  & 19.49 & 19.49 & 19.49 & 0.00  & 19.49 & 19.49 & $\infty$   & 19.49 & $\infty$ \\
		6     & 16.37 & $\infty$   & 0.00  & 16.37 & 16.37 & 16.37 & 0.00  & 16.37 & 16.37 & $\infty$ \\
\hline
	\end{tabular}%
	\label{apeB.bridge}%
  \end{scriptsize}
\end{table}%

\begin{table}[htbp]
	\centering
	\begin{scriptsize}
		\caption{Continuation - observed data of $n=100$ bridge systems.}
		\begin{tabular}{c|cc|cc|cc|cc|cc}
			\hline
			System ID  & \multicolumn{2}{c|}{Component 1} & \multicolumn{2}{c|}{Component 2} & \multicolumn{2}{c|}{Component 3} & \multicolumn{2}{c|}{Component 4} & \multicolumn{2}{c}{Component 5} \\
			& l     & u     & l     & u     & l     & u     & l     & u     & l     & u \\
			\hline
			51    & 18.12 & $\infty$   & 18.12 & 18.12 & 0.00  & 18.12 & 0.00  & 18.12 & 18.12 & $\infty$ \\
			52    & 17.66 & $\infty$   & 0.00  & 17.66 & 17.66 & 17.66 & 0.00  & 17.66 & 17.66 & $\infty$ \\
			53    & 14.93 & 14.93 & 0.00  & 14.93 & 0.00  & 14.93 & 14.93 & $\infty$   & 14.93 & $\infty$ \\
			54    & 19.76 & $\infty$   & 0.00  & 19.76 & 0.00  & 19.76 & 19.76 & 19.76 & 19.76 & $\infty$ \\
			55    & 0.00  & 21.56 & 21.56 & $\infty$   & 21.56 & $\infty$   & 0.00  & 21.56 & 21.56 & 21.56 \\
			56    & 13.89 & $\infty$   & 0.00  & 13.89 & 13.89 & 13.89 & 0.00  & 13.89 & 13.89 & $\infty$ \\
\hline
		\end{tabular}%
		\label{apeB.bridge1}%
	\end{scriptsize}
\end{table}%

\begin{table}[htbp]
	\centering
	\caption{Device-G observed data.}
	\begin{tabular}{c|cc|cc}
		\hline
		Device-G ID  & \multicolumn{2}{c|}{Component W} & \multicolumn{2}{c}{Component S} \\
		& l     & u     & l     & u \\
		\hline
		1     & 275   & 275   & 275   & $\infty$ \\
		2     & 13    & $\infty$   & 13    & 13 \\
		3     & 147   & 147   & 147   & $\infty$ \\
		4     & 23    & $\infty$   & 23    & 23 \\
		5     & 181   & 181   & 181   & $\infty$ \\
		6     & 30    & $\infty$   & 30    & 30 \\
		7     & 65    & $\infty$   & 65    & 65 \\
		8     & 10    & $\infty$   & 10    & 10 \\
		9     & 300   & $\infty$   & 300   & $\infty$ \\
		10    & 173   & $\infty$   & 173   & 173 \\
		11    & 106   & $\infty$   & 106   & 106 \\
		12    & 300   & $\infty$   & 300   & $\infty$ \\
		13    & 300   & $\infty$   & 300   & $\infty$ \\
		14    & 212   & 212   & 212   & $\infty$ \\
		15    & 300   & $\infty$   & 300   & $\infty$ \\
		16    & 300   & $\infty$   & 300   & $\infty$ \\
		17    & 300   & $\infty$   & 300   & $\infty$ \\
		18    & 2     & $\infty$   & 2     & 2 \\
		19    & 261   & $\infty$   & 261   & 261 \\
		20    & 293   & 293   & 293   & $\infty$ \\
		21    & 88    & $\infty$   & 88    & 88 \\
		22    & 247   & $\infty$   & 247   & 247 \\
		23    & 28    & $\infty$   & 28    & 28 \\
		24    & 143   & $\infty$   & 143   & 143 \\
		25    & 300   & $\infty$   & 300   & $\infty$ \\
		26    & 23    & $\infty$   & 23    & 23 \\
		27    & 300   & $\infty$   & 300   & $\infty$ \\
		28    & 80    & $\infty$   & 80    & 80 \\
		29    & 245   & 245   & 245   & $\infty$ \\
		30    & 266   & 266   & 266   & $\infty$ \\
		\hline
	\end{tabular}%
	\label{deviceG_data}%
\end{table}%

\chapter{Datasets - Chapter 4} 
\label{apeC}

\begin{table}[htbp]
	\centering
	\tiny
	\caption{Observed data of $n=300$ $2$-out-of-$3$ systems in masked scenario (continues on the next pages).}
	\vspace{-0.2cm}
	\begin{tabular}{c|c|cc|cc|cc}
		\hline
		System ID & t & $\delta_1$    & $\upsilon_1$    & $\delta_2$    & $\upsilon_2$     & $\delta_3$   & $\upsilon_3$ \\
		\hline
	    1     & 16.91 & 3     & 0     & 1     & 0     & 2     & 0 \\
	2     & 19.31 & 2     & 0     & 3     & 0     & 1     & 0 \\
	3     & 17.12 & 3     & 0     & 2     & 0     & 1     & 0 \\
	4     & 19.17 & -     & 1     & 2     & 0     & -     & 1 \\
	5     & 15.58 & 3     & 0     & 1     & 0     & 2     & 0 \\
	6     & 18.04 & 3     & 0     & 1     & 0     & 2     & 0 \\
	7     & 17.89 & 1     & 0     & 3     & 0     & 2     & 0 \\
\hline
	\end{tabular}%
	\label{2de3_masked_ex}%
\end{table}%

\begin{table}[htbp]
	\centering
	\tiny
	\caption{Continuation of Table \ref{2de3_masked_ex} (continues on the next pages).}
	\vspace{-0.2cm}
	\begin{tabular}{c|c|cc|cc|cc}
		\hline
		System ID & t & $\delta_1$    & $\upsilon_1$    & $\delta_2$    & $\upsilon_2$     & $\delta_3$   & $\upsilon_3$ \\
		\hline
		    41    & 19.87 & -     & 1     & 2     & 0     & -     & 1 \\
		42    & 15.52 & 3     & 0     & 1     & 0     & 2     & 0 \\
		43    & 16.28 & 3     & 0     & 1     & 0     & 2     & 0 \\
		44    & 19.72 & 3     & 0     & 2     & 0     & 1     & 0 \\
		45    & 18.44 & 3     & 0     & 1     & 0     & 2     & 0 \\
		46    & 19.92 & 3     & 0     & 2     & 0     & 1     & 0 \\
		47    & 17.24 & 1     & 0     & 3     & 0     & 2     & 0 \\
\hline
	\end{tabular}%
	\label{cont_2de3_mask}%
\end{table}%

\begin{table}[htbp]
	\centering
	\tiny
	\caption{Continuation of Table \ref{2de3_masked_ex} (continues on the next pages).}
	\vspace{-0.2cm}
	\begin{tabular}{c|c|cc|cc|cc}
		\hline
		System ID & t & $\delta_1$    & $\upsilon_1$    & $\delta_2$    & $\upsilon_2$     & $\delta_3$   & $\upsilon_3$ \\
		\hline
	136   & 16.61 & 3     & 0     & 1     & 0     & 2     & 0 \\
	137   & 21.86 & 3     & 0     & 2     & 0     & 1     & 0 \\
	138   & 18.65 & 3     & 0     & 1     & 0     & 2     & 0 \\
	139   & 17.09 & -     & 1     & -     & 1     & 2     & 0 \\
    140   & 17.23 & 3     & 0     & 2     & 0     & 1     & 0 \\
141   & 18.78 & -     & 1     & -     & 1     & 2     & 0 \\
142   & 19.94 & -     & 1     & 2     & 0     & -     & 1 \\
\hline
	\end{tabular}%
	\label{cont_2de3_mask2}%
\end{table}%

\begin{table}[htbp]
	\centering
	\tiny
	\caption{Continuation of Table \ref{2de3_masked_ex}.}
	\vspace{-0.2cm}
	\begin{tabular}{c|c|cc|cc|cc}
		\hline
		System ID & t & $\delta_1$    & $\upsilon_1$    & $\delta_2$    & $\upsilon_2$     & $\delta_3$   & $\upsilon_3$ \\
		\hline
		    231   & 15.69 & -     & 1     & -     & 1     & 2     & 0 \\
		232   & 15.98 & 3     & 0     & 2     & 0     & 1     & 0 \\
		233   & 17.80 & -     & 1     & 2     & 0     & -     & 1 \\
		234   & 16.55 & 3     & 0     & 2     & 0     & 1     & 0 \\
		235   & 20.64 & -     & 1     & 2     & 0     & -     & 1 \\
		236   & 19.07 & 3     & 0     & 1     & 0     & 2     & 0 \\
		237   & 16.41 & -     & 1     & -     & 1     & 2     & 0 \\
\hline
	\end{tabular}%
	\label{2de3_mask_cont2}%
\end{table}%

\begin{table}[htbp]
	\centering
	\tiny
	\caption{Observed data of $n=100$ systems as \ref{fig:sistemachines} in masked scenario.}
	\vspace{-0.2cm}
	\begin{tabular}{c|c|cc|cc|cc|cc|cc}
		\hline
		System ID & t & $\delta_1$    & $\upsilon_1$    & $\delta_2$    & $\upsilon_2$     & $\delta_3$   & $\upsilon_3$ & $\delta_4$   & $\upsilon_4$ & $\delta_5$   & $\upsilon_5$\\
		\hline
		1     & 11.39 & 3     & 0     & 1     & 0     & 2     & 0     & 2     & 0     & 2     & 0 \\
		2     & 10.12 & 2     & 0     & 3     & 0     & 2     & 0     & 3     & 0     & 3     & 0 \\
		3     & 11.87 & 2     & 0     & 2     & 0     & 2     & 0     & 1     & 0     & 3     & 0 \\
		4     & 11.64 & 1     & 0     & 3     & 0     & -     & 1     & -     & 1     & -     & 1 \\
		5     & 10.01 & 3     & 0     & 2     & 0     & 3     & 0     & -     & 1     & 1     & 1 \\
		6     & 11.25 & 3     & 0     & 2     & 0     & 2     & 0     & 3     & 0     & 3     & 0 \\
		7     & 9.30  & 3     & 0     & 3     & 0     & 2     & 0     & 2     & 0     & 2     & 0 \\
\hline
	\end{tabular}%
	\label{chin_masked_ex}%
\end{table}%

\begin{table}[htbp]
	\centering
	\tiny
	\caption{Observed data of $n=172$ computer hard-drive data (continues on the next page).}
	\begin{tabular}{c|c|cc|cc|cc}
		\hline
		System ID & t & $\delta_1$    & $\upsilon_1$    & $\delta_2$    & $\upsilon_2$     & $\delta_3$ & $\upsilon_3$ \\
		\hline
		1     & 0.018 & 1     & 0     & 2     & 0     & 2     & 0 \\
		2     & 0.036 & 1     & 0     & 2     & 0     & 2     & 0 \\
		3     & 0.043 & -     & 1     & 2     & 0     & -     & 1 \\
		4     & 0.045 & 1     & 0     & 2     & 0     & 2     & 0 \\
		5     & 0.074 & -     & 1     & -     & 1     & -     & 1 \\
		6     & 0.119 & 1     & 0     & 2     & 0     & 2     & 0 \\
		7     & 0.131 & -     & 1     & -     & 1     & -     & 1 \\
\hline
	\end{tabular}%
	\label{computer_mask}%
\end{table}%

\begin{table}[htbp]
	\centering
	\tiny
	\caption{Continuation - observed data of $n=172$ computer hard-drive data.}
	\begin{tabular}{c|c|cc|cc|cc}
		\hline
System ID & t & $\delta_1$    & $\upsilon_1$    & $\delta_2$    & $\upsilon_2$     & $\delta_3$ & $\upsilon_3$ \\
		\hline
		93    & 2.410 & 2     & 0     & 1     & 0     & 2     & 0 \\
		94    & 2.430 & -     & 1     & -     & 1     & -     & 1 \\
		95    & 2.490 & 2     & 0     & 2     & 0     & 1     & 0 \\
		96    & 2.520 & -     & 1     & 2     & 0     & -     & 1 \\
		97    & 2.570 & -     & 1     & -     & 1     & -     & 1 \\
		98    & 2.640 & 2     & 0     & 2     & 0     & 1     & 0 \\
		99    & 2.660 & -     & 1     & -     & 1     & -     & 1 \\
\hline
	\end{tabular}%
	\label{computer_mask_cont}%
\end{table}%

\backmatter \singlespacing   
\bibliographystyle{plainnat-ime} 

\index{TBP|see{periodicidade região codificante}}
\index{DSP|see{processamento digital de sinais}}
\index{STFT|see{transformada de Fourier de tempo reduzido}}
\index{DFT|see{transformada discreta de Fourier}}
\index{Fourier!transformada|see{transformada de Fourier}}

\printindex   

\end{document}